\documentclass[aps,prb,twocolumn,superscriptaddress,10pt,longbibliography]{revtex4-2}
\usepackage[dvipdfmx]{graphicx}
\usepackage{amsmath,amsbsy,amssymb}
\usepackage{bm}
\usepackage{mathrsfs}
\usepackage{textgreek}
\usepackage{mathrsfs}
\usepackage{mleftright}
\usepackage{braket}
\usepackage{physics}
\usepackage{hyperref}
\usepackage{mathtools}
\usepackage[percent]{overpic}
\usepackage{verbatim}
\graphicspath{{./figures/}}

\usepackage{color}

\newcommand{\diff}{\mathrm{d}}

\newcommand{\imu}{\mathrm{i}}
\newcommand{\epn}{e}

\newcommand{\ua}{\uparrow}
\newcommand{\da}{\downarrow}
\newcommand{\dg}{\dagger}
\newcommand{\la}{\langle}
\newcommand{\ra}{\rangle}
\newcommand{\al}{\alpha}
\newcommand{\sg}{\sigma}
\newcommand{\gm}{\gamma}

\hypersetup{colorlinks=true,linkcolor=blue,citecolor=blue,urlcolor=blue}

\begin{document}

\title{
Semiclassical representation of the Hubbard model
}

\author{Yuki Yamasaki }
\affiliation{Department of Physics, Saitama University, Saitama 338-8570, Japan}
\affiliation{Department of Physics, Chiba University, Chiba 263-8522, Japan}

\author{Hidemaro Suwa}
\affiliation{Department of Physics, The University of Tokyo, Tokyo 113-0033, Japan}

\author{Cristian D. Batista}
\affiliation{Department of Physics and Astronomy, The University of Tennessee, Knoxville, TN 37996, USA}
\affiliation{Quantum Condensed Matter Division and Shull-Wollan Center, Oak Ridge National Laboratory, Oak Ridge, TN 37831, USA}

\author{Shintaro Hoshino}
\email{shinaro.hoshino@chiba-u.jp}
\affiliation{Department of Physics, Chiba University, Chiba 263-8522, Japan}

\date{\today}

\begin{abstract}

By revisiting the path-integral formulation of the Hubbard model, we propose a theoretical approach based on a semiclassical approximation employing an unconventional coherent-state representation. 
Within this framework, a subset of the dynamical variables is treated as static, yielding a nonperturbative scheme that is applicable at finite temperature, incorporates intersite correlations, and can be naturally extended to multiorbital systems. 
We assess the validity of the approximation by comparing its results with exact solutions for one- and two-site systems, focusing in particular on the particle number, double occupancy, hopping amplitude, and spin correlations, and find that the present approach qualitatively reproduces the exact behavior. 
Quantitatively, deviations arise, which is associated with 
the continuum (non-discretized) character of the underlying density of states.
Furthermore, we derive the exact transformation associated with the coherent-state construction, thereby providing additional insight into the representation of the Hubbard model.

\end{abstract}

\maketitle

\section{Introduction}

As one of the most fundamental models of strongly correlated electron systems, 
the Hubbard model is widely recognized \cite{Arovas22}. 
Its Hamiltonian is given by
\begin{align}
    \mathscr{H} 
    &= - t \sum_{\langle ij \rangle \sigma} 
    \left( c_{i\sigma}^\dagger c_{j\sigma} + \mathrm{H.c.} \right)
    - \mu \sum_{i\sigma} n_{i\sigma}
    + U \sum_i n_{i\uparrow} n_{i\downarrow}
    \label{eq:hubbard_original} \\
    &\equiv \sum_{\langle ij \rangle} \mathscr{H}_t (i,j) 
    + \sum_i \mathscr{H}_{\mathrm{loc}} (i),
    \label{eq:hubbard_original2} 
\end{align}
where $t$ and $U$ denote the transfer integral 
and on-site interaction strength, respectively, 
and $\mu$ is the chemical potential. 
The fermionic operator $c_{i\sigma}$ annihilates an electron with spin $\sigma$ at site $i$, 
and $n_{i\sigma} = c_{i\sigma}^\dagger c_{i\sigma}$ is the corresponding number operator. 
The summation $\langle ij \rangle$ runs over 
pairs of sites. 
In the context of real materials, the Hubbard model in Eq.~\eqref{eq:hubbard_original} 
is often regarded as an effective low-energy model for copper-oxide high-temperature superconductors. 
Despite its apparent simplicity, it exhibits a remarkably rich variety of correlation-driven quantum phenomena, 
including Fermi-liquid behavior, magnetism, the Mott metal–insulator transition, and superconductivity. 
However, as a genuine quantum many-body problem, obtaining quantitatively reliable solutions, either analytically or numerically, is notoriously challenging. 
Consequently, a wide range of approximate approaches has been developed, each providing complementary insights into its physics.

Standard textbooks on quantum mechanics typically introduce three basic approximation schemes: perturbation theory, the variational method, and the semiclassical approximation \cite{Griffiths_book}. 
These methods are highly versatile and are established as fundamental methodologies across many areas of physics. 
In the context of the Hubbard model \cite{Fazekas_book}, perturbative approaches involve expansions either in the interaction strength $U$ or in the hopping amplitude $t$. 
The former corresponds to an expansion around the noninteracting Fermi-gas limit, while the latter amounts to a strong-coupling expansion about the atomic (localized-electron) limit. 
The variational approach assumes a trial many-body wave function containing variational parameters, which are determined by minimizing the energy expectation value. 
By contrast, the semiclassical approximation for the Hubbard model is typically formulated within the path-integral representation, where the imaginary-time fluctuations of the auxiliary bosonic fields introduced via the Hubbard-Stratonovich (HS) transformation are neglected \cite{Okamoto05,Lee08,Lee13,Park23,Dubi07,Schulz90,Mukherjee14,Chern18}. 
These three approaches are computationally less demanding and, owing to their clear underlying principles, admit systematic refinements through the incorporation of fluctuations or higher-order corrections \cite{Auerbach_book}.

In the semiclassical approximation, however, there is some arbitrariness in the way the HS field is introduced \cite{Castellani79}, which can affect the physical states that can be described. This issue becomes particularly nontrivial in multi-orbital systems, where the choice of auxiliary field is not obvious.
In this work, we propose an alternative formulation of the semiclassical approximation based on an unconventional coherent-state representation. 
The construction is guided by a simple principle: minimizing the use of Grassmann variables. 
As a consequence, the resulting scheme is nonperturbative, applicable at finite temperature, and readily extendable to multiorbital systems, without imposing 
further 
physical constraints at the level of approximation. 
An additional
advantage of this construction is that it naturally provides
semiclassical degrees of freedom for both the spin and charge sectors
of the local Hilbert space.  As a consequence, the resulting framework
is well suited for describing intertwined orders involving simultaneous
spin and charge symmetry breaking, such as the coexistence of magnetic
order with superconducting or charge-density-wave correlations.
Although the original Hamiltonian in Eq.~\eqref{eq:hubbard_original} contains only single-particle hopping between sites, the present representation reformulates the intersite hopping term in terms of effective many-body interactions.

Our construction of the semiclassical scheme may be viewed as an extension of path-integral formalisms developed for the $t$--$J$ model \cite{Wiegmann88,Ferraz11} and for multiorbital Mott insulators \cite{Zhang21,Dahlbom22,Iwazaki23,Pohle25}, where either doubly occupied states or fermionic degrees of freedom are projected out of the Hilbert space. 
In contrast, the present formulation applies directly to the Hubbard model, in which both doubly occupied states and fermionic degrees of freedom are retained.

\begin{figure}[t]
    \centering
    \includegraphics[width=85mm]{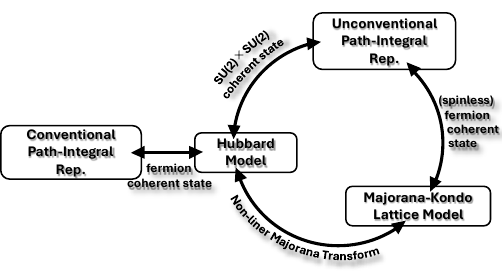}
    \caption{Overview on the represenations of the Hubbard model considered in this paper.}
    \label{fig:overview}
\end{figure}

We benchmark the performance of the method by comparing it with exact solutions for one- and two-site systems. 
The results qualitatively reproduce the exact behavior. 
In particular, at half filling in the two-site system, where the low-energy sector reduces effectively to a Heisenberg spin model, our approach yields results corresponding to the classical-spin limit. 
Furthermore, by deriving the operator-form Hamiltonian corresponding to the path-integral representation constructed from the unconventional coherent states, we demonstrate its equivalence to a nonlinear fermionic transformation \cite{Ostlund91,Ostlund92,Ostlund06,Ostlund07,Kumar08,Scharnhorst11,Bazzanella14,Shinjo21}. 
A schematic overview of the model transformations encountered in this paper is shown in Fig.~\ref{fig:overview}.

The following of this paper is organized as follows. In Sec.~\ref{sec:path-int}, we introduce the path-integral formulation based on an unconventional coherent-state representation. In Sec.~\ref{sec:semiclassical}, we develop the semiclassical approximation. Sections~\ref{sec:one_site} and \ref{sec:two_site} present numerical results for the one-site and two-site models, respectively. 
We show
that the semiclassical approximation emerges as the large-$M$ limit of
a generalized Hubbard model in Sec.~\ref{sec:controlled_semiclassical_limit}, and then describe how the coherent-state
construction generalizes to multiorbital systems in Sec.~\ref{sec:multiorb}.
In Sec.~\ref{sec:mapping}, we demonstrate the mapping of the Hubbard model onto a Kondo-lattice-like model composed of itinerant fermions and localized spins. 
Finally, Sec.~\ref{sec:summary} summarizes the paper. 
The Appendix~\ref{sec:solution_one_body} describes the method for solving the generic one-body problems encountered in evaluating partition functions.

\section{Path-integral formalism}
\label{sec:path-int}

\subsection{Conventional path-integral representation of the Hubbard model}

Before introducing the unconventional approach, we briefly review the conventional path-integral formulation \cite{Negele_book}. 
The standard fermionic coherent state at site $i$ is defined as
\begin{align}
    |\psi_i \rangle = \prod_{\sigma} e^{- \psi_{i\sigma} c_{i\sigma}^\dagger } |0\rangle,
    \label{eq:coherent_conventional}
\end{align}
where $\psi_{i\sigma}$ is a Grassmann variable and $\bar\psi_{i\sigma}$ denotes its conjugate. 
The partition function is then written as
\begin{align}
    Z &= \int \Big( \prod_{i\sigma} \mathcal D \bar\psi_{i\sigma} \mathcal D \psi_{i\sigma} \Big)
    e^{-S},
\end{align}
with the imaginary-time action
\begin{align}
    S &= \int \diff \tau \Bigg[
    \sum_{i\sigma} \bar\psi_{i\sigma} (\partial_\tau - \mu) \psi_{i\sigma}
    - t \sum_{\langle ij\rangle \sigma}
    (\bar\psi_{i\sigma} \psi_{j\sigma} + \bar\psi_{j\sigma} \psi_{i\sigma})
    \nonumber \\
    &\qquad\qquad
    + U \sum_i
    \bar\psi_{i\uparrow} \bar\psi_{i\downarrow}
    \psi_{i\downarrow} \psi_{i\uparrow}
    \Bigg].
\end{align}
The quartic Grassmann term arising from the interaction renders the functional integral nontrivial, and approximations are generally required. 
To this end, one introduces auxiliary Hubbard-Stratonovich (HS) fields,
\begin{align}
    Z &\propto
    \int \Big( \prod_i \mathcal D \bm b_i \Big)
    \Big( \prod_{i\sigma} \mathcal D \bar\psi_{i\sigma} \mathcal D \psi_{i\sigma} \Big)
    e^{-S_{\rm eff}},
\end{align}
where an unimportant proportionality constant has been omitted and we have assumed $U>0$. 
The resulting effective action, which depends on both Grassmann and complex fields, reads
\begin{align}
    S_{\rm eff} &= 
    \int \diff \tau \Bigg[
    \sum_{i\sigma\sigma'} 
    \bar\psi_{i\sigma}
    \Big\{
    (\partial_\tau - \mu)\delta_{\sigma\sigma'}
    - \sqrt{\frac{U}{3}}\, \bm b_i \cdot \bm \sigma_{\sigma\sigma'}
    \Big\}
    \psi_{i\sigma'}
    \nonumber \\
    &\qquad
    - t\sum_{\langle ij\rangle \sigma}
    (\bar\psi_{i\sigma} \psi_{j\sigma} + \bar\psi_{j\sigma} \psi_{i\sigma})
    + \frac{1}{2} \sum_i \bm b_i^2
    \Bigg].
\end{align}
Although a fully quantum-mechanical treatment of this representation remains challenging, a substantial simplification arises if one focuses on static correlations and neglects the imaginary-time dependence of the auxiliary fields, i.e., $\bm b_i(\tau) \simeq \bm b_i$. 
Under this static approximation, the problem reduces to a one-body Hamiltonian subject to configurations of the bosonic fields $\{\bm b_i\}$ distributed according to a Gaussian weight. 
Such approaches have been explored in Refs.~\cite{Okamoto05,Lee08,Lee13,Park23}, and in Ref.~\cite{Dubi07} for the attractive Hubbard model with pairing auxiliary fields. 
Because the HS field introduced here is local, its static treatment effectively acts as a site-dependent magnetic field, analogous to a disorder potential arising from magnetic impurities. 
Consequently, this approximation does not capture nonlocal correlations required, for example, to describe $d$-wave superconductivity.
This limitation motivates us to reconsider the construction of the coherent-state representation.

\subsection{
Graded coherent states with 
\texorpdfstring{$SU(2)_{\eta}\times SU(2)_{\mathrm{spin}}$}
{SU(2)eta times SU(2)spin}
structure
}

Having established that the standard semiclassical coherent-state path-integral
representation of the Hubbard model does not readily capture physics
arising from intersite correlations, we now consider an alternative
construction adapted to the graded structure of the local Hilbert
space.  The key idea is to represent the four local states
$\{|0\rangle,|\uparrow\rangle,|\downarrow\rangle,|\uparrow\downarrow\rangle\}$
by a coherent state containing a single Grassmann coordinate that
mediates transitions between the even- and odd-parity sectors.
The bosonic degrees of freedom encode the orientation of the spin and
charge sectors, while the Grassmann variable keeps track of the
fermion parity.  This construction reduces the number of Grassmann
variables compared with the conventional fermionic coherent state
and leads to a path-integral representation in which the interaction
term becomes bilinear in the Grassmann fields.  We therefore introduce
a generalized coherent state at site $i$ defined as
\begin{align}
    |\Omega_i,\psi_i\rangle 
    &= C_1(\Omega_i) |0\rangle 
    - C_2(\Omega_i)\, \psi_i c_{i\uparrow}^\dagger |0\rangle 
    \nonumber \\
    &\quad 
    - C_3(\Omega_i)\, \psi_i c_{i\downarrow}^\dagger |0\rangle
    + C_4(\Omega_i)\, c_{i\uparrow}^\dagger c_{i\downarrow}^\dagger |0\rangle
    \nonumber \\
    &\equiv |\alpha_i\rangle + \psi_i |\beta_i\rangle .
    \label{eq:generic}
\end{align}
Here, the complex coefficients $C_\alpha(\Omega_i)$ depend on a set of parameters collectively denoted by $\Omega_i$, whose explicit form will be specified later. 
Importantly, the Grassmann variable $\psi_i$ carries no spin index. 
We have introduced the shorthand notation $|\alpha_i\rangle$ and $|\beta_i\rangle$, which separate the even- and odd-fermion-parity sectors: $|\alpha_i\rangle$ consists of states with an even number of fermions, whereas $|\beta_i\rangle$ consists of states with an odd number of fermions. 
The conjugate state is also defined as
\begin{align}
    \langle \Omega_i,\psi_i| 
    &= \langle 0| C_1^*(\Omega_i)
    - \langle 0| c_{i\uparrow} \bar\psi_i C_2^*(\Omega_i)
    \nonumber \\
    &\quad 
    - \langle 0| c_{i\downarrow} \bar\psi_i C_3^*(\Omega_i)
    + \langle 0| c_{i\downarrow} c_{i\uparrow} C_4^*(\Omega_i).
\end{align}
From these definitions, we see that $C_1$ and $C_4$ correspond to the holon and doublon amplitudes, respectively, while $C_2$ and $C_3$ correspond to singly occupied states with spin $\uparrow$ and $\downarrow$.

The present coherent state satisfies a useful relation under the action of the annihilation operator of the composite fermions introduced in Ref.~\cite{Mancini04}.
The operators are defined as
\begin{align}
    \xi_{i\sigma} &= (1 - n_{i\bar\sigma}) c_{i\sigma},
    \\
    \eta_{i\sigma} &= n_{i\bar\sigma} c_{i\sigma},
\end{align}
where $\bar\sigma$ denotes the spin opposite to $\sigma$. 
The fermionic operator can then be decomposed as
\begin{align}
    c_{i\sigma} = \xi_{i\sigma} + \eta_{i\sigma}.
    \nonumber
\end{align}
Within this formulation, the following relations hold:
\begin{align}
    \xi_{i\sigma} |\alpha_i\rangle = 0,
    \ \ \ 
    \eta_{i\sigma} |\beta_i\rangle = 0.
\end{align}
These properties follow directly from the definitions of the composite fermion operators.
The operator $\eta_{i\sigma}$ annihilates an electron on a doubly occupied site and thus describes the degrees of freedom associated with the upper Hubbard band in the Mott-insulating regime \cite{Mancini04}.
In contrast, $\xi_{i\sigma}$ annihilates an electron on a singly occupied site and corresponds to the lower Hubbard band.
Consequently, $\xi_{i\sigma}$ acts trivially on the even-parity sector $|\alpha_i\rangle$, while $\eta_{i\sigma}$ acts trivially on the odd-parity sector $|\beta_i\rangle$.

Next, we consider the integral over the Grassmann variables.
A path-integral representation can be constructed once a resolution of the identity is available. 
We therefore postulate the following
resolution of unity
\begin{align}
    1 &= \mathcal C \int \diff \Omega_i \, \diff \bar\psi_i \, \diff \psi_i \,
    e^{-\bar\psi_i \psi_i}
    |\Omega_i, \psi_i\rangle 
    \langle \Omega_i, \psi_i |,
    \label{eq:resolve_unity}
\end{align}
where $\mathcal C$ is a normalization constant. 
The explicit parametrization of $\Omega_i$ is given in terms of, e.g., an $SU(4)$ coherent state with $|C_1|^2\!+\!|C_2|^2\!+\!|C_3|^2\!+\!|C_4|^2\! = \!1$ \cite{Nemoto00,Zhang21,Iwazaki23}. 
The path-integral formulation then follows by inserting the resolution of unity in Eq.~\eqref{eq:resolve_unity} at each imaginary-time slice.

The $SU(4)$ coherent state, however, leads to a complicated path-integral representation due to its normalization depending on $\Omega_i$.
To render the resulting path integral analytically and numerically tractable, we further impose the normalization condition
\begin{align}
    \langle \Omega_i, \psi_i | \Omega_i, \psi_i \rangle 
    = e^{\bar\psi_i \psi_i},
\end{align}
which fixes the normalization of the coefficients $C_\alpha(\Omega_i)$:
\begin{align}
    |C_1(\Omega_i)|^2 \!+\! |C_4(\Omega_i)|^2 
    = |C_2(\Omega_i)|^2 \!+\! |C_3(\Omega_i)|^2 
    = 1.
\end{align}
This structure reflects the $SU(2)_{\eta}\times SU(2)_{\rm spin}$
symmetry of the half filled single-orbital Hubbard model, corresponding to the
doublon–holon (charge) and spin sectors, respectively. Correspondingly, the bosonic coordinates can be written as
$\Omega_i=(\Omega_i^{c},\Omega_i^{s})$, which parametrize two
independent Bloch spheres associated with the charge and spin sectors,
so that the bosonic manifold is $S_c^2\times S_s^2$.
Under this condition, we obtain the orthonormal relations
\begin{equation}
\begin{aligned}
    \langle \alpha_i | \alpha_i \rangle &= 1, 
    \qquad 
    \langle \beta_i | \beta_i \rangle = 1,
    \\
    \langle \alpha_i | \beta_i \rangle &= 0,
    \qquad
    \langle \beta_i | \alpha_i \rangle = 0,
\end{aligned}
\end{equation}
which simplify subsequent analytical calculations.

It is useful to rewrite the coherent state in Eq.~(8) in an exponential form that makes explicit its interpretation as the action of local generators on the vacuum. On the patch $C_{i1}(\Omega_i)\neq 0$, one can write
\begin{widetext}
\begin{equation}
\label{eq:coherent_exponential_i}
|\Omega_i,\psi_i\rangle
=
C_{i1}(\Omega_i)\,
\exp\!\left[
\frac{C_{i4}(\Omega_i)}{C_{i1}(\Omega_i)}\,c_{i\uparrow}^\dagger c_{i\downarrow}^\dagger
-
\frac{\psi_i}{C_{i1}(\Omega_i)}
\Big(
C_{i2}(\Omega_i)\,c_{i\uparrow}^\dagger
+
C_{i3}(\Omega_i)\,c_{i\downarrow}^\dagger
\Big)
\right]|0\rangle_i .
\end{equation}
\end{widetext}
Because $(c_{i\uparrow}^\dagger c_{i\downarrow}^\dagger)^2=0$ and $\psi_i^2=0$, the exponential truncates and reproduces Eq.~\eqref{eq:generic} exactly.

It is also convenient to express the same state in a form that makes contact with a generalized construction that will be presented in Sec.~\ref{sec:multiorb}. Introducing the normalized spinor
$
\theta_{i\sigma} = C_{i\sigma}(\Omega_i)/\sqrt{|C_{i2}|^2+|C_{i3}|^2}
$
with $\sigma=\uparrow,\downarrow$, and defining $z_i=C_{i4}/C_{i1}$, the coherent state can be written as
\begin{equation}
|\Omega_i,\psi_i\rangle
=
C_{i1}(\Omega_i)\,
\exp\!\left(
z_i\, c_{i\uparrow}^\dagger c_{i\downarrow}^\dagger
+
\psi_i\,\theta_{i}\,c_{i}^\dagger
\right) |0\rangle_i
, \label{eq:coherentnew}
\end{equation}
where the scalar product of the spinors is given by $\theta_i c_i^\dg = \sum_\sg \theta_{i\sg} c_{i\sg}^\dg$. 

To summarize, a generic normalized state of the local Hilbert space, Eq.~\eqref{eq:generic}, need not be a standard spin-dependent fermionic coherent state of a graded algebra.  Coherent states instead form a particular submanifold of the projective Hilbert space corresponding to the orbit of a reference state under the action of the algebra.  For the four-dimensional local Hilbert space of the single-orbital Hubbard model, a completely general normalized state modulo an overall phase is parametrized by the complex projective space $\mathbb{CP}^3$.  The coherent states introduced above, however, do not span this entire space.  Rather, they are constructed to respect the $\mathbb Z_2$ grading associated with fermion parity, which separates the local Hilbert space into even and odd sectors connected by a single Grassmann coordinate.  The bosonic parameters therefore describe independent rotations within the even (doublon–holon) and odd (spin) sectors, each of which is two-dimensional and naturally associated with a Bloch sphere.  Consequently, the bosonic manifold of the coherent states reduces from $\mathbb{CP}^3$ to the product $S^2_c \times S^2_s$.  In this way, the fermionic nature of the problem—encoded in the $\mathbb Z_2$ parity grading of the Hubbard Hilbert space—selects a graded coherent-state manifold adapted to the spin–charge structure of the local degrees of freedom.

\subsection{Partition function}

Now we derive the path-integral representation of the partition function.
We consider the Suzuki-Trotter decomposition as
\begin{align}
    Z = \lim_{N\to \infty} {\rm Tr\,} \prod_{n=1}^N \epn^{-\varDelta \tau \mathscr H}
\end{align} 
where $\varDelta \tau = \beta /N$.
We insert the resolution of unity and obtain
\begin{widetext}
\begin{align}
    &Z = \int \qty( \prod_{i}\prod_{n=1}^N \mathcal C \diff \Omega_{in} \diff \bar\psi_{in} \diff \psi_{in} \epn^{-\bar \psi_{in} \psi_{in}}  
    \la \Omega_{in},\psi_{in} | \Omega_{i,n-1},\psi_{i,n-1} \ra  ) 
    \nonumber \\
    &\times
    \exp \bigg( 
    - \varDelta\tau \sum_{in}  \frac{\la \Omega_{in},\psi_{in} | \mathscr H_{\rm loc}(i) | \Omega_{i,n-1},\psi_{i,n-1} \ra }{
    \la \Omega_{in},\psi_{in} | \Omega_{i,n-1},\psi_{i,n-1} \ra
    }
    - \varDelta\tau \sum_{\la ij\ra n} \frac{ \la \Omega_{jn},\psi_{jn} | \la
    \Omega_{in},\psi_{in} |  \mathscr H_{t}(i,j) | \Omega_{i,n-1},\psi_{i,n-1} \ra | \Omega_{j,n-1},\psi_{j,n-1} \ra 
    }{ \la \Omega_{in},\psi_{in} | \Omega_{i,n-1},\psi_{i,n-1} \ra  \la \Omega_{jn},\psi_{jn} | \Omega_{j,n-1},\psi_{j,n-1} \ra }
    \bigg)
\end{align}
The terms involving inner product are rewritten as
\begin{align}
    &\epn^{-\bar \psi_{in} \psi_{in}} \la \Omega_{in},\psi_{in} | \Omega_{i,n-1},\psi_{i,n-1} \ra
= 
    \exp \bigg( \ln \la \al_{in} | \al_{i,n-1} \ra - \bar \psi_{in} (\psi_{in} - \psi_{i,n-1} ) 
 + \bar \psi_{in} \psi_{i,n-1}  \frac{\la \beta_{in} | \beta_{i,n-1} \ra - \la \al_{in} | \al_{i,n-1} \ra}{ \la \al_{in} | \al_{i,n-1} \ra } \bigg)
 ,
\\
&\la \Omega_{in},\psi_{in} | \Omega_{i,n-1},\psi_{i,n-1} \ra^{-1}
=   \la \al_{in} | \al_{i,n-1} \ra^{-1} 
    \bigg( 1
 - \bar \psi_{in} \psi_{i,n-1}  \frac{\la \beta_{in} | \beta_{i,n-1} \ra}{ \la \al_{in} | \al_{i,n-1} \ra } \bigg)
 .
\end{align}
The matrix elements of the local and intersite Hamiltoinains defined in Eq.~\eqref{eq:hubbard_original2} are also evaluated as
\begin{align}
    &\la \Omega_{in},\psi_{in} | \mathscr H_{\rm loc}(i) | \Omega_{i,n-1},\psi_{i,n-1} \ra
    = 
    -\mu \bar \psi_{in}\psi_{i,n-1} \big[ C_2^*(\Omega_{in})C_2(\Omega_{i,n-1})+
    C_3^*(\Omega_{in})C_3(\Omega_{i,n-1}) \big]
    +(U-2\mu) C_4^*(\Omega_{in})C_4(\Omega_{i,n-1})
    ,
    \\
    & \la \Omega_{jn},\psi_{jn} | \la \Omega_{in},\psi_{in} |  \mathscr H_{t}(i,j) | \Omega_{i,n-1},\psi_{i,n-1} \ra | \Omega_{j,n-1},\psi_{j,n-1} \ra 
    \nonumber \\
    &= -t \sum_\sg
     \Big( \la \al_{in} | c_{i\sg}^\dg |\beta_{i,n-1}\ra  \psi_{i,n-1} 
     +  \la \beta_{in} | c_{i\sg}^\dg |\al_{i,n-1}\ra \bar \psi_{in} \Big) 
     \Big( \la \al_{jn} | c_{j\sg} |\beta_{j,n-1}\ra  \psi_{j,n-1} 
     +  \la \beta_{jn} | c_{j\sg} |\al_{j,n-1}\ra \bar \psi_{jn} \Big) 
      + {\rm conj.}
\end{align}
\end{widetext}
In the above, We do not use any approximation.

We now take the continuum limit of the Suzuki-Trotter decomposition, retaining only the slowly varying components. 
It can be shown that, even after this procedure, the original Hubbard model is recovered as demonstrated in Sec.~\ref{sec:mapping}.
We assume that coherent states at neighboring imaginary-time slices overlap almost perfectly, namely,
$\langle \alpha_{i n} | \alpha_{i,n-1} \rangle \simeq 1$
and 
$\langle \beta_{i n} | \beta_{i,n-1} \rangle \simeq 1
$.
Expanding the overlap to leading order in $\varDelta\tau$, we obtain
\begin{align}
    \langle \alpha_{i n} | \alpha_{i,n-1} \rangle 
    \simeq 
    \exp\!\left[
    - \varDelta\tau \,
    \langle \alpha_i | \partial_\tau | \alpha_i \rangle
    + O(\varDelta\tau^2)
    \right],
\end{align}
where we have introduced the imaginary-time representation $\tau = n \varDelta\tau$. 
An analogous expression holds for 
$\langle \beta_{i n} | \beta_{i,n-1} \rangle$.

Taking the continuum limit, the partition function becomes
\begin{align}
    Z \sim \int 
    \left( \prod_i \mathcal D \Omega_i 
    \mathcal D \bar\psi_i 
    \mathcal D \psi_i \right)
    e^{-\mathscr S_{\rm eff}},
\end{align}
with the effective action
\begin{align}
    \mathscr S_{\rm eff} 
    &=  \sum_i \int_0^\beta \diff \tau 
    \Big[
    \langle \alpha_i | \partial_\tau | \alpha_i \rangle 
    + \bar\psi_i \partial_\tau \psi_i
    \nonumber \\
    &\qquad+  \bar\psi_i \psi_i 
    \big(
    \langle \alpha_i | \partial_\tau | \alpha_i \rangle 
    - \langle \beta_i | \partial_\tau | \beta_i \rangle
    \big)
    \nonumber \\
    &\qquad
    - \mu \bar\psi_i \psi_i 
    + (U - 2\mu) (1 - \bar\psi_i \psi_i) 
    |C_4(\Omega_i)|^2
    \Big] 
    \nonumber \\
    &\hspace{-5mm}
    - t \int_0^\beta \diff \tau 
    \sum_{\langle ij \rangle \sigma}
    \Big[
    (Y_{i\sigma}^* \psi_i + X_{i\sigma}^* \bar\psi_i)
    (X_{j\sigma} \psi_j + Y_{j\sigma} \bar\psi_j)
    + \text{c.c.}
    \Big].
    \label{eq:Seff}
\end{align}
Here we have defined
\begin{align}
&\langle \alpha_i | \partial_\tau | \alpha_i \rangle 
=
C_1^*(\Omega_i) \partial_\tau C_1(\Omega_i)
+ C_4^*(\Omega_i) \partial_\tau C_4(\Omega_i),
\\
&\langle \beta_i | \partial_\tau | \beta_i \rangle 
=
\sum_{\sigma}
C_\sigma^*(\Omega_i) 
\partial_\tau C_\sigma(\Omega_i),
\\
&X_{i\sigma} 
=
\langle \alpha_i | c_{i\sigma} | \beta_i \rangle
=
\langle \alpha_i | \xi_{i\sigma} | \beta_i \rangle
=
C_1^*(\Omega_i) C_\sigma(\Omega_i),
\\[1mm]
&Y_{i\sigma}
=
\langle \beta_i | c_{i\sigma} | \alpha_i \rangle
=
\langle \beta_i | \eta_{i\sigma} | \alpha_i \rangle
=
(-1)^{\sigma}
C_{\bar\sigma}^*(\Omega_i) C_4(\Omega_i).
\end{align}
We have introduced the notation $C_\uparrow = C_2$ and $C_\downarrow = C_3$, with 
$(-1)^{\uparrow} = +1$ and 
$(-1)^{\downarrow} = -1$.
Importantly, the resulting action is bilinear in the spinless Grassmann variables, in contrast to the conventional path-integral formulation of the Hubbard model. 
Note that no approximation has been made for the Grassmann fields themselves; in particular,
$\bar\psi_i \partial_\tau \psi_i$ 
should be understood as the continuum notation corresponding to 
$\bar\psi_{i n} (\psi_{i n} - \psi_{i,n-1})/\varDelta\tau$ in the discretized representation.

\subsection{Physical quantities}

We represent physical operators such as 
$c_{i\uparrow}^\dagger c_{j\downarrow}$ ($i \neq j$) 
by $\mathscr{A}_{ij}$. 
Local operators of the form 
$c_{i\sigma}^\dagger c_{i\sigma}$ 
are included in the same notation by formally suppressing the site index $j$. 
The expectation value of $\mathscr{A}_{ij}$ can be evaluated within the functional-integral formalism through the auxiliary quantity
\begin{align}
    A_{ij}[\bm{\Omega}, \bm{\psi}] 
    = 
    \frac{
    \langle \Omega_j \psi_j | 
    \langle \Omega_i \psi_i | 
    \mathscr{A}_{ij} 
    | \Omega_i \psi_i \rangle 
    | \Omega_j \psi_j \rangle
    }{
    \langle \Omega_i \psi_i | \Omega_i \psi_i \rangle
    \langle \Omega_j \psi_j | \Omega_j \psi_j \rangle
    } ,
    \label{eq:phys_quant_gen}
\end{align}
where we have introduced the shorthand notations 
$\bm{\Omega} = \{ \Omega_i \} = \{ \Omega_i^s, \Omega_i^c \}$ and 
$\bm{\psi} = \{ \psi_i \}$. 
For brevity, the imaginary-time dependence of these variables is suppressed.

For example, the local number operator 
$n_{i\sigma} = c_{i\sigma}^\dagger c_{i\sigma}$ 
is represented as
\begin{align}
    n_{i\sigma}[\bm{\Omega}, \bm{\psi}] 
    = \bar{\psi}_i \psi_i \, |C_\sigma(\Omega_i)|^2
    + \left(1 - \bar{\psi}_i \psi_i \right) |C_4(\Omega_i)|^2 .
    \label{eq:def_of_number}
\end{align}
The spin-raising operator 
$S_i^{+} = c_{i\uparrow}^\dagger c_{i\downarrow}$ 
and the $z$-component 
$S_i^z = (n_{i\ua} - n_{i\da})/2$
are represented as
\begin{align}
    S_i^{+}[\bm{\Omega}, \bm{\psi}] 
    &= C_\uparrow^*(\Omega_i) C_\downarrow(\Omega_i) \,
    \bar{\psi}_i \psi_i 
    ,
    \label{eq:def_splus}
    \\
    S_i^z [\bm{\Omega}, \bm{\psi}]
    &= \frac{1}{2} \big( |C_{\ua}(\Omega_i)|^2 - |C_{\da}(\Omega_i)|^2 \big) \bar \psi_i \psi_i
    .
    \label{eq:def_sz}
\end{align}
The intersite spin correlation $\bm S_i\cdot \bm S_j$ can be evaluated by the combinations of the above representations for spin.
The double-occupancy operator 
$D_i = n_{i\uparrow} n_{i\downarrow}$ 
takes the form
\begin{align}
    D_i[\bm{\Omega}, \bm{\psi}] 
    = \left(1 - \bar{\psi}_i \psi_i \right) 
    |C_4(\Omega_i)|^2 .
    \label{eq:def_of_Docc}
\end{align}
Similarly, the holon (vacant-state) occupancy 
$H_i = (1 - n_{i\uparrow})(1 - n_{i\downarrow})$ 
is given by
\begin{align}
    H_i[\bm{\Omega}, \bm{\psi}] 
    = \left(1 - \bar{\psi}_i \psi_i \right) 
    |C_1(\Omega_i)|^2 .
\end{align}
One can also consider the local pair operator 
$P_{{\rm loc},i} = c_{i\uparrow}^\dagger c_{i\downarrow}^\dagger$, which
is expressed in the path-integral formulation as
\begin{align}
    P_{{\rm loc},i}[\bm{\Omega}, \bm{\psi}] 
    = \left(1 - \bar{\psi}_i \psi_i \right) 
    C_4^*(\Omega_i) C_1(\Omega_i) .
\end{align}
This quantity becomes relevant for the attractive interaction $U<0$.
The representation for the intersite hopping amplitude $c_{i\sigma}^\dagger c_{j\sigma}$ has already been formulated as the term proportional to $t$ in Eq.~\eqref{eq:Seff}.

\subsection{Symmetry operations}
\label{sec:sym_op}

Here we summarize the symmetry operations within our unconventional coherent-state representation. 
The local $\mathrm{U}(1)$ gauge transformation of the electron operator is defined as
\begin{align}
    \tilde c_{i\sigma} = e^{i \theta_i} c_{i\sigma} .
\end{align}
In the path-integral representation, this transformation is expressed as
\begin{align}
    C_\sg(\tilde \Omega_i) &= e^{i \theta_i} C_\sg(\Omega_i) ,
    \\
    C_4(\tilde \Omega_i) &= e^{2 i \theta_i} C_4(\Omega_i) .
\end{align}
Next, the spin rotation is defined by
\begin{align}
    \tilde c_{i\sigma} = \sum_{\sg'} U_{i\sigma\sigma'} c_{i\sigma'} ,
\end{align}
where $U_i$ is a unitary matrix acting in spin space. 
The corresponding transformation in the path-integral representation reads
\begin{align}
    C_\sigma(\tilde \Omega_i) 
    = \sum_{\sg'} U_{i\sigma\sigma'} C_{\sigma'}(\Omega_i) .
\end{align}
Thus, the gauge and spin degrees of freedom of the original electron operator are inherited by 
$C_\sigma(\Omega_i)$ and $C_4(\Omega_i)$.

\section{Semiclassical approximation}
\label{sec:semiclassical}

The partition function and action introduced in the previous subsection 
still describe a genuine quantum many-body problem, whose exact solution 
remains highly nontrivial. 
Here we introduce a semiclassical approximation for the bosonic variables,
\begin{align}
    \Omega_i(\tau) \longrightarrow \Omega_i ,
\end{align}
i.e., the imaginary-time dependence, which encodes quantum fluctuations, 
is neglected. 
This semiclassical approximation can be justified in a manner similar to the large-$S$ expansion of the quantum spin model, as discussed in Sec.~\ref{sec:controlled_semiclassical_limit}.

Within this approximation, the partition function reduces to
\begin{align}
    Z_{\rm cl} 
    = \int \left( \prod_i \diff \Omega_i 
    \mathcal{D}\bar{\psi}_i \mathcal{D}\psi_i \right)
    e^{-S_{\rm cl}} ,
\end{align}
where the semiclassical action is given by
\begin{align}
    S_{\rm cl} 
    &= \beta (U - 2\mu) \sum_i |C_4(\Omega_i)|^2
    \nonumber \\
    &\quad
    + \sum_i \int_0^\beta \diff \tau 
    \Big[
    \bar{\psi}_i \partial_\tau \psi_i
    - \Big(
    \mu + (U - 2\mu)|C_4(\Omega_i)|^2
    \Big)
    \bar{\psi}_i \psi_i
    \Big]
    \nonumber \\
    &\quad
    - t \int_0^\beta \diff \tau 
    \sum_{\langle ij \rangle \sigma}
    \Big[
    (Y_{i\sigma}^*\psi_i + X_{i\sigma}^* \bar{\psi}_i)
    (X_{j\sigma}\psi_j + Y_{j\sigma} \bar{\psi}_j)
    \nonumber \\
    &\hspace{28mm}
    +
    (Y_{j\sigma}^*\psi_j + X_{j\sigma}^* \bar{\psi}_j)
    (X_{i\sigma}\psi_i + Y_{i\sigma} \bar{\psi}_i)
    \Big] .
    \label{eq:def_hop}
\end{align}
Since the Grassmann number part of the action is bilinear, 
the partition function can be evaluated exactly within the Hamiltonian formalism. 
For a fixed spatial configuration 
$\bm{\Omega} = \{ \Omega_i \}$, 
the resulting effective one-body Hamiltonian reads
\begin{align}
    H_{\rm eff}[\bm{\Omega}, \bm{d}]
    &= - \sum_i 
    \Big[
    \mu + (U - 2\mu) |C_4(\Omega_i)|^2
    \Big]
    d_i^\dagger d_i
    \nonumber \\
    &\quad
    + \sum_{\langle ij \rangle}
    \left(
    T_{ij} d_i^\dagger d_j
    + \Delta_{ij} d_i^\dagger d_j^\dagger
    \right)
    + \mathrm{H.c.},
\end{align}
where $d_i$ annihilates a spinless fermion at site $i$. 
The transfer matrix 
is defined as
\begin{align}
    T_{ij} 
    &= -t \sum_\sigma 
    \left(
    X_{i\sigma}^* X_{j\sigma}
    - Y_{i\sigma} Y_{j\sigma}^*
    \right)
    = T_{ji}^*
    \nonumber \\
    &= -t \sum_\sigma 
    \left(
    C_{1i} C_{1j}^* C_{\sigma i}^* C_{\sigma j}
    - C_{4i} C_{4j}^* C_{\sigma i}^* C_{\sigma j}
    \right) ,
    \label{eq:def_of_T_ij}
\end{align}
and pairing potential as
\begin{align}
    &\Delta_{ij} = -t \sum_\sg (X_{i\sg}^*Y_{j\sg} - Y_{i\sg} X_{j\sg}^* ) = -\Delta_{ji}    
    \nonumber \\
    &= -t \sum_\sg(-1)^\sg (C_{\sg i}^*C_{1i} C^*_{\bar \sg j}C_{4 j} - C^*_{\sg j}C_{1j}C_{\bar \sg i}^* C_{4 i})
    . \label{eq:def_of_Delta_ij}
\end{align}
The partition function is then given by 
\begin{align}
    Z_{\rm cl} &= \int \diff \bm \Omega \, \epn^{-\beta (U-2\mu) \sum_i|C_4(\Omega_i)|^2} {\rm Tr\,} \epn^{-\beta H_{\rm eff}[\bm \Omega,\bm d]}
    ,
\end{align}
where the trace ${\rm Tr}$ is taken for the fermions $\bm d = \{d_{i}\}$.
We introduce the shorthand notation 
$C_{\sigma i} \equiv C_{\sigma}(\Omega_i)$; 
the same convention applies to $C_{1i}$ and $C_{4i}$. 
Numerically, the partition function can be evaluated by solving 
the resulting one-body eigenvalue problem (see Appendix~\ref{sec:solution_one_body}). 
Therefore, no error associated with a Suzuki–Trotter decomposition arises.

A physical quantity represented by the operator $\mathscr A$ is generally expressed as
\begin{align}
    \langle \mathscr{A} \rangle 
    &= \frac{1}{Z_{\rm cl}} 
    \int \diff \bm{\Omega} \,
    \big\langle A[\bm{\Omega}, \bm{d}] \big\rangle_d
    \nonumber \\
    &\quad \times 
    e^{-\beta (U - 2\mu) \sum_i |C_4(\Omega_i)|^2}
    \,
    \mathrm{Tr}\,
    e^{-\beta H_{\rm eff}[\bm{\Omega}, \bm{d}]} ,
\end{align}
where the average with respect to the effective Hamiltonian is defined as
\begin{align}
    \langle \cdots \rangle_d
    =
    \frac{
    \mathrm{Tr}\, 
    \left[
    (\cdots)\,
    e^{-\beta H_{\rm eff}[\bm{\Omega}, \bm{d}]}
    \right]
    }{
    \mathrm{Tr}\,
    e^{-\beta H_{\rm eff}[\bm{\Omega}, \bm{d}]}
    } .
\end{align}
The explicit form of $A$ is given in Eq.~\eqref{eq:phys_quant_gen}.
Within the semiclassical approximation, Wick's theorem applies, 
which substantially simplifies the evaluation of physical observables. 
The explicit integration over the fermionic degrees of freedom $d_i$ 
can be carried out using the method described in Appendix~\ref{sec:solution_one_body}.

In the numerical simulation, a concrete representation of the coherent state is required, for which we employ the following parametrization:
\begin{equation}
\begin{aligned}
    C_1(\Omega_i) &= \cos \frac{\theta_{1i}}{2}, \\
    C_4(\Omega_i) &= \epn^{\imu \varphi_{1i}} \sin \frac{\theta_{1i}}{2}, \\
    C_\ua(\Omega_i) &= \cos \frac{\theta_{2i}}{2}, \\
    C_\da(\Omega_i) &= \epn^{\imu \varphi_{2i}} \sin \frac{\theta_{2i}}{2},
\end{aligned}
\end{equation}
where $\theta_{1i}, \theta_{2i} \in [0,\pi]$ and $\varphi_{1i},\varphi_{2i} \in [0,2\pi)$. 
The integration measure is then given by
\begin{align}
    \diff \bm{\Omega}
    = \prod_i 
    \sin \theta_{1i}\,\sin \theta_{2i}\,
    \diff \theta_{1i}\,\diff \theta_{2i}\,
    \diff \varphi_{1i}\,\diff \varphi_{2i},
\end{align}
which is identical to the product of the standard measures on two Bloch spheres at each site.

\section{Semiclassical results for the Hubbard atom}
\label{sec:one_site}

\subsection{Partition function}

In this section, we analyze the performance of the semiclassical approximation by considering a simple atomic limit and comparing the results with the exact solution.
Taking the $d$-fermion integral,
the semiclassical partition function is given by
\begin{align}
    Z_{\rm cl}(\beta; U',\mu) 
    &= \int \diff \Omega \epn^{-\beta U'|C_4(\Omega)|^2}
    \qty( 1 + \epn^{\beta \mu + \beta U'|C_4(\Omega)|^2} )
    \label{eq:atomic_z}
\end{align}
where $U'=U-2\mu$.
It can be evaluated to be 
\begin{align}
    Z_{\rm cl}(\beta; U',\mu) &= (4\pi)^2 \qty(\frac{1- \epn^{-\beta U'}}{\beta U'} + \epn^{\beta \mu}).
\end{align}
For comparison, we also show the exact solutions:
\begin{align}
    Z_{\rm exact}(\beta;U',\mu) &= 1+ 2\epn^{\beta \mu} + \epn^{-\beta U'} .
\end{align}
This is composed of the three Boltzman factors, each of which corresponds to vacant, one- and two-particle states.

\subsection{Density of states}

First, we consider the density of states $W_{\rm cl}(E)$ defined through
\begin{align}
    Z_{\rm cl}(\beta) 
    = \int_{-\infty}^{\infty} dE \,
    W_{\rm cl}(E)\,
    e^{-\beta E}.
\end{align}
It is obtained from the partition function $Z_{\rm cl}(\beta)$ by the inverse Laplace transformation, yielding
\begin{align}
    W_{\rm cl}(E) 
    &= 
    (4\pi)^2 \qty( \delta(E+\mu) 
    + 
\frac{1}{|U'|}\theta\big( E(U'-E) \big)
    )
    ,
    \label{eq:cl_dos}
\end{align}
where $\theta(x)$ denotes the Heaviside step function. 
A notable feature is that the states are distributed continuously and uniformly over the energy range $E \in [0,U'] $ for $U'>0$ or $E \in [U',0] $ for $U'<0$, reflecting the semiclassical variable $\Omega$.

For comparison, we show the exact density of states given by
\begin{align}
    W_{\rm exact}(E) 
    &= 2\delta(E+\mu) 
    + \delta(E) 
    + \delta(E-U'),
    \label{eq:exa_dos}
\end{align}
which consists solely of delta-function peaks.
In contrast to Eq.~\eqref{eq:cl_dos}, the exact spectrum is therefore discrete rather than continuous.
Despite this difference, the total spectral weight is conserved up to an overall normalization factor:
\begin{align}
    \int W_{\rm exact}(E)\, \diff E 
    =
    \int \frac{W_{\rm cl}(E)}{8\pi^2}\, \diff E 
    = 4.
\end{align}
In particular, at half filling with $U' = U - 2\mu = 0$, the semiclassical and exact densities coincide:
\begin{align}
    W_{\rm exact}(E) 
    = \frac{W_{\rm cl}(E) }{8\pi^2}
    = 2\delta(E+\mu) + 2\delta(E).
\end{align}
This agreement arises because the quantum $d$ fermion is retained explicitly within the semiclassical approximation and governs the 
physics of the Hubbard atom at half filling.

\begin{figure*}[t]
    \centering
    
    \begin{minipage}[t]{0.49\textwidth}\centering
        \large\bfseries Semiclassical
    \end{minipage}
    \begin{minipage}[t]{0.49\textwidth}\centering
        \large\bfseries Exact     
    \end{minipage}
    \par
   
    \begin{minipage}[t]{0.24\textwidth}\centering
        \includegraphics[width=\linewidth]{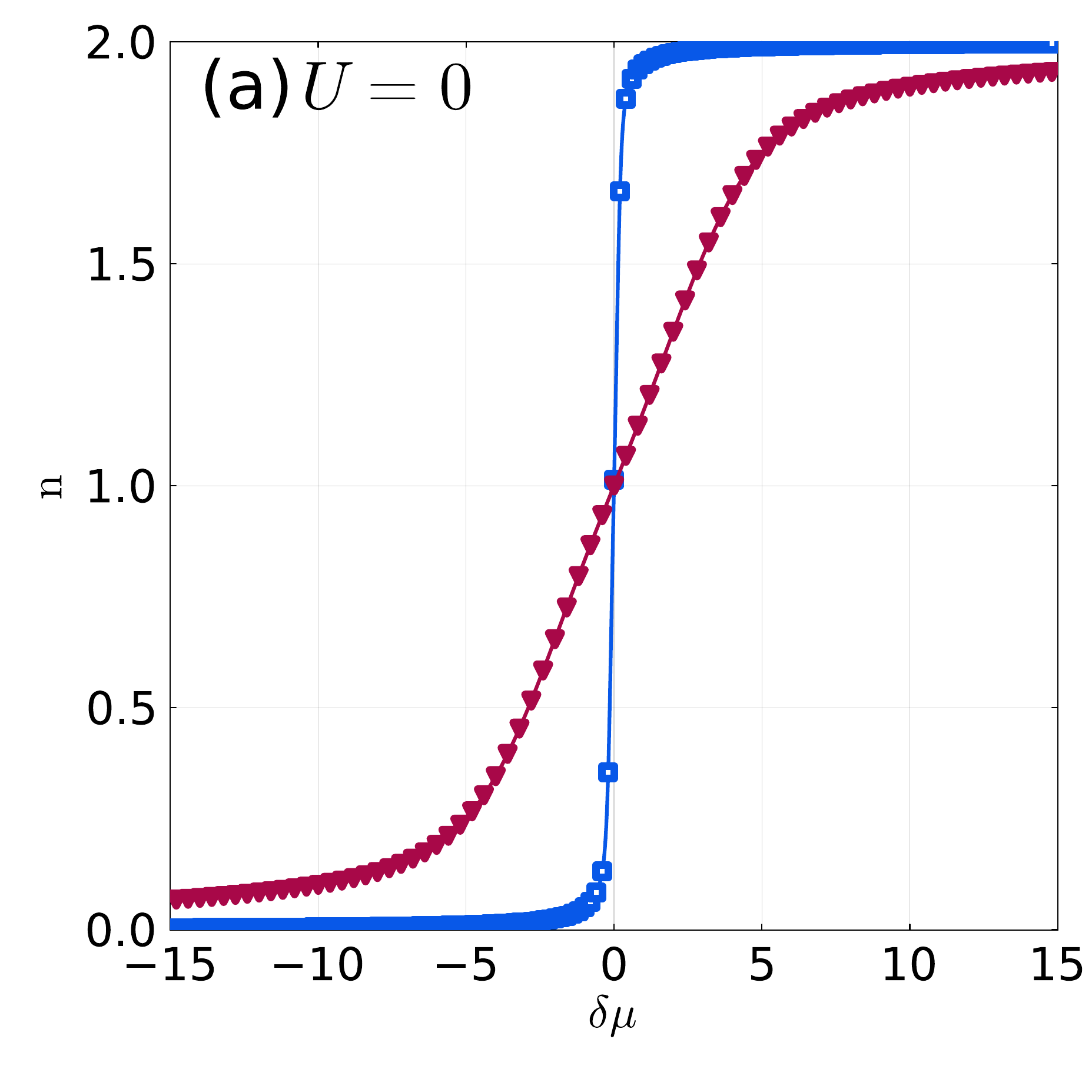}
    \end{minipage}
    \begin{minipage}[t]{0.24\textwidth}\centering
        \includegraphics[width=\linewidth]{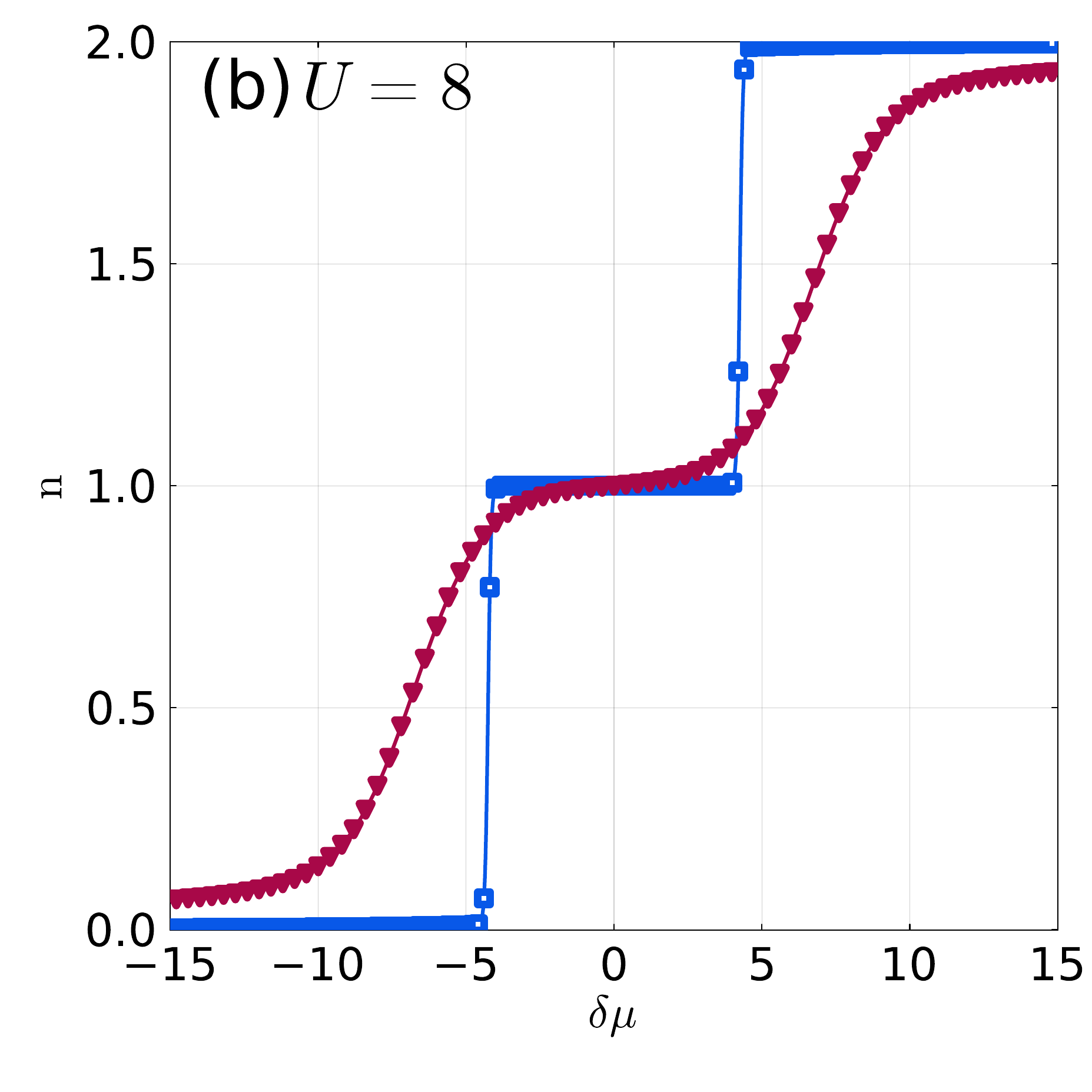}
    \end{minipage}
    \begin{minipage}[t]{0.24\textwidth}\centering
        \includegraphics[width=\linewidth]{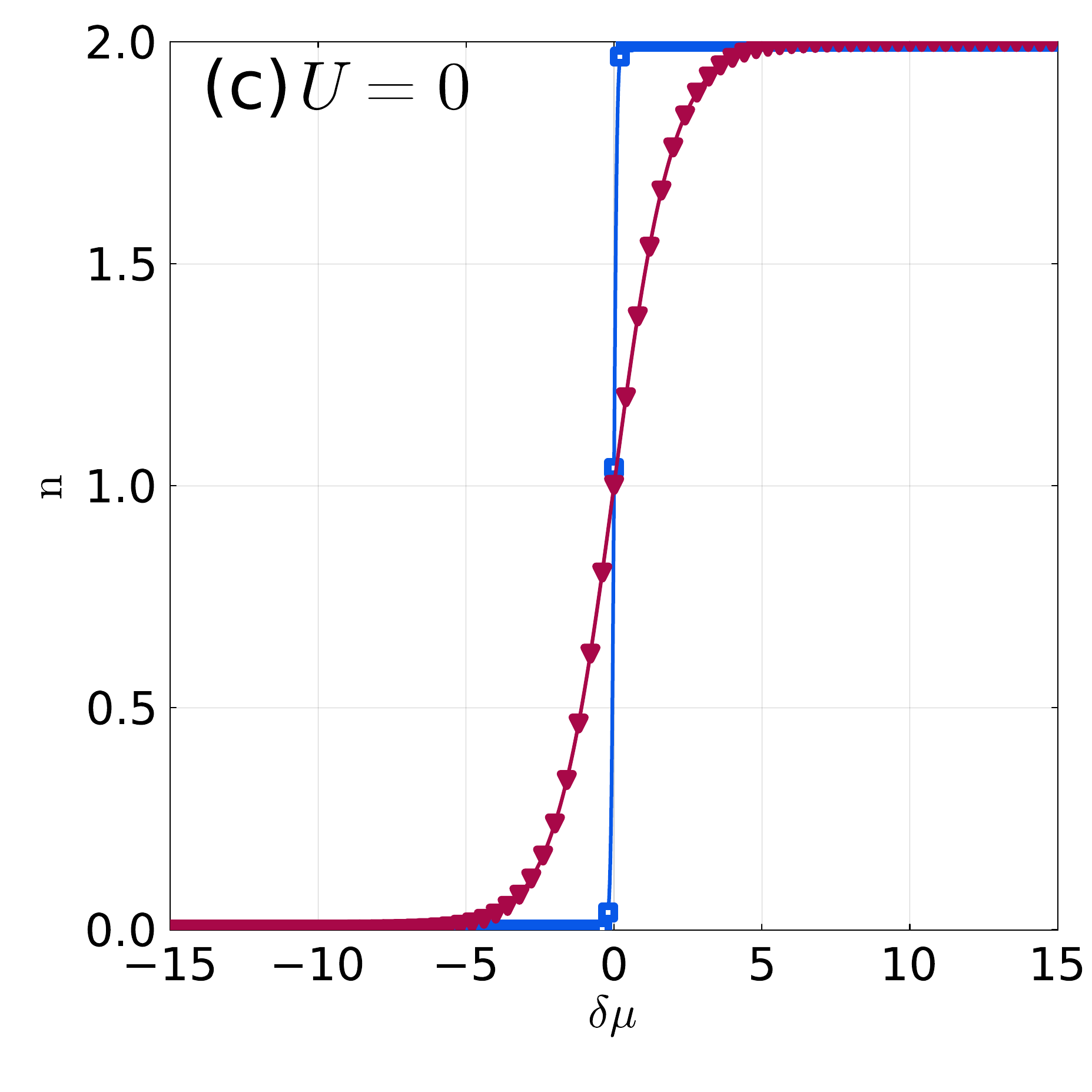}
    \end{minipage}
    \begin{minipage}[t]{0.24\textwidth}\centering
        \includegraphics[width=\linewidth]{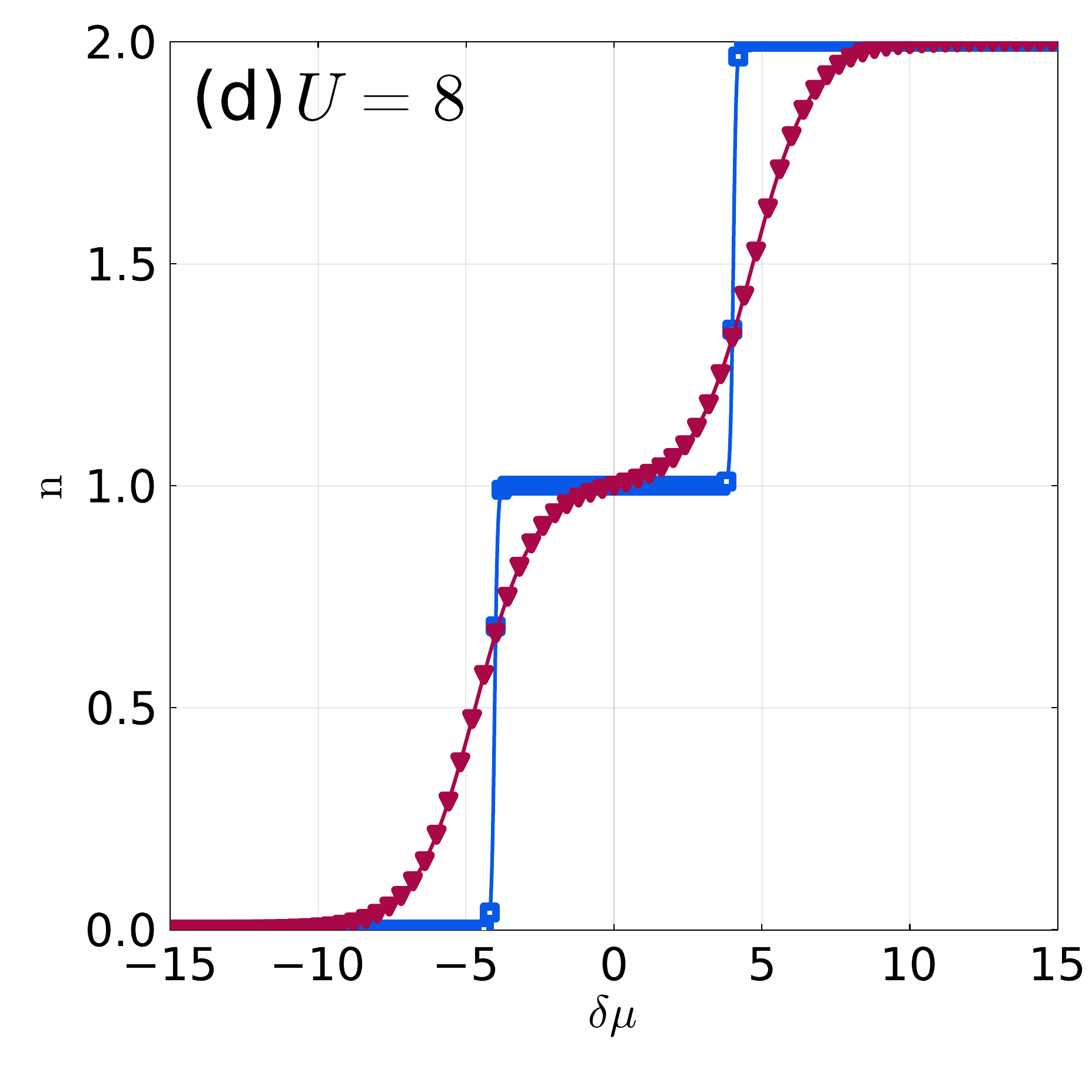}
    \end{minipage}
    \par
    \vspace{-2mm}
    
    \begin{minipage}[t]{0.245\textwidth}\centering
        \hspace*{0.3mm}
        \includegraphics[width=\linewidth]{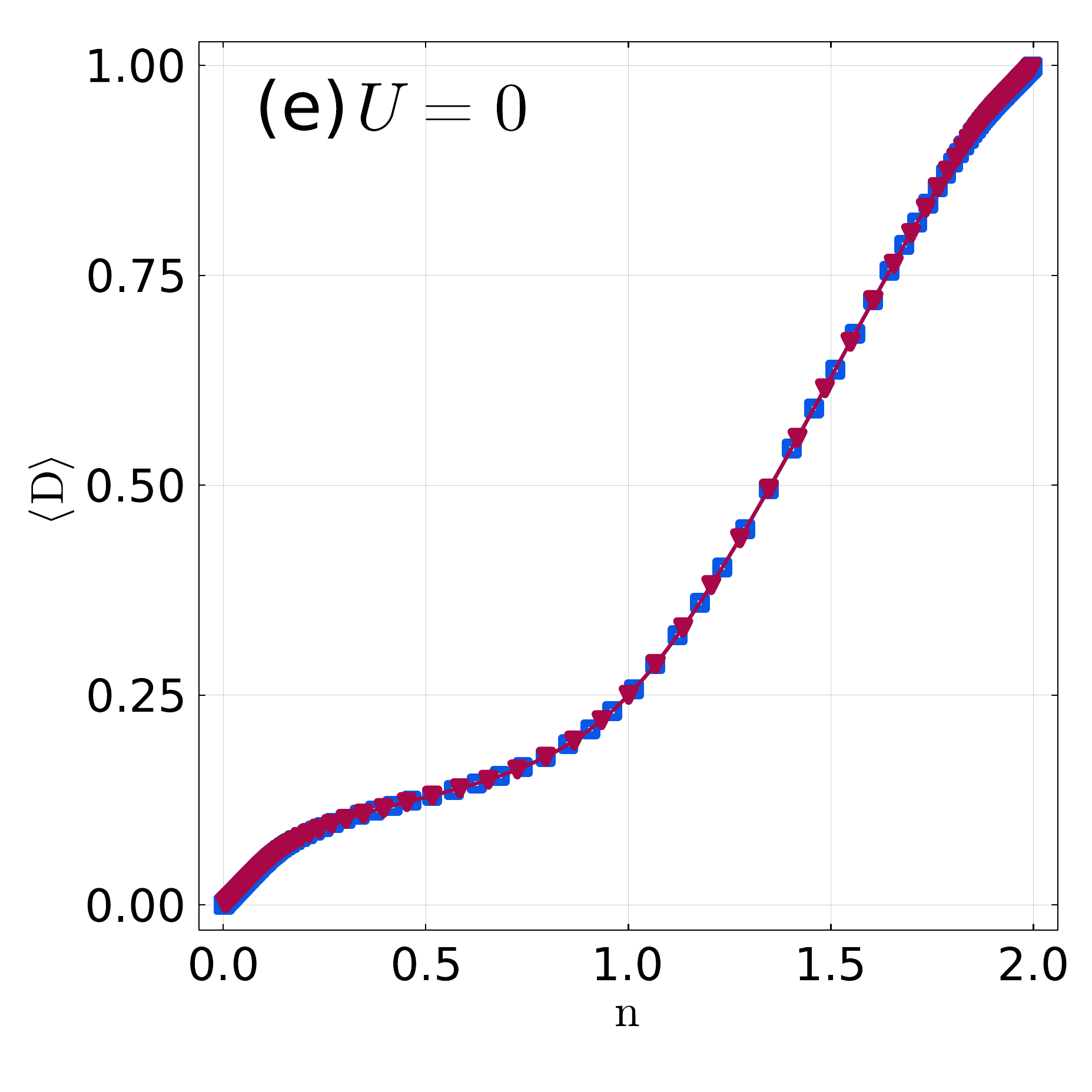}
    \end{minipage}
    \begin{minipage}[t]{0.245\textwidth}\centering
        \hspace*{-2.05mm}
        \includegraphics[width=\linewidth]{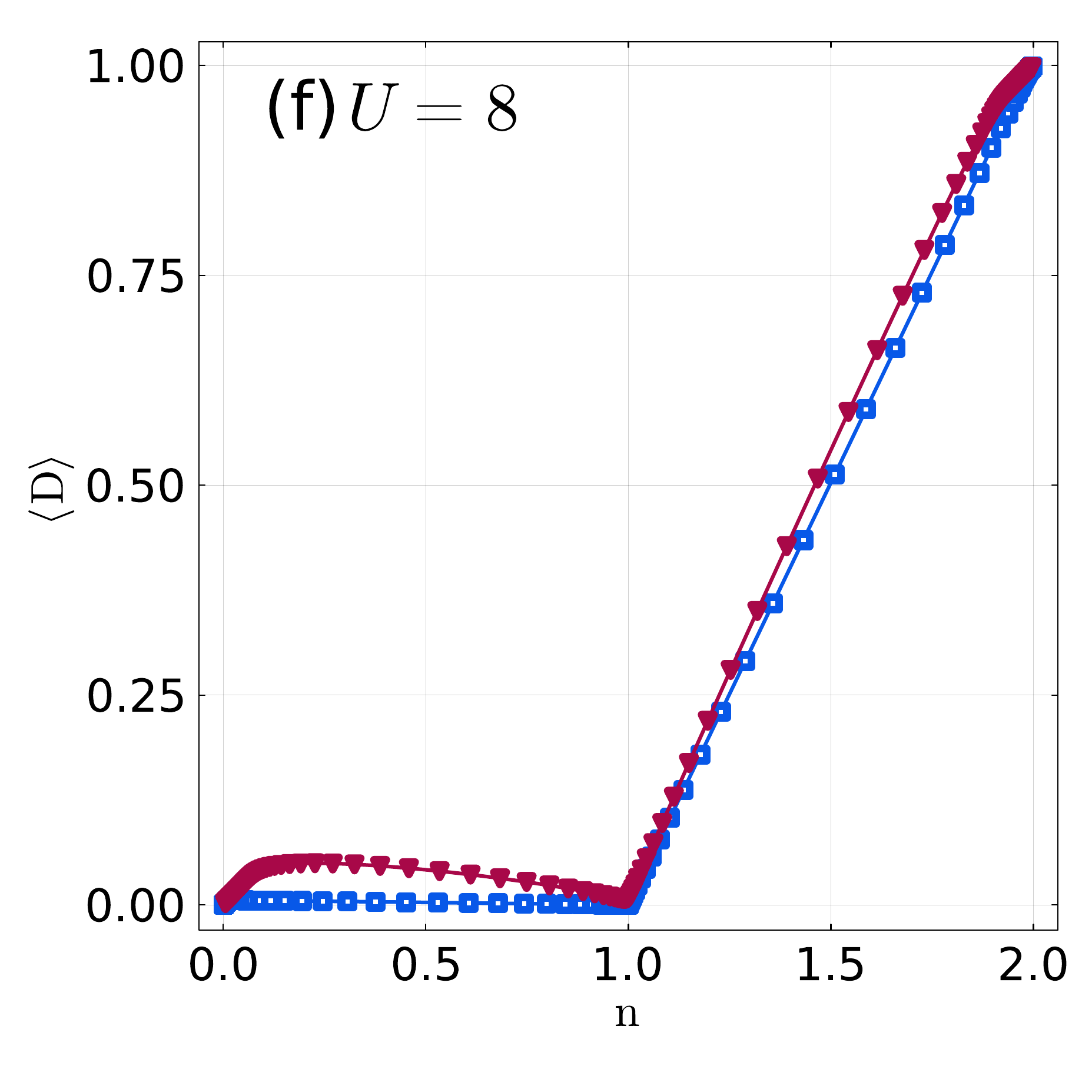}
    \end{minipage}
    \begin{minipage}[t]{0.245\textwidth}\centering
        \hspace*{-1.78mm}
        \includegraphics[width=\linewidth]{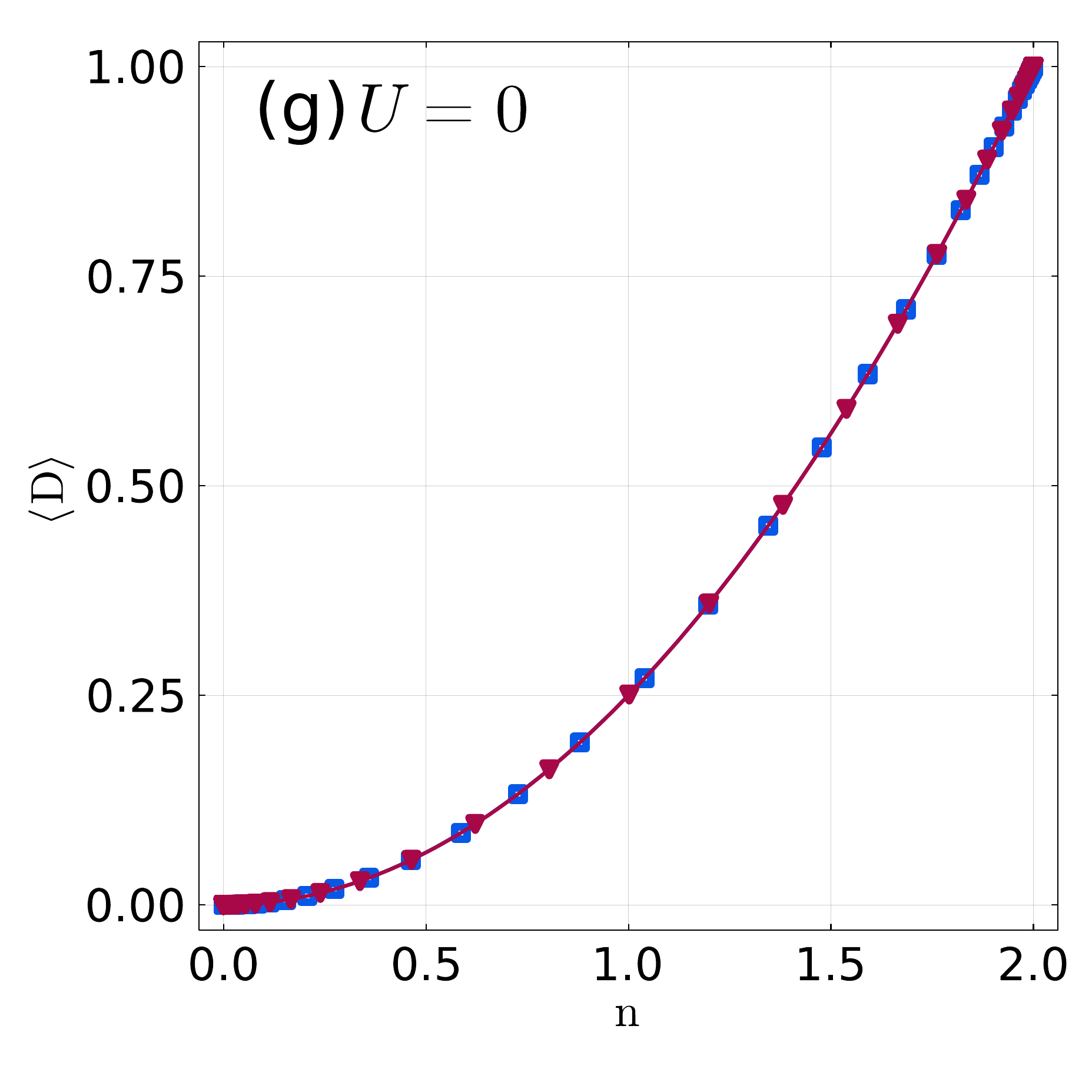}
    \end{minipage}
    \begin{minipage}[t]{0.245\textwidth}\centering
        \hspace*{-2.8mm}
        \includegraphics[width=\linewidth]{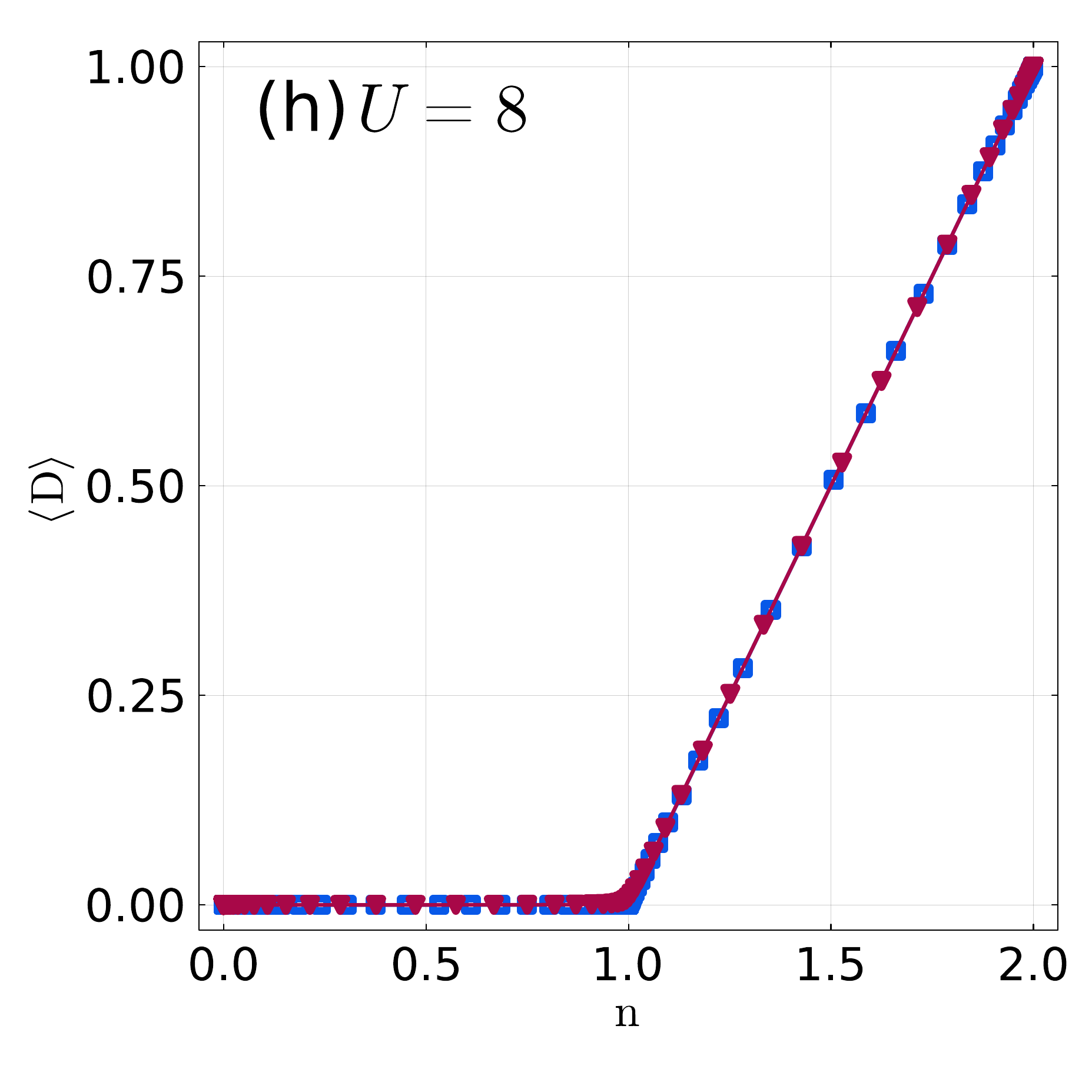}
    \end{minipage}\par

    \includegraphics[width=0.3\textwidth]{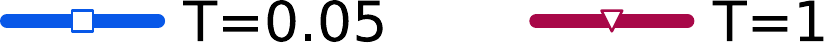}

    \caption{(Top row) Filling versus chemcal potential $\delta \mu = \mu - U/2$ for the (a,b) semiclassical and (c,d) exact results at temperatures $T=0.05, 1.0$. 
    (Bottom row) Double occupancy versus filling for the (e,f) semiclassical and (g,h) exact results. 
    The interaction parameters are chosen as (a,c,e,g) $U=0$ and (b,d,f,h) $U=8.0$.
    }
    \label{fig:atom_lim}
\end{figure*}

\subsection{Electron number}

In this subsection, we consider the electron number.
Evaluating Eq.~\eqref{eq:def_of_number} in the semiclassical approximation, we obtain the following results:
\begin{align}
    \la n \ra
    &= \frac{\displaystyle 
    \ \ \ 
      \epn^{\beta \mu} +  2\qty(\frac{1- \epn^{-\beta U'}}{(\beta U')^2} - \frac{\epn^{-\beta U'}}{\beta U'})
    \ \ \ 
    }{\displaystyle 
    \frac{1- \epn^{-\beta U'}}{\beta U'} + \epn^{\beta \mu}
    }
    ,
    \\[2mm]
    \la n \ra_{\rm exact} 
    &= \frac{2\epn^{\beta \mu} + 2\epn^{-\beta U'}}{1 + 2\epn^{\beta \mu} + \epn^{-\beta U'}}
    .
\end{align}
Here, we have also evaluated the exact solution for comparison. 
Since the analytical expressions are not particularly transparent, we plot the functional behavior graphically and discuss the results. 
The upper panels of Fig.~\ref{fig:atom_lim} show the particle number of the Hubbard atom (one-site system) plotted as a function of the chemical potential 
$\delta\mu = \mu - \frac{U}{2} = -\frac{U'}{2}$, 
obtained from (a,b) the semiclassical approximation and (c,d) the exact solution. 
For the Coulomb interaction strength, we adopt (a,c) $U=0$ and (b,d) $U=8$. 
In each panel, the results for two temperatures, $T=0.05$ and $T=1$, are displayed.

By comparing the semiclassical and exact results, we observe that the qualitative behavior, such as the overall increase and decrease of the particle number, is reproduced correctly. 
Quantitatively, however, the dependence on the chemical potential $\delta\mu$ and the temperature $T$ differs. 
In particular, the asymptotic behavior is algebraic in the semiclassical approximation, whereas it is exponential in the exact solution. 
More specifically, fixing the temperature $T = 1/\beta$ and the interaction $U$, and considering the large-$\delta\mu$ limit, the particle number behaves as
\begin{align}
    \langle n \rangle 
    &\longrightarrow 
    2 - \frac{1}{\beta \delta\mu}, 
    \\[2mm]
    \langle n \rangle_{\rm exact} 
    &\longrightarrow 
    2 - 2 e^{-2\beta \delta\mu},
\end{align}
for $\delta\mu \to +\infty$. 
This difference reflects the distinct nature of the energy distributions in Eqs.~\eqref{eq:cl_dos} and \eqref{eq:exa_dos}, 
namely, whether the states are distributed continuously or discretely.

\subsection{Double occupancy}

Next, we consider the double occupancy in Eq.~\eqref{eq:def_of_Docc}.
Within the semiclassical approximation, we obtain the following results:
\begin{align}
    \la D \ra 
    &= \frac{\displaystyle 
    \ \ \ 
        \frac{1- \epn^{-\beta U'}}{(\beta U')^2} - \frac{\epn^{-\beta U'}}{\beta U'}
    \ \ \ 
    }{\displaystyle 
    \frac{1- \epn^{-\beta U'}}{\beta U'} + \epn^{\beta \mu}
    }
    , \\[2mm]
    \la D \ra_{\rm exact} 
    &= \frac{\epn^{-\beta U'}}{1 + 2\epn^{\beta \mu} + \epn^{-\beta U'}}
    ,
\end{align}
where we have also provided the exact solution.
We confirm that $\la D\ra$ and $\la D \ra_{\rm exact}$ coincies at half filling ($\mu=U/2$ or $U'=0$).

The lower panels of Fig.~\ref{fig:atom_lim} show the double occupancy plotted as a function of the particle number, where the semiclassical results are located in (e,f) and the exact ones in (g,h). 
The parameters are the same as in the upper panels, and the conversion between the particle number and the chemical potential is performed based on the results shown in the upper panels of Fig.~\ref{fig:atom_lim}. 
As seen in Fig.~\ref{fig:atom_lim}(f), at high temperatures and for $n<1$, the double occupancy exhibits a nonmonotonic behavior. 
This arises because, at finite temperature, the semiclassical approximation acquire contributions from the continuously distributed density of states, leading to this artifact. 
We note, however, that this spurious behavior remains small in magnitude and disappears at low temperatures for $U\neq 0$.
More specifically, for fixed temperature $T=1/\beta$ and interaction $U$, the double occupancy behaves as
$    \langle D \rangle 
    \to  \frac{n}{2}$
in the limit $n \to 0$. 
The range over which this asymptotic behavior is observed becomes narrower as $T/U \to 0$ as shown in Fig.~\ref{fig:atom_lim}(f).

At $U=0$, the relation $\langle D \rangle_{\rm exact} = \langle n \rangle_{\rm exact}^2 / 4$ holds for the exact solution, reflecting the absence of correlations in the noninteracting limit. 
In contrast, this factorization property is not satisfied within the semiclassical approximation, leading to an artificial correlation. 
The deviation can be estimated for $U=0$ as
\begin{align}
    0 \le 
    \langle D \rangle 
    - \frac{\langle n \rangle^2}{4}
    \le 
    0.07957,
\end{align}
where the lower bound is attained at $\delta \mu = 0$ (half filling), and the upper bound is reached at $\beta \delta \mu = \pm 4.356$ away from half filling.

\subsection{Single-particle Green's function}

Let us also consider the single-particle Green's function in the atomic limit, as a representative example of the dynamical quantity:
\begin{align}
    G_\sg (\tau) &= - \la c_{\sg}(\tau) c_\sg^\dg(0) \ra
\end{align}
for $0<\tau<\beta$, and
\begin{align}
    G_\sg (\tau) &= \la c_\sg^\dg(0)  c_{\sg}(\tau)\ra
\end{align}
for $-\beta <\tau<0$.
The weight of the Green's function is given by $G_\sg(-0)- G_\sg(+0)$, which represents a jump at $\tau=0$ and becomes unity for the exact calculation.
However, 
within the semiclassical approximation, one obtains
\begin{align}
G_\sg(-0)- G_\sg(+0)
&= \la |C_\sg|^2 \ra \leq  1
.
\end{align}
Namely, the weight becomes smaller compared to the fully quantum mechanical case.
This weight loss behaviors are also seen in the spin correlation and hopping amplitude as discussed in the next section.
We also note that the limit $G_\sg(\tau\to -0)$ does not approach to $\la n_\sg\ra$, since the $\tau$-dependence of $C_{1,2,3,4}$ is neglected in the semiclassical approximation.
Thus, one must notice that $\tau$ dependence of the Green's functions reflect the artifact of the semiclassical approximation.

\begin{figure*}[t]
        
    \begin{minipage}[t]
    {0.4\textwidth}\centering
        \hspace{-3mm}
        \large\bfseries Semiclassical
    \end{minipage}
    \begin{minipage}[t]{0.4\textwidth}\centering
        \hspace{30mm}
        \large\bfseries Exact     
    \end{minipage}
    \par
    
    \centering 
    \begin{minipage}[t]{0.49\textwidth}\centering
        \includegraphics[width=\linewidth]{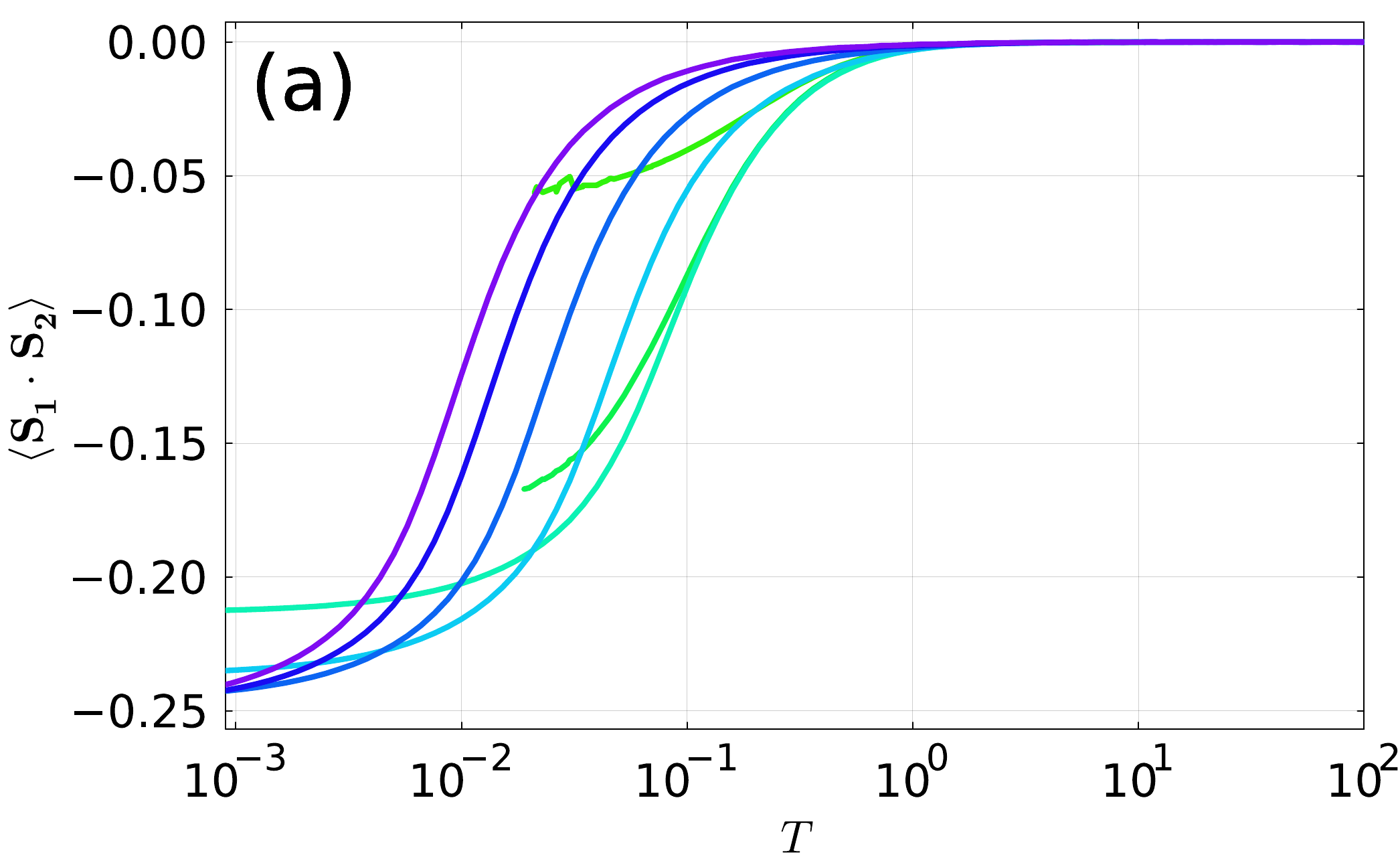}
    \end{minipage}
    \hspace{-2mm}
    \hfill
    \begin{minipage}[t]{0.49\textwidth}\centering
        \includegraphics[width=\linewidth]{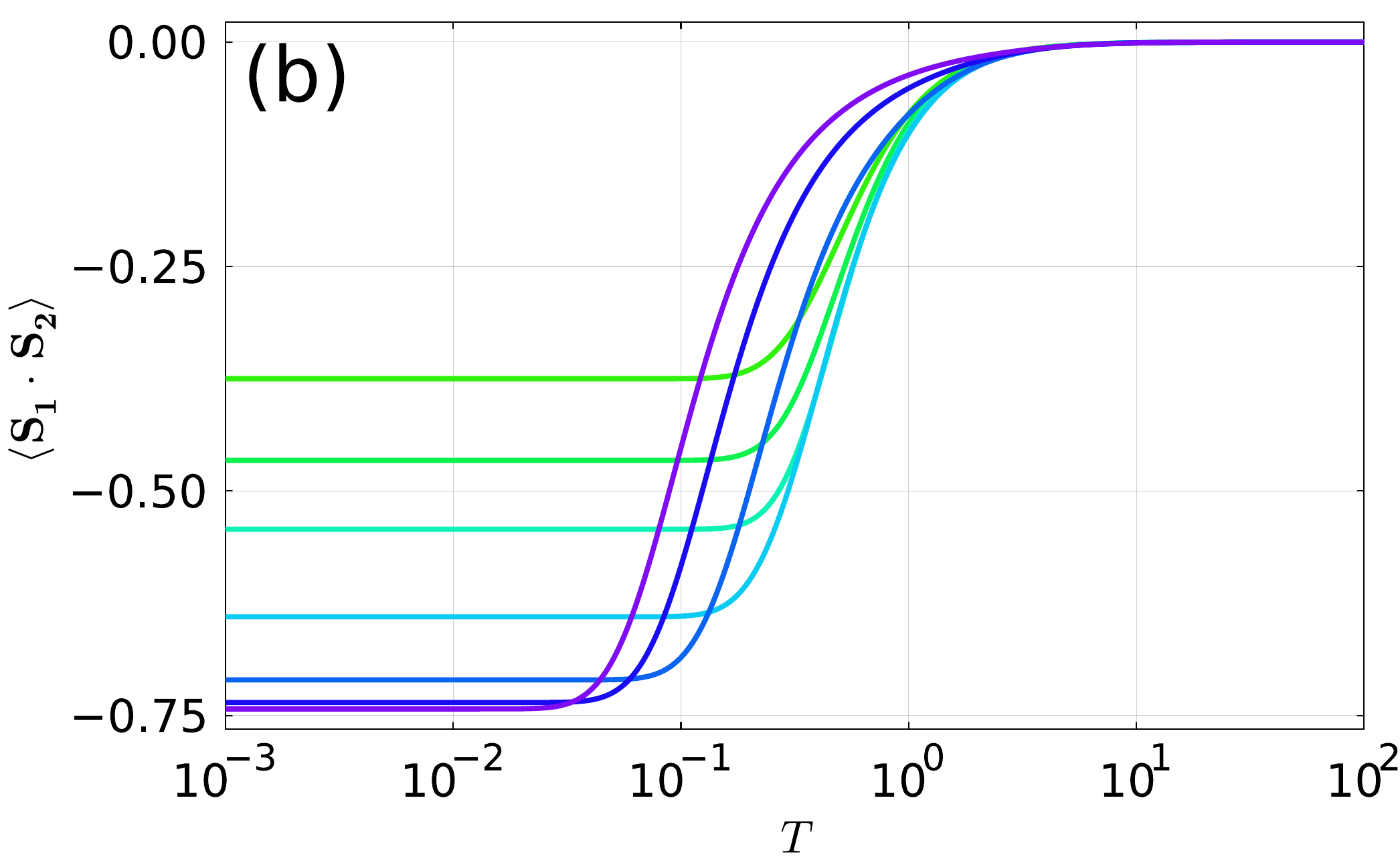}
    \end{minipage}
    \par
    \begin{minipage}[t]{0.49\textwidth}\centering
        \includegraphics[width=\linewidth]{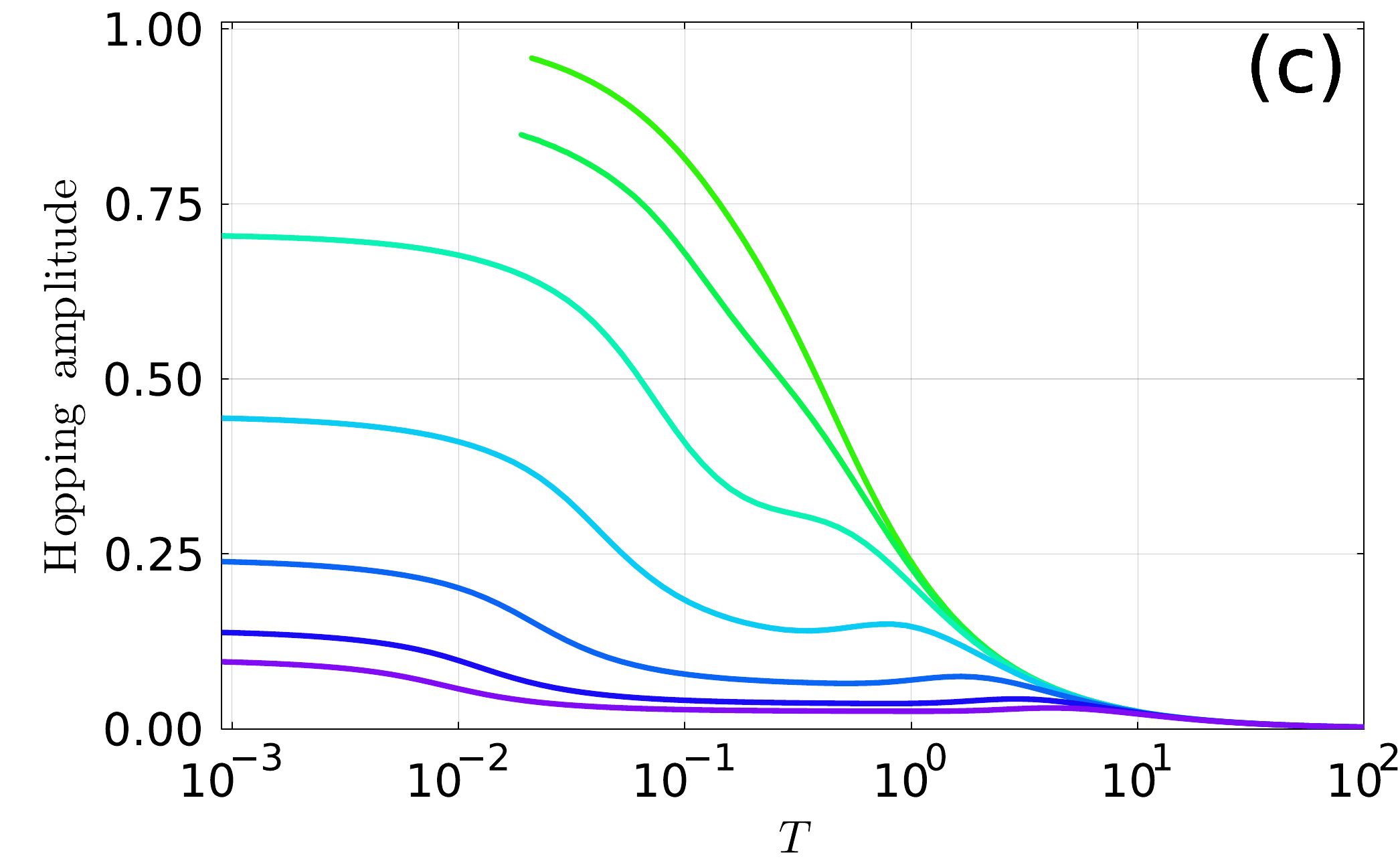}
    \end{minipage}
    \hspace{-2mm}
    \hfill
    \begin{minipage}[t]{0.49\textwidth}\centering
        \includegraphics[width=\linewidth]{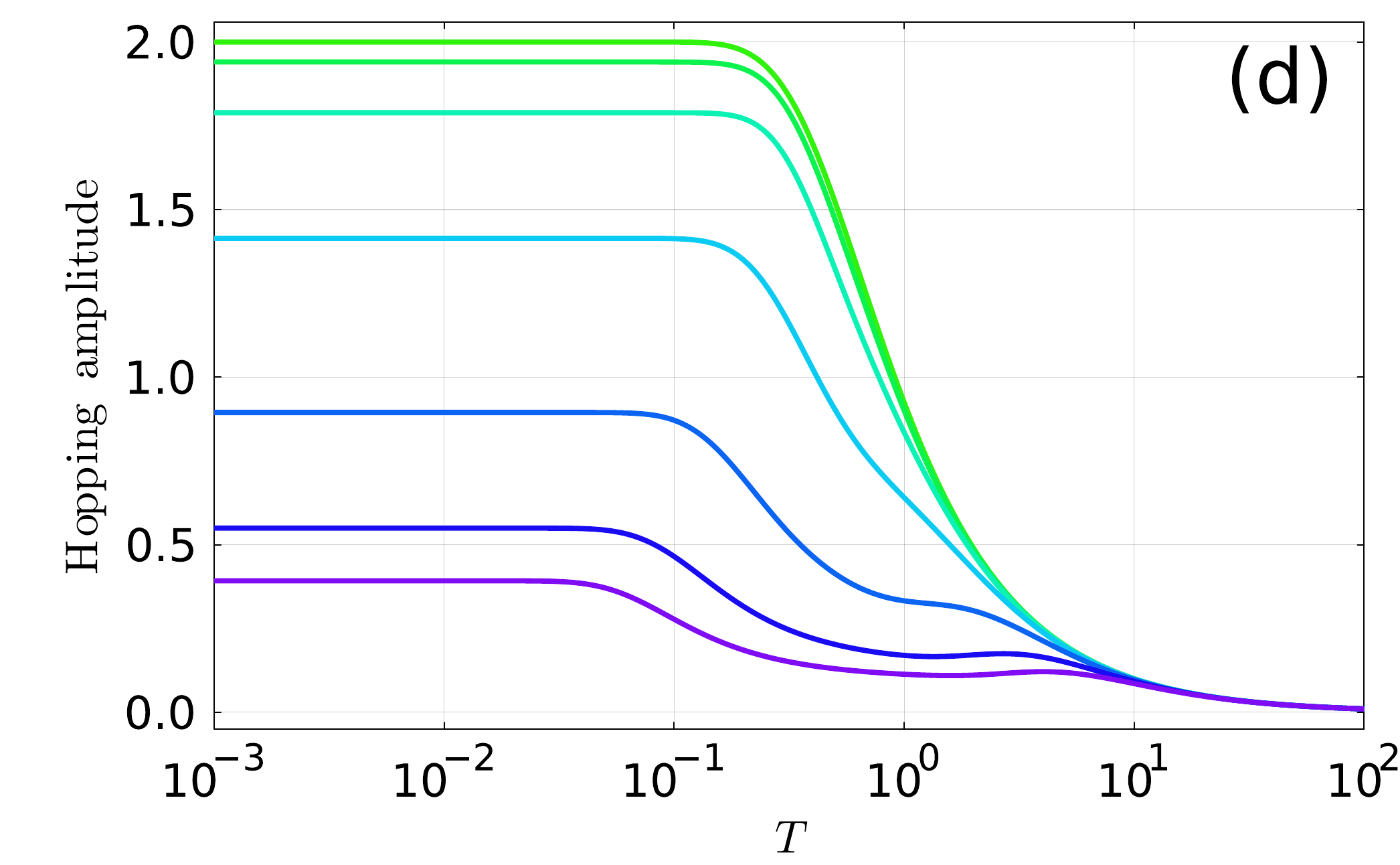}
    \end{minipage}
    \hspace{-1.5mm}
    \par
    \begin{minipage}[t]{0.5\textwidth}\centering
        \includegraphics[width=\linewidth]{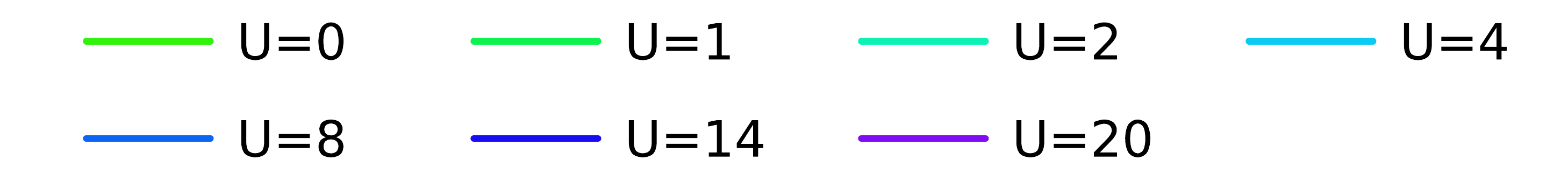}
    \end{minipage}
    \caption{
    (Top row) Temperature dependence of the two-site spin correlation 
$\langle \bm S_1 \cdot \bm S_2 \rangle$ 
for several values of $U$, obtained by (a) the semiclassical method and (b) exact calculation. 
(Bottom row) Same as in the top row, but for the hopping amplitude: 
(c) semiclassical method and (d) exact calculation.
    }
    \label{fig:two_site_T}
\end{figure*}

\section{Semiclassical numerical simulation for the two-site model}
\label{sec:two_site}

\subsection{Simulation algorithm}

For the two-site model, 
we rely on Monte Carlo simulations to evaluate the physical quantities.
For the evaluation of the partition function, we also need the partial partition function, ${\rm Tr}\, \epn^{-\beta H_{\rm eff}}$, for a given configuration of $\bm \Omega$, which is computed using the method described in Appendix~\ref{sec:solution_one_body}. 
Since this weight function is positive, importance sampling can be applied. 
Specifically, we employ the Metropolis algorithm to update $\bm \Omega \to \bm \Omega'$ at each Monte Carlo step. 
More sophisticated update algorithms could improve simulation efficiency, but we leave such developments for future work. 
The primary purpose of this study is to assess how accurately the present framework captures intersite correlation effects in the two-site Hubbard model.

\subsection{Temperature dependence of spin correlation}

First of all, we consider the spin correlation, 
    $\langle \bm S_1 \cdot \bm S_2 \rangle$
which is calculated through Eqs.~(\ref{eq:def_splus}) and (\ref{eq:def_sz}). 
This quantity becomes nonzero only when $t \neq 0$, and quantifies the degree of activated spin.

The upper panels of Fig.~\ref{fig:two_site_T} show the temperature dependence of the two-site spin correlation at half filling for several values of $U$. 
Panel (a) displays the results obtained using the semiclassical approximation, which is the focus of this work, while panel (b) shows the corresponding exact solution for comparison. 
In the semiclassical calculation, the results are obtained via Monte Carlo simulations; however, for small $U$ at low temperatures, the calculations become unstable, and only data from regions with a sufficiently high signal-to-noise ratio are plotted, resulting in curves that are truncated at low temperatures. 
The high-temperature behavior is similar between the two approaches, but at low temperatures, the exact solution asymptotically approaches the zero-temperature limit exponentially, whereas the semiclassical results follow a power-law behavior. 
This difference arises because, as discussed in the previous section, the coherent state in the semiclassical approximation is continuously distributed with respect to energy.

In particular, in the limit $U \to \infty$, the system is expected to be mapped onto a Heisenberg-type interaction with localized spins. 
In the exact solution, the two quantum spins $S=1/2$ forms a singlet, yielding 
$\langle \bm S_1 \cdot \bm S_2 \rangle_{\rm exact} = - S(S+1) = -3/4$ at zero temperature. 
By contrast, in the classical-spin picture, $\bm S_{1,2}$ are treated as classical vectors of length $S=1/2$, so that their correlation satisfies 
$\langle \bm S_1 \cdot \bm S_2 \rangle = -S^2 = -1/4$. 
Therefore, the reduced magnitude of the spin correlation observed in Fig.~\ref{fig:two_site_T}(a) can be attributed to the classical treatment of the spins.

\begin{figure}[t]
    \centering
    \includegraphics[width=85mm]{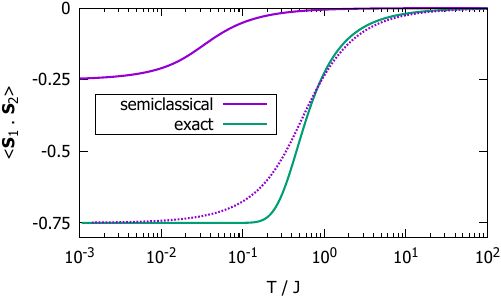}
    \caption{Temperature depnendece of the spin correlation for semiclassical and quantum two-site spin models. The dotted line is a rescaled version of the semiclassical result. }
    \label{fig:S1S2}
\end{figure}

\subsection{Effective spin model}

Figure~\ref{fig:two_site_T}(a) and (b) show that the characteristic temperature scale of the semiclassical approximation appears smaller than that of the exact solution. 
To discuss this point more clearly, we compare the two-site model in the quantum case with a classical-spin model. 
At large $U$ and half filling, the two-site Hubbard model can be effectively mapped onto a spin model, described by
\begin{align}
    \mathscr H_{\rm eff} &= J \, \bm S_1 \cdot \bm S_2.
\end{align}  
For the quantum-mechanical (qm) case, the interaction parameter is given by $J=4t^2/U$, and the spin operator is defined by
\begin{align}
    \bm S_i^{\rm qm} &= \frac 1 2 \sum_{\sg\sg'} |\sg\rangle_i \, \bm \sigma_{\sg\sg'} \, {}_i\langle \sg' |,
\end{align}
with $i=1,2$. 
The exact partition function for the quantum case is
\begin{align}
    Z_{\rm exact} &= \epn^{3\beta J/4} + 3 \epn^{-\beta J/4}.
\end{align}
The spin correlation is then obtained from the derivative of the free energy with respect $J$ as
\begin{align}
    \langle \bm S_1 \cdot \bm S_2 \rangle_{\rm exact} 
    &= - \frac{3}{4} \frac{\epn^{\beta J} - 1}{\epn^{\beta J} + 3}.
    \label{eq:S1S2_exact}
\end{align}
Hence, the low temperature limit of the spin correlation is given by $-3/4$ in the quantum two-spin problem.

Next, we derive the effective semiclassical spin model.
We begin with the partition function at half filling:
\begin{align}
    Z_{\rm cl} &= \int \diff \bm \Omega \ {\rm Tr\,}
    \epn^{-H_{\rm eff}[\bm \Omega,\bm d]}
    , \\
    H_{\rm eff} &= -\mu(d_1^\dg d_1 + d_2^\dg d_2)
    + \qty( T_{12} d_1^\dg d_2 + \Delta_{12} d_1^\dg d_2^\dg + {\rm H.c.})
    ,
\end{align}
where $T_{12}$ and $\Delta_{12}$ have been defined in Eqs.~\eqref{eq:def_of_T_ij} and \eqref{eq:def_of_Delta_ij}.
Note that we have used $U'=U-2\mu = 0$.
To proceed further, it is convenient to define the $SU(2)$ spinors by
\begin{align}
    u_i = \begin{pmatrix}
        C_{\ua i} \\ C_{\da i}
    \end{pmatrix}
    ,\ \ \ 
    w_i = \begin{pmatrix}
        C_{1 i} \\ C_{4 i}
    \end{pmatrix}    
    ,
\end{align}
which satisfies
the normalization conditions $|u_i|^2 = |w_i|^2 = 1$.
We also introduce the unit vectors
\begin{align}
\hat {\bm n}_i = u_i^\dg \bm \sg u_i
,\ \ \ 
\hat {\bm \tau}_i = w_i^\dg \bm \sg w_i
.
\end{align}
The $d$-fermion trace can be explicitly performed by the procedure in Appendix~\ref{sec:solution_one_body_2}, and we obtain
\begin{align}    
Z_{\rm cl} &= 2\epn^{\beta \mu} \int \diff \bm \Omega \qty[\cosh(\beta |T_{12}|) + \cosh ( \beta \sqrt{\mu^2 + |\Delta_{12}|^2}) ]
,
\label{eq:Z_two_site_HF}
\end{align}
where
the absolute values of $T_{12},\Delta_{12}$ are written in a simple manner:
\begin{align}
    &|T_{\rm 12}| = t \sqrt{\frac{1+x}{2}} \sqrt{\frac{1+y}{2}} ,
    \\
    &|\Delta_{\rm 12}| = t \sqrt{\frac{1-x}{2}} \sqrt{\frac{1-y}{2}} ,
    \\[2mm]
    &x = \hat {\bm n}_1 \cdot  \hat {\bm n}_2
, \ \ \ 
y= \hat {\bm \tau}_1 \cdot  \hat {\bm \tau}_2 .
\end{align}
Here defined $x$ and $y$ represent SU(2) invariants.
This expression is useful for the integration $\int \diff \bm \Omega$, since the integration with respect to the vector $\hat{\bm n_1}$ ($\hat{\bm \tau_1}$) can be performed under the situation with fixed $\hat {\bm n_2}$ ($\hat{\bm \tau_2}$).
Equation \eqref{eq:Z_two_site_HF} then simplifies to
\begin{align}
    &Z_{\rm cl} = (8\pi)^2 \int_{-1}^1\diff x \int_{-1}^1\diff y \ \mathcal F(x,y)
    , \\
    &\mathcal F(x,y) = 2\epn^{\beta \mu}\bigg[ \cosh\qty(\frac{\beta t}{2} \sqrt{(1+x)(1+y)})
    \nonumber \\
    &\hspace{15mm} + \cosh\qty(\beta \sqrt{\mu^2 + \frac{t^2}{4}(1-x)(1-y)}) \bigg]
    .
    \label{eq:half_filled_F_prior} 
\end{align}
Now we expand the expression up to the second order of the hopping $t$ and integrate over the variable $y$.
The explicit parition function is given by
\begin{align}
    &Z_{\rm cl} =(8\pi^2)^2  \mathcal F_0 \int_{-1}^1 \diff x \exp[ - \beta \mathcal J(\beta,U)\frac{x}{4} + O(t^4) ]
    ,\\
    &\mathcal J(\beta, U) = \frac{t^2}{U} \tanh \frac{\beta U}{4} - \frac{\beta t^2}{2(1+\cosh \frac{\beta U}{2})}
    ,
    \label{eq:half_filled_F} 
\end{align}
where ${\mathcal F}_0$ is an $x$-independent constant.
We note that $x/4$ is identical to the $\bm S^{\rm cl}_1\cdot \bm S^{\rm cl}_2$ with $\bm S^{\rm cl}_i$ being the semiclassical spin vector
\begin{align}
    \bm S_i^{\rm cl} &= \frac 1 2 \Big(
    \hat{\bm x} \sin\theta_i \cos \varphi_i
    + \hat{\bm y} \sin\theta_i \sin \varphi_i
    + \hat{\bm z} \cos \theta_i \Big),
\end{align}
with $\theta_i \in [0,\pi]$ and $\varphi_i \in [0,2\pi)$.
Assuming the low temperature case with $\beta U \gg 1$, we identify the semiclassical spin interaction parameter $\mathcal J(\beta , U) \to t^2/U \equiv J_{\rm cl}$, which is four times smaller than the quantum model with $J=4t^2/U$.

The partition function can now be evaluated explicitly for the semiclassical spin model, and the spin correlation function follows as
\begin{align}
    \langle \bm S_1 \cdot \bm S_2 \rangle_{\rm cl} 
    &= \frac{1}{\beta J_{\rm cl}} - \frac{1}{4} \coth \frac{\beta J_{\rm cl}}{4}.
    \label{eq:S1S2_semiclass}
\end{align}
We note that this spin correlation function does not coincide with the result shown in Fig.~\ref{fig:two_site_T} even in the limit $t/U \ll 1$, although the temperature dependence of the two results is similar.
This difference arises because the spin correlation in Fig.~\ref{fig:two_site_T} is calculated from Eqs.~\eqref{eq:def_splus} and \eqref{eq:def_sz}, which involve fermionic operators ($d,d^\dagger$), whereas Eq.~\eqref{eq:S1S2_semiclass} contains only semiclassical variables ($\theta,\varphi$).

The temperature dependence for $\la \bm S_1\cdot \bm S_2\ra$ is shown in Fig.~\ref{fig:S1S2} together with the exact result in Eq.~\eqref{eq:S1S2_exact}.
Although both the quantum and semiclassical results are characterized by the same energy scale $t^2/U$, the plots in Fig.~\ref{fig:S1S2} appear to show different characteristic temperature scales. 
This discrepancy arises from a factor of $1/16$ in the energy scale of the semiclassical expression. 
Indeed, if we rescale the axes by multiplying the vertical axis by $3$ and the horizontal axis by $16$, 
 the characteristic energy scale visually aligns with that of the exact result, as shown in  Fig.~\ref{fig:S1S2} with the dotted line. 
Thus, the semiclassical approximation effectively reduces the energy scale by a constant factor.

\subsection{Temperature dependence of hopping amplitude}

Let us also compare the temperature dependence of the hopping amplitude
$\displaystyle  \sum_\sigma \big\langle c_{1\sigma}^\dagger c_{2\sigma} + {\rm H.c.} \big\rangle$,
which can be evaluated from the terms proportional to the hopping integral $t$
in Eq.~\eqref{eq:def_hop}. 
This quantifies the degree of electron delocalization between the two sites.
Figures~\ref{fig:two_site_T}(c) and (d) show the results obtained by the semiclassical approximation and the exact solution, respectively. 
As in the case of the spin correlation, the low-temperature behavior exhibits a difference, being either power-law or exponential, and the characteristic energy scale is lower in the semiclassical approximation. 
It should be noted, however, that such differences in characteristic energy scales between the two methods do not necessarily apply to local observables. 
This is evident from the fact that local quantities on a single site at half-filling exactly agree with the exact solution.

The low-temperature limit of the semiclassical result is smaller than that of the exact solution, which is analogous the behavior observed for the spin correlation shown in Figs.~\ref{fig:two_site_T}(a) and (b). 
In particular, in the noninteracting limit $U\to 0$, the exact solution yields a hopping amplitude that asymptotically approaches $2$, whereas in the semiclassical approximation it appears to approach exactly half of this value, $1$. 
This can be attributed to the fact that in the semiclassical approximation the hopping is carried by spinless fermions, whereas in the exact solution the spinful fermions have two internal degrees of freedom. 
The fact that the hopping amplitude approaches unity in the $U \to 0$ limit can also be understood from the analytic result based on Eq.~\eqref{eq:half_filled_F_prior} with 
$\mu = U/2 = 0$.

\begin{figure*}[t]
    \begin{minipage}[t]{0.49\textwidth}\centering
        \large\bfseries Semiclassical
    \end{minipage}
    \begin{minipage}[t]{0.49\textwidth}\centering
        \large\bfseries Exact     
    \end{minipage}
    \par
    \centering    
    \begin{minipage}[t]{0.23\textwidth}\centering
        \includegraphics[width=\linewidth]{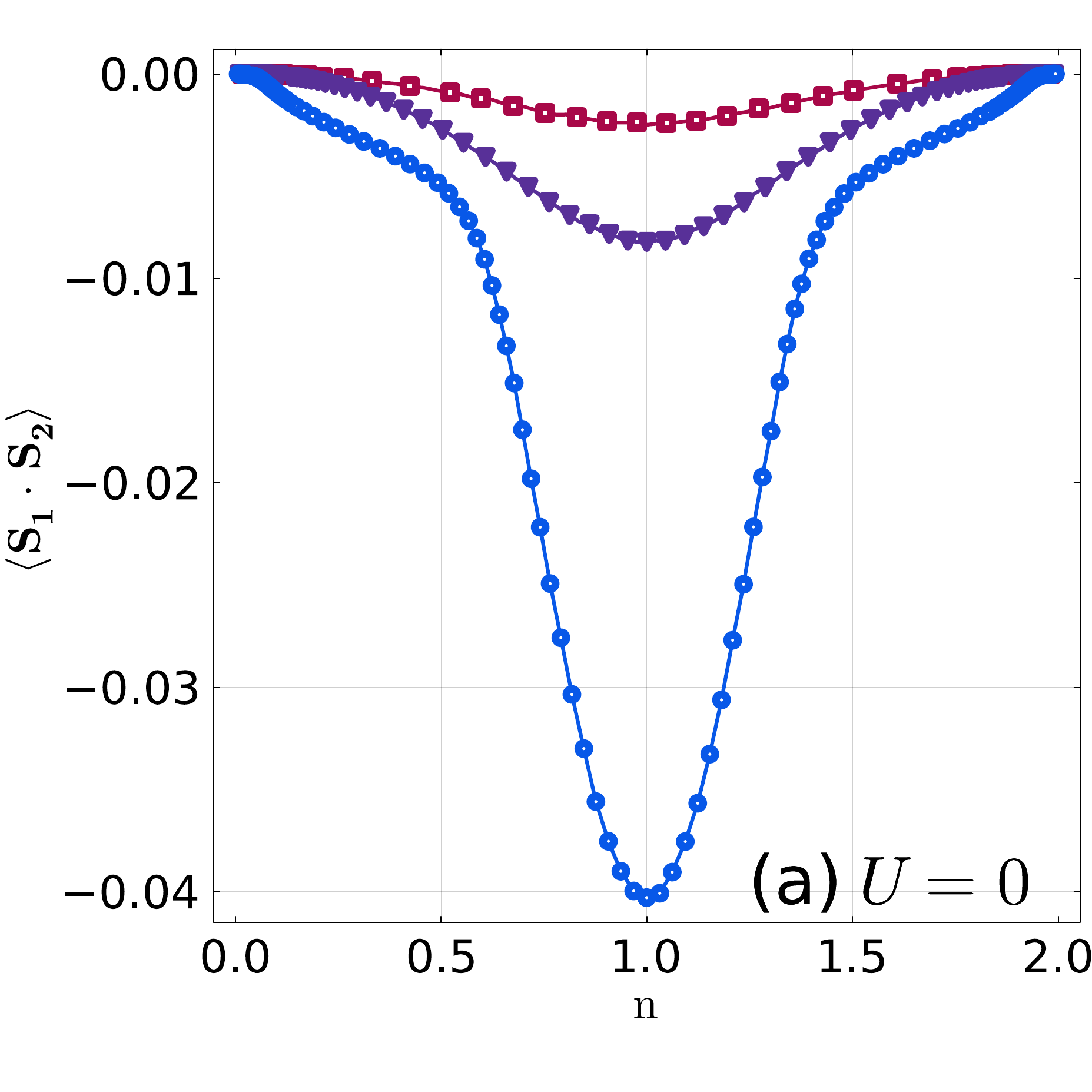}
    \end{minipage}
    \begin{minipage}[t]{0.23\textwidth}\centering
        \includegraphics[width=\linewidth]{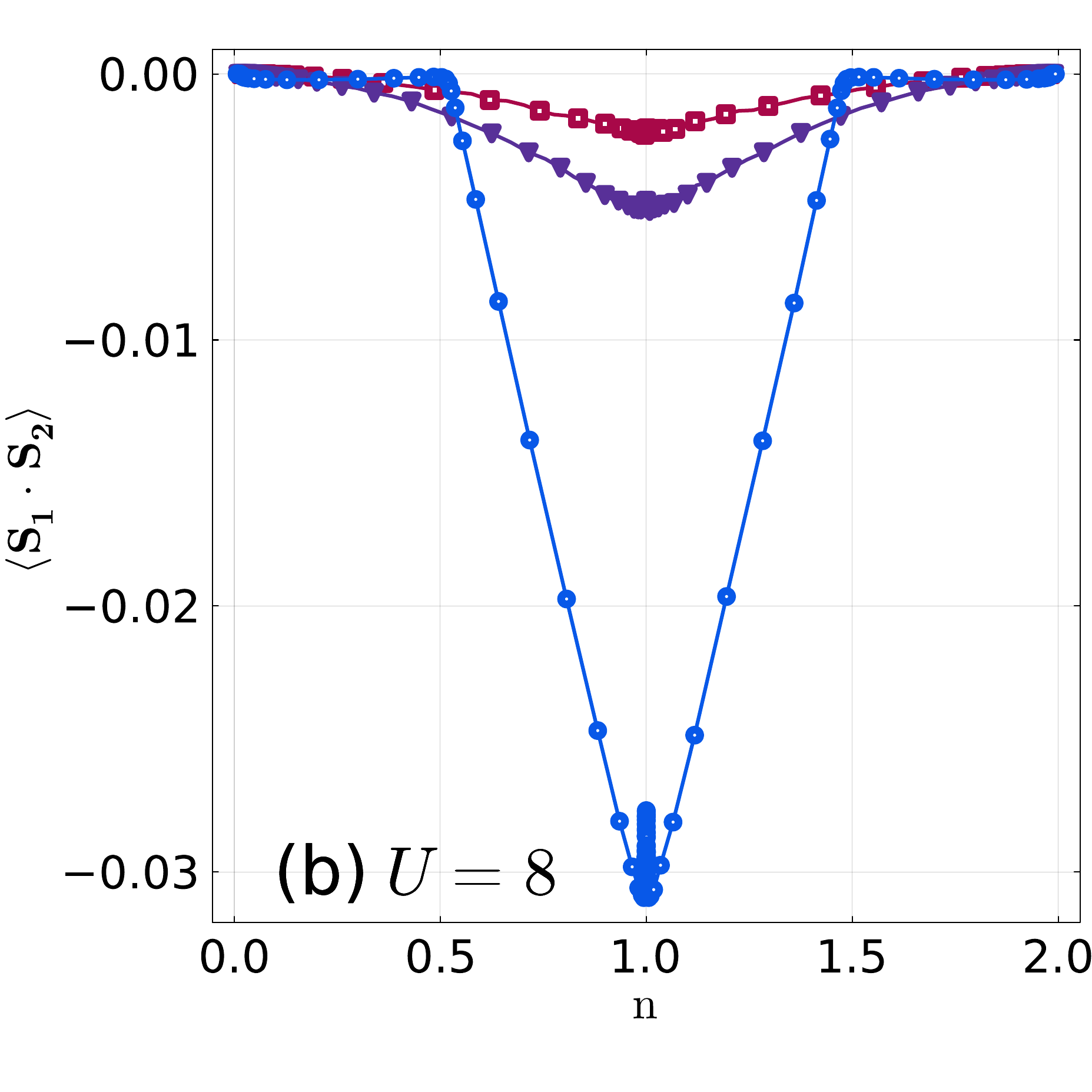}
    \end{minipage}
    \begin{minipage}[t]{0.23\textwidth}\centering
        \includegraphics[width=\linewidth]{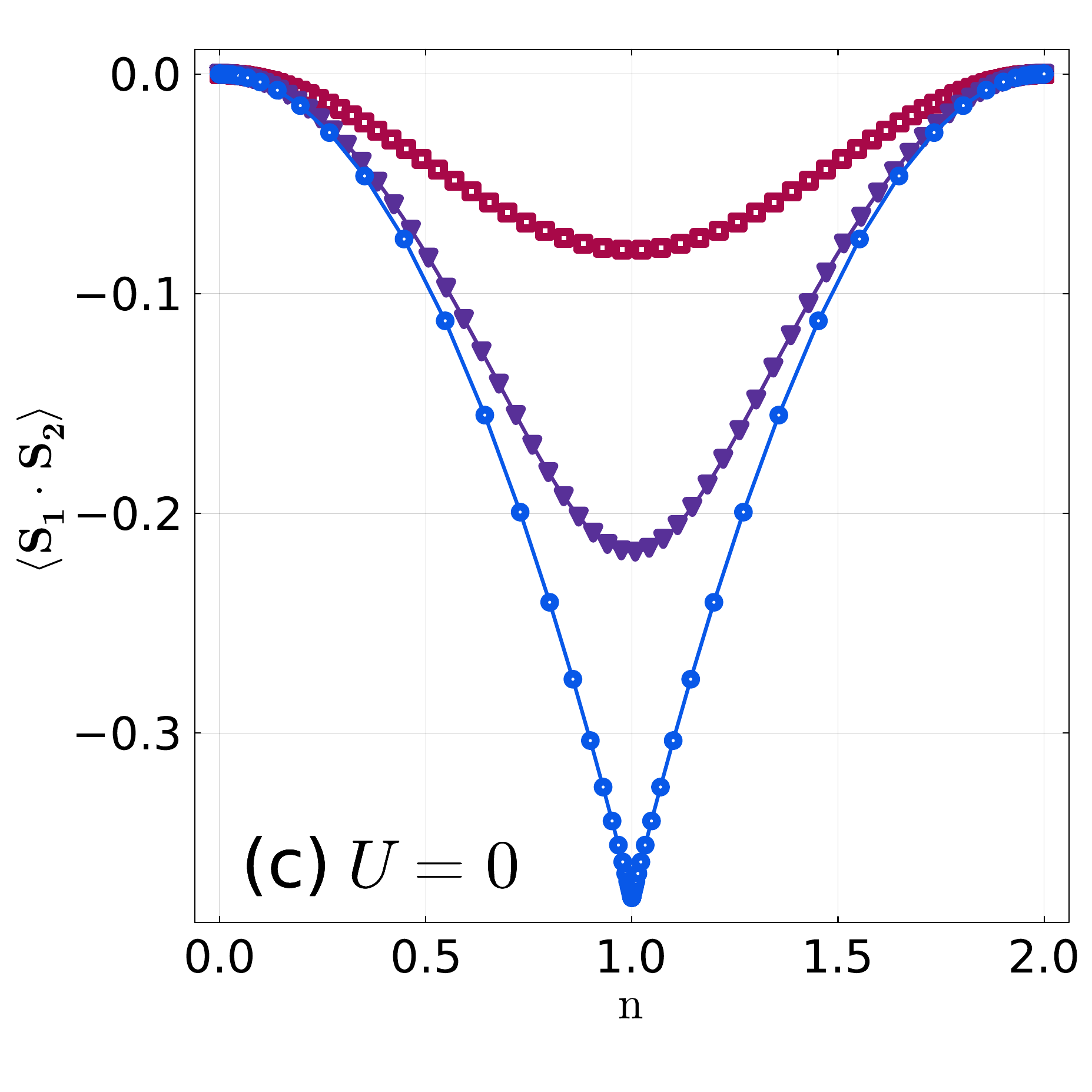}
    \end{minipage}
    \begin{minipage}[t]{0.23\textwidth}\centering
        \includegraphics[width=\linewidth]{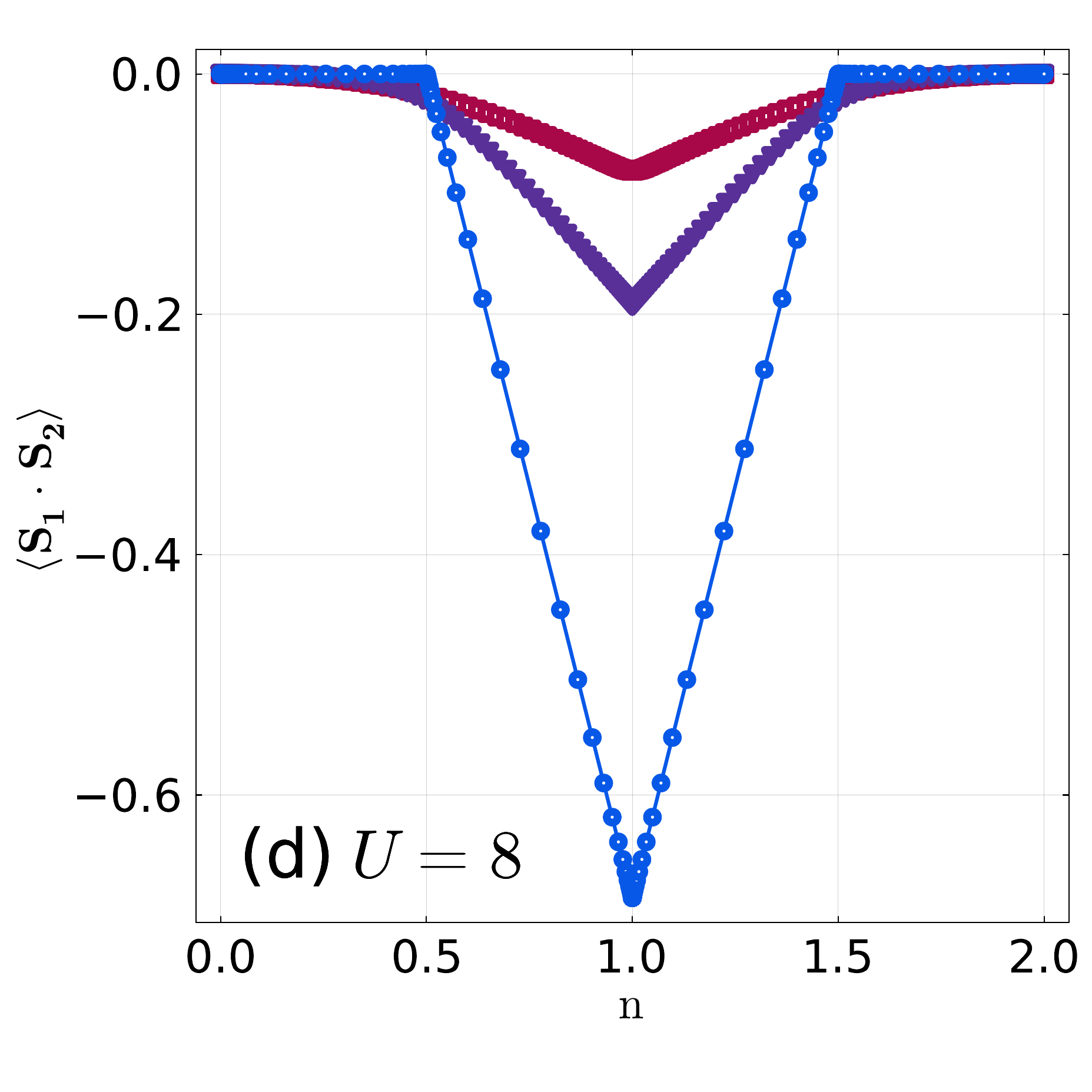}
    \end{minipage}\par 
    \vspace{-2mm}
    \begin{minipage}[t]{0.23\textwidth}\centering
        \includegraphics[width=\linewidth]{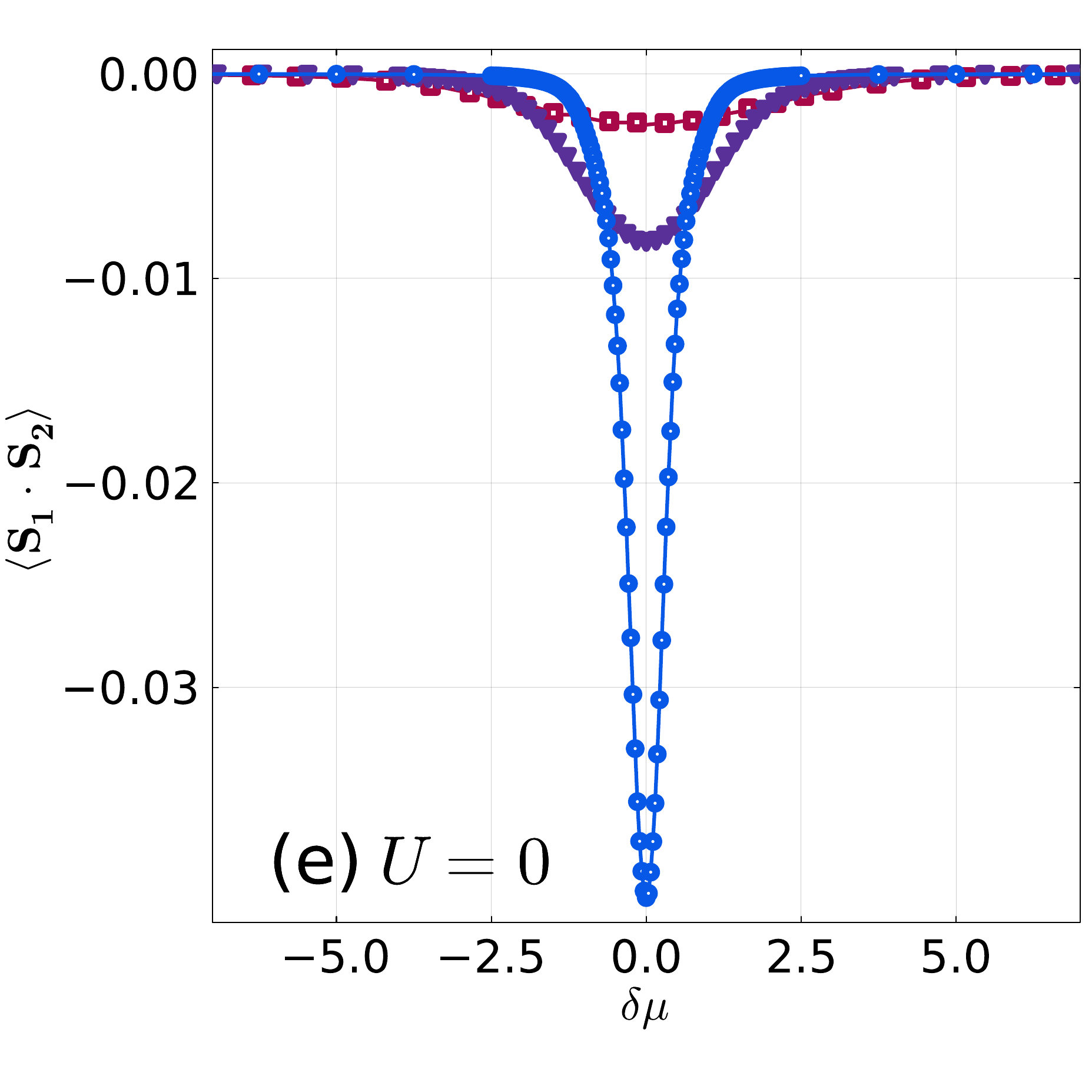}
    \end{minipage}
    \begin{minipage}[t]{0.23\textwidth}\centering
        \includegraphics[width=\linewidth]{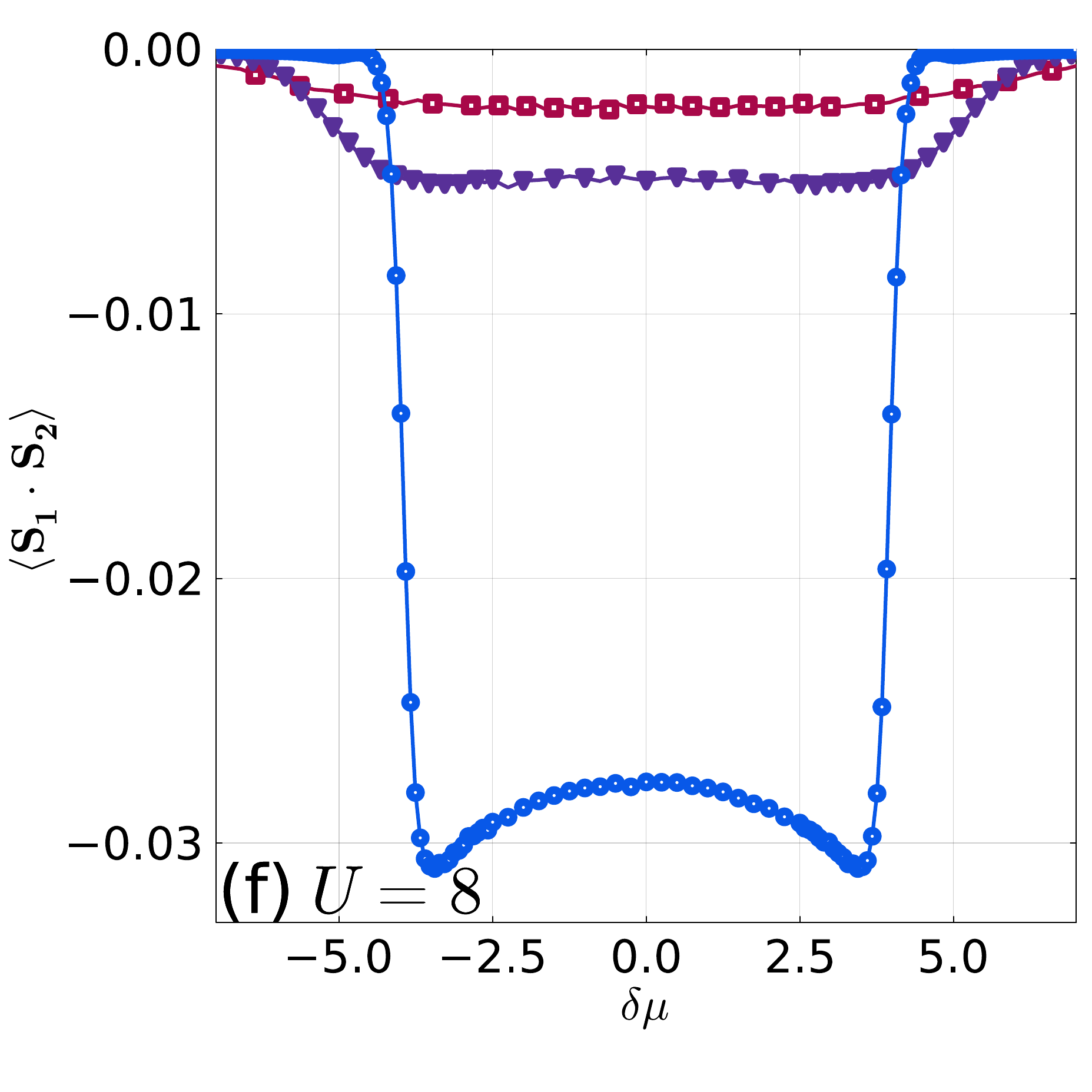}
    \end{minipage}
    \begin{minipage}[t]{0.23\textwidth}\centering
        \includegraphics[width=\linewidth]{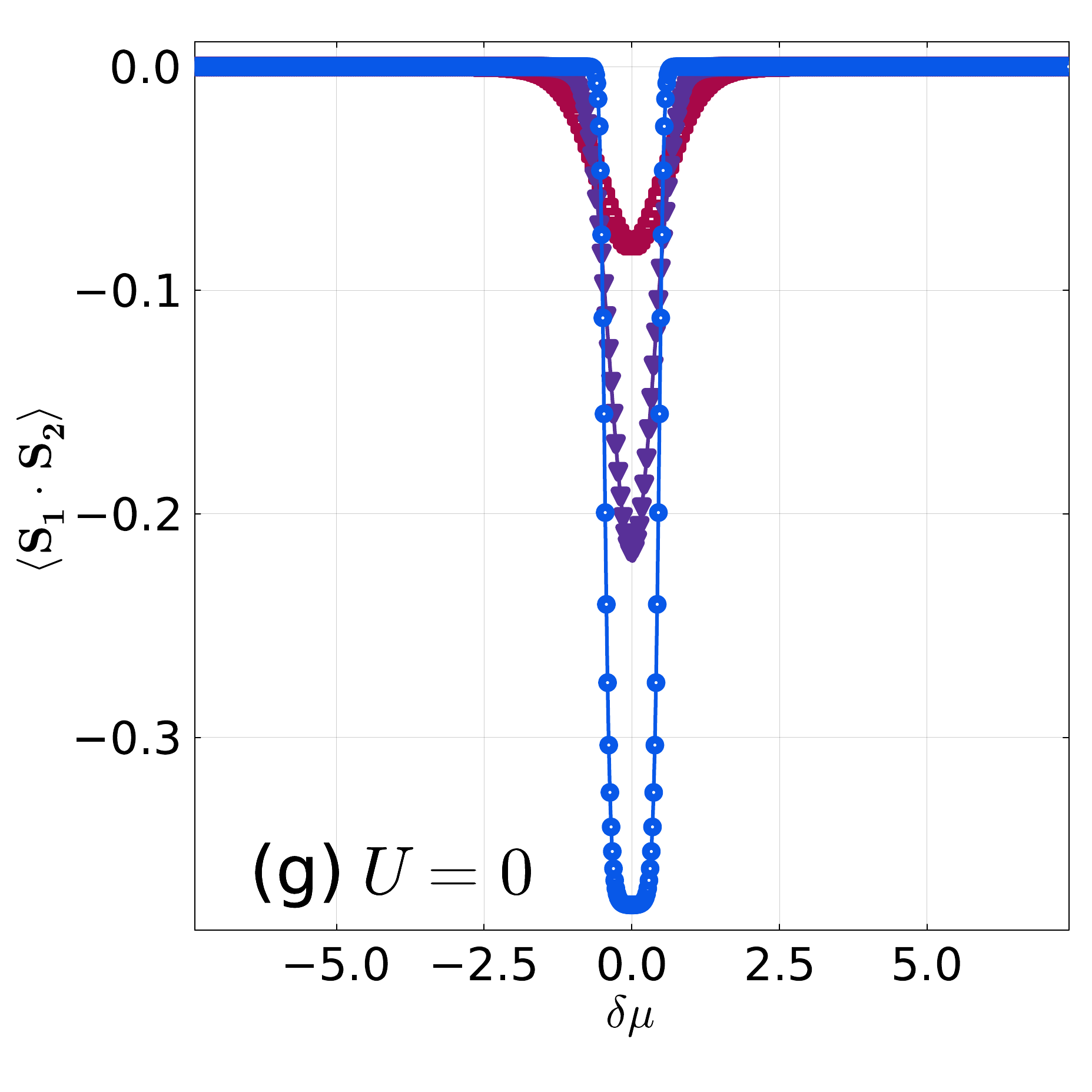}
    \end{minipage}
    \begin{minipage}[t]{0.23\textwidth}\centering
        \includegraphics[width=\linewidth]{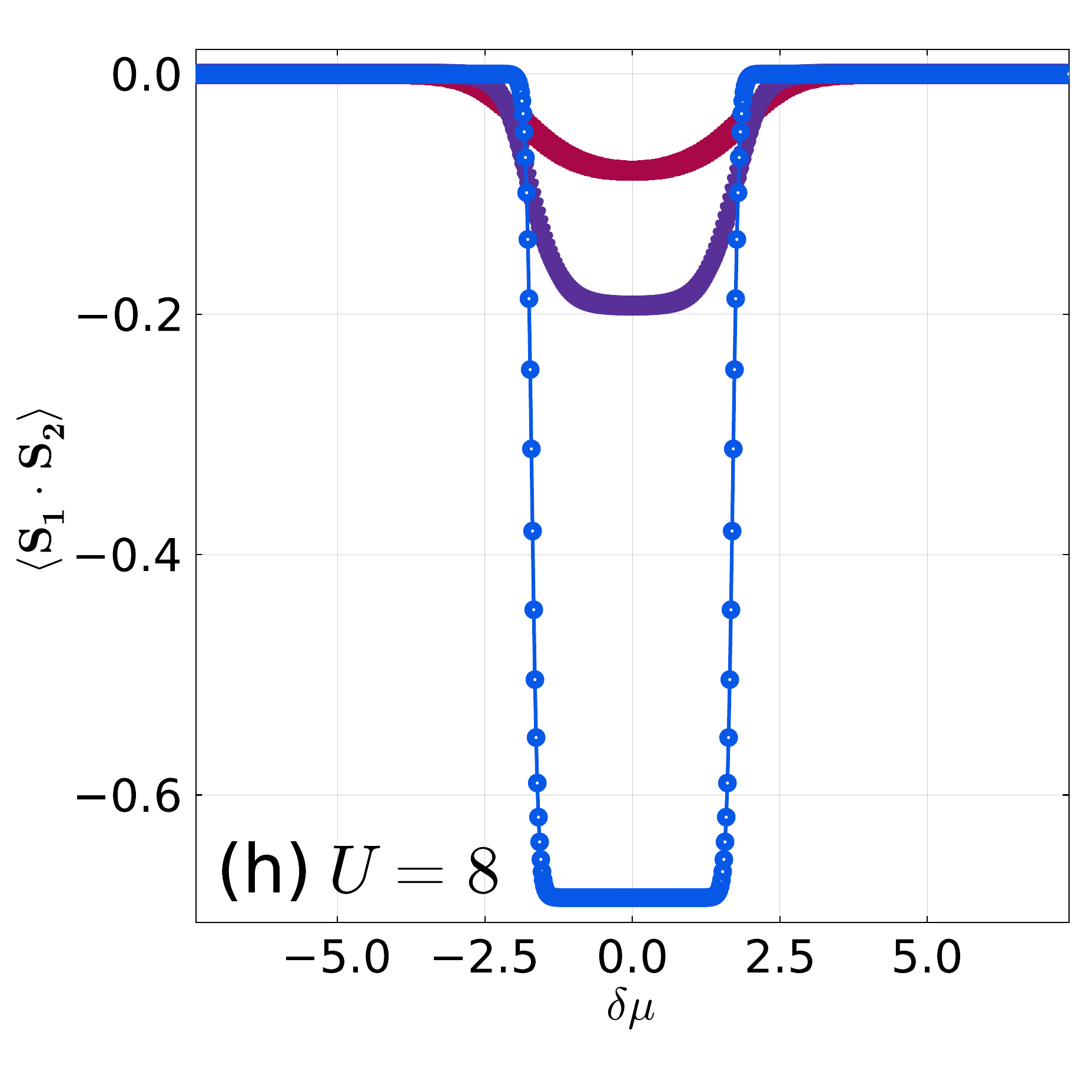}
    \end{minipage}\par 
    \vspace{-2mm}
    \begin{minipage}[t]{0.23\textwidth}\centering
        \hspace*{0.5mm}
        \includegraphics[width=\linewidth]{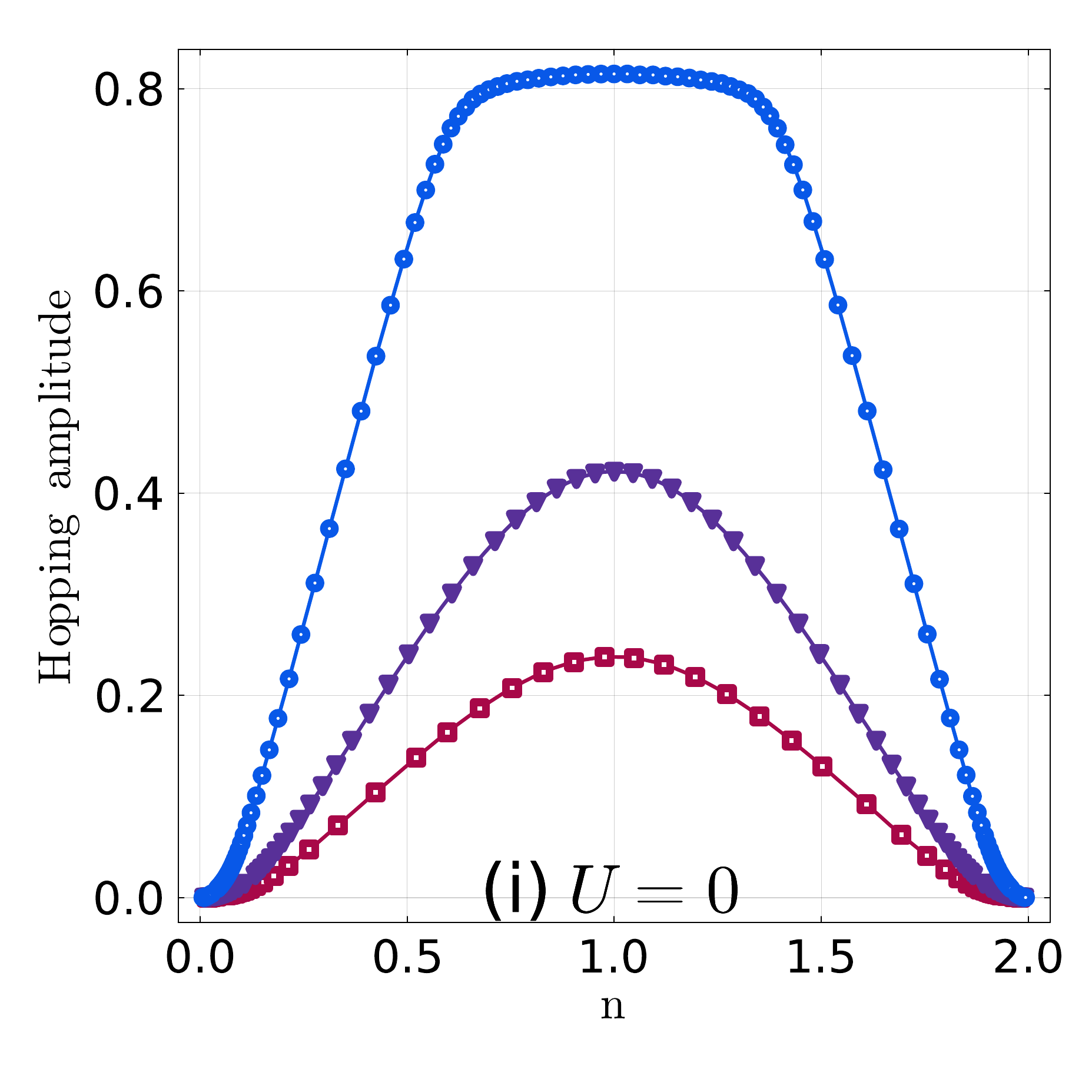}
    \end{minipage}
    \begin{minipage}[t]{0.23\textwidth}\centering
        \hspace*{0.5mm}
        \includegraphics[width=\linewidth]{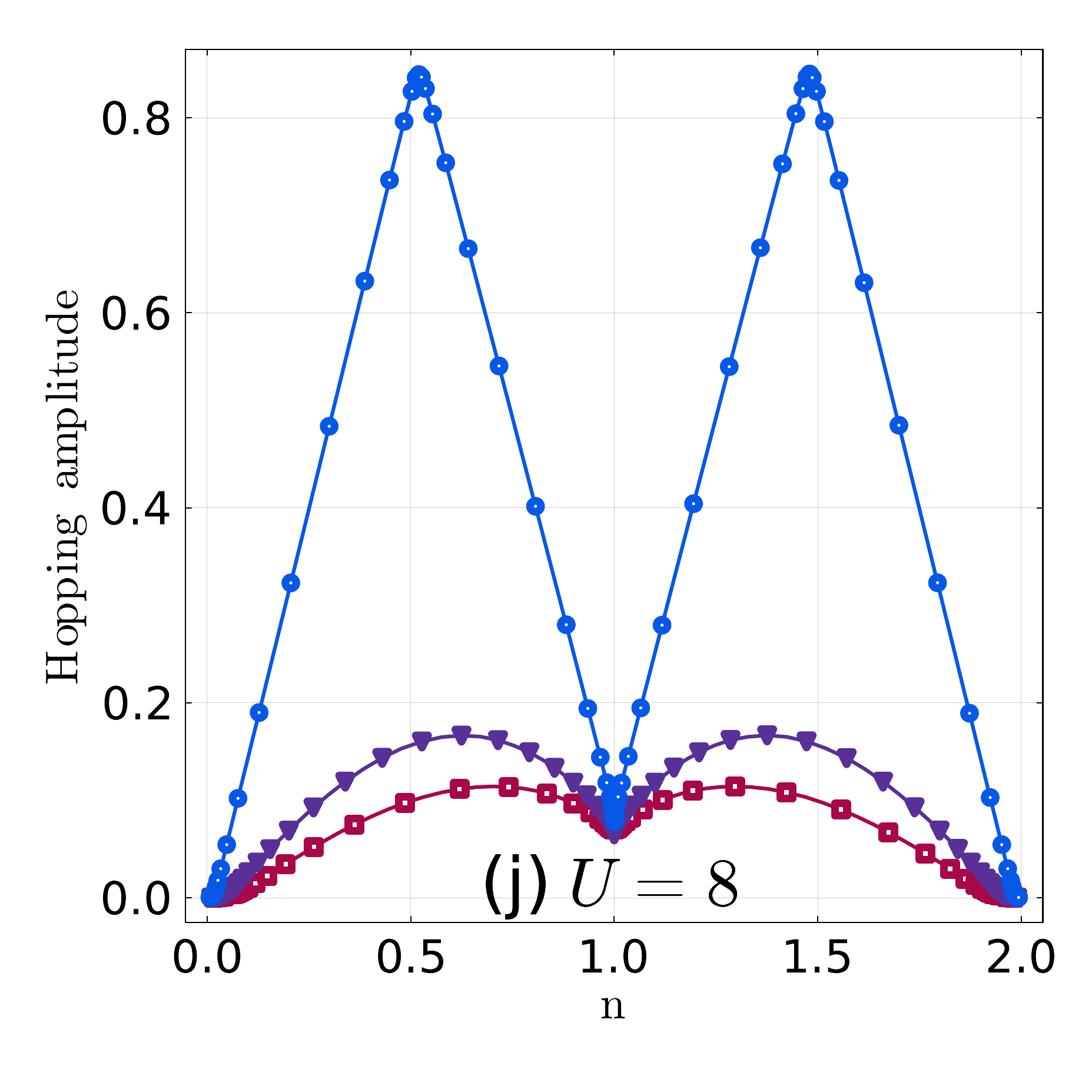}
    \end{minipage}
    \begin{minipage}[t]{0.23\textwidth}\centering
        \hspace*{0.0mm}
        \includegraphics[width=\linewidth]{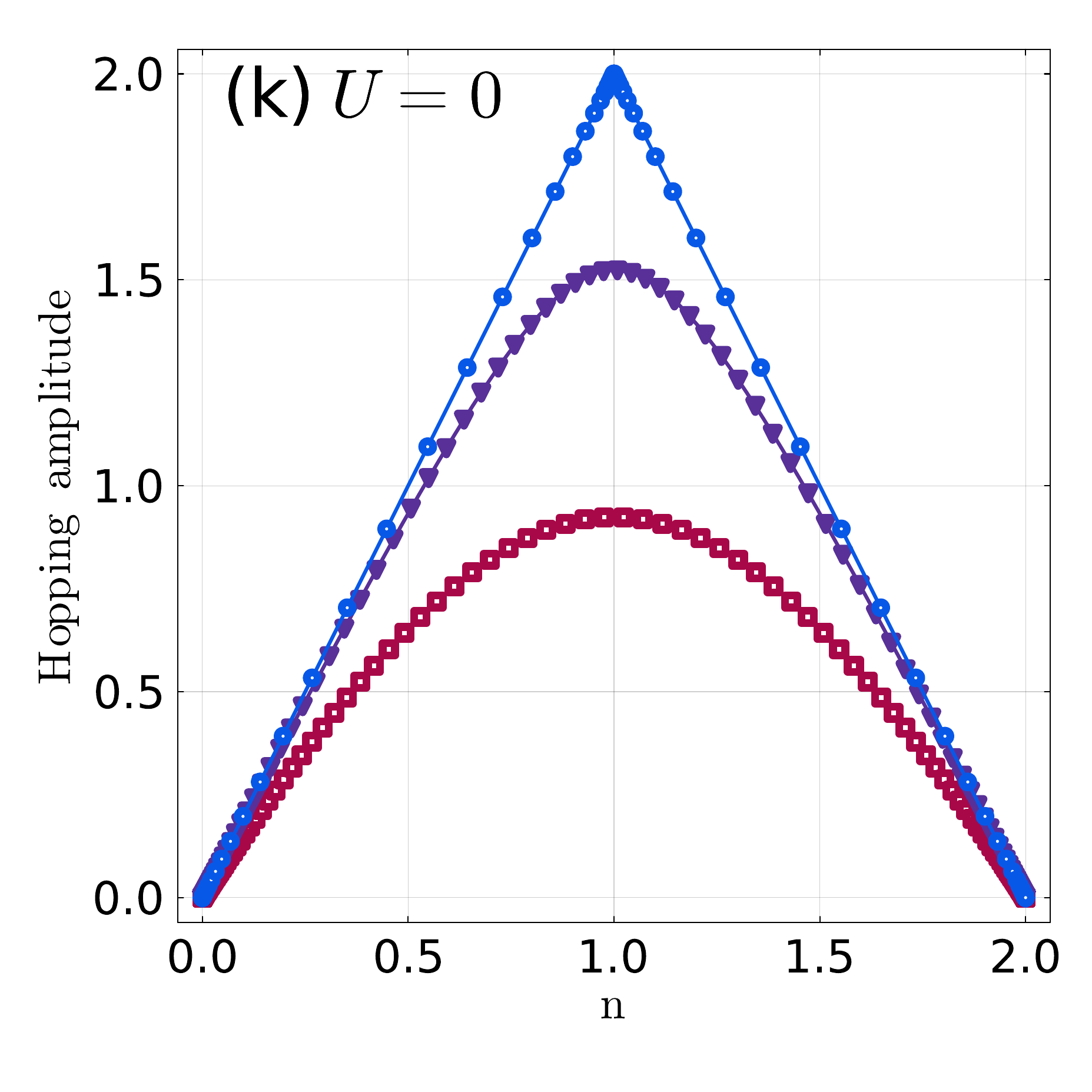}
    \end{minipage}
    \begin{minipage}[t]{0.23\textwidth}\centering
        \hspace*{-0.5mm}
        \includegraphics[width=\linewidth]{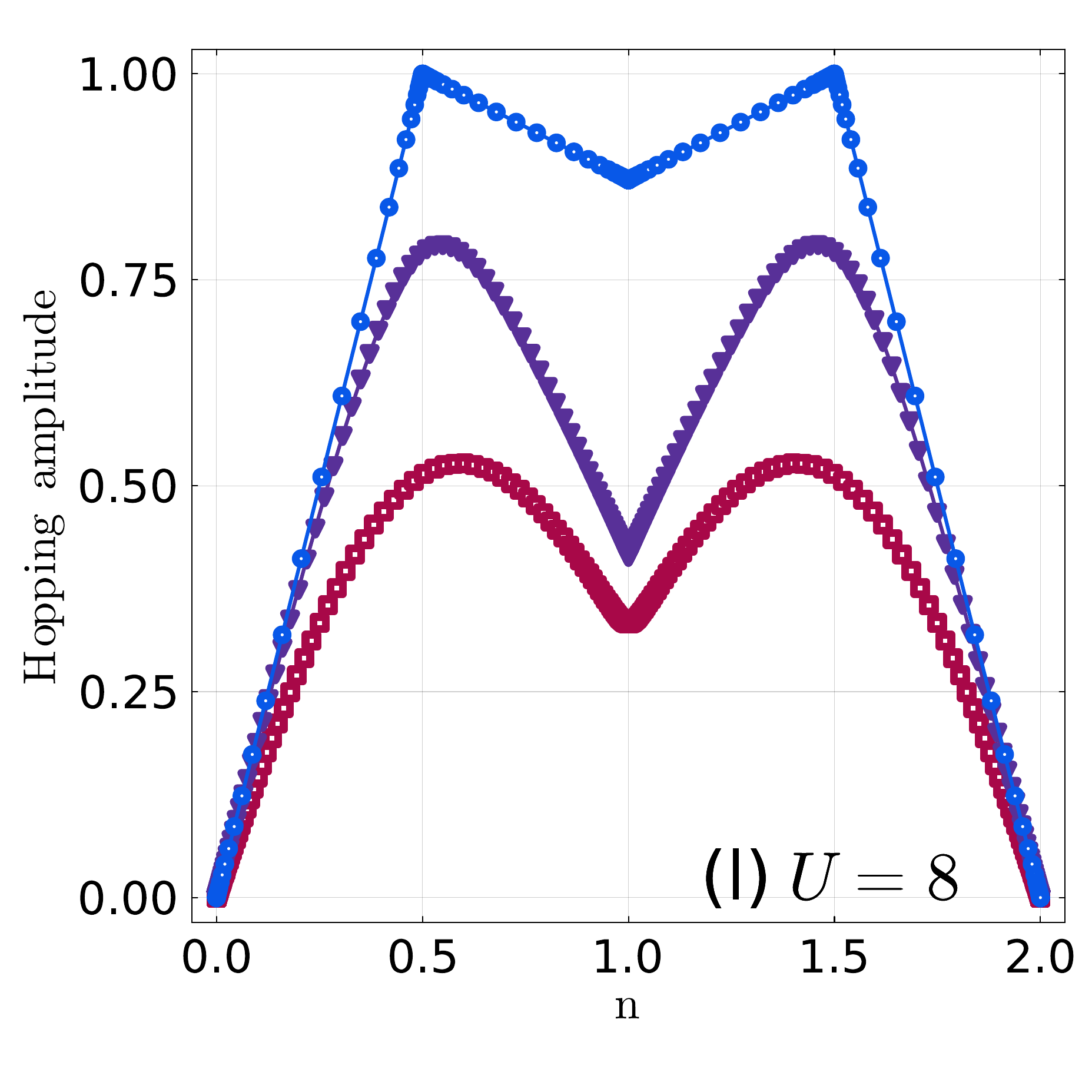}
    \end{minipage}
    \includegraphics[width=0.5\linewidth]                       {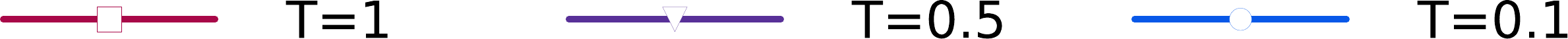}
    \caption{
(Top row) Filling dependence of the two-site spin correlation at several temperatures, 
obtained by (a,b) the semiclassical method and (c,d) exact calculation. 
The interaction parameter is $U=0$ in (a,c) and $U=8$ in (b,d). 
\ \ 
(Middle row) Same as in the top row, but with the horizontal axis replaced by the chemical potential, defined as $\delta\mu = \mu - U/2$. Panels (e,f,g,h) correspond to panels (a,b,c,d), respectively.
(Bottom row) Analogous to the top row, but with the vertical axis replaced by the hopping amplitude. The calculation methods and interaction strengths in panels (i,j,k,l) correspond to those in panels (a,b,c,d), respectively.
    }
    \label{fig:two_site_N}
\end{figure*}

\subsection{Filling dependence}

In this subsection, we discuss the filling dependence of quantities specific to the two-site system. 
Figures~\ref{fig:two_site_N}(a) and (b) show the filling dependence of the spin correlation $\langle {\bm S}_1 \cdot {\bm S}_2 \rangle$ obtained within the semiclassical approximation for $U=0$ and $U=8$, respectively. 
The corresponding exact results are shown in Figs.~\ref{fig:two_site_N}(c) and (d). 
Overall, the functional form is similar between the two methods. 
In particular, at half-filling $n=1$, the probability of having one electron with spin degrees of freedom is maximized, leading to a peak in the spin correlation, which then decreases monotonically as the filling deviates from half-filling. 
The smaller magnitude of the semiclassical results on the vertical axis reflects the differences in the characteristic energy scale and low-temperature limit discussed in the previous subsection.

Figure~\ref{fig:two_site_N}(b) exhibits a peculiar behavior at $n=1$ for $T=0.1$. 
To examine this in more detail, the middle row of Fig.~\ref{fig:two_site_N} [panels (e,f,g,h)] shows the same data as the top row but with the horizontal axis replaced by the chemical potential $\delta \mu = \mu - U/2$. 
From Fig.~\ref{fig:two_site_N}(f), it is evident that the singular behavior at $n=1$ is related to the $\mu$ dependence of the spin correlation. 
This behavior originates from the continuous distribution of states characteristic of the semiclassical approximation, which are sampled by the finite-temperature Boltzmann factor.
At low temperatures, however, such behavior is suppressed. 
While the convex region near $n=1$ at high temperatures appears to be an artifact of the semiclassical approximation, the deviation of its value compared to the nonconvex case remains at roughly the 10\% level.

Finally, the bottom row of Fig.~\ref{fig:two_site_N}  [panels (i,j,k,l)] shows the filling dependence of the hopping amplitude. 
For small $U$, the hopping amplitude is enhanced at $n=1$, where the probability of having a single electron per site is maximized, and decreases upon electron or hole doping. 
In contrast, for large $U$ at half-filling, strong on-site repulsion suppresses hopping amplitude, resulting in a small value. Upon doping with electrons or holes, the hopping gradually increases. 
For fillings around $n=0.5$ or $n=1.5$, the total number of electrons or holes 
falls below one, reducing the probability of double occupancy, rendering interaction effects irrelevant, and leading to a turnover from increasing to decreasing behavior in the filling dependence. 
These qualitative trends are captured even within the semiclassical approximation.

The above results complete the numerical benchmark for one- and two-site systems.
Before discussing alternative operator representations of the
Hubbard model (Sec.~\ref{sec:mapping}), it is useful to place the semiclassical approximation
introduced above on a controlled footing and to discuss its extension
to multi-orbital systems.  In the following two sections we first show
that the semiclassical approximation emerges as the large-$M$ limit of
a generalized Hubbard model, and then describe how the coherent-state
construction naturally generalizes to systems with multiple
spin–orbital flavors.

\section{Controlled semiclassical limit: large-\texorpdfstring{$M$}
{M} construction}

\label{sec:controlled_semiclassical_limit}

The semiclassical approximation introduced in Sec.~\ref{sec:semiclassical} can be
placed on a controlled footing by embedding the Hubbard model
into a sequence of theories whose local Hilbert spaces realize
increasingly large irreducible representations of the underlying
graded algebra.  In general, classical limits of quantum systems
associated with Lie algebras are obtained by considering a family
of highest-weight irreducible representations whose Dynkin labels
scale uniformly with an integer parameter $M$,
$\lambda^{(M)} = M\lambda^{(1)}$, and then taking the limit
$M\to\infty$.  In this limit the expectation values of the
generators become extensive while their commutators are suppressed
by $1/M$, so that the operator algebra reduces to the Poisson
algebra on the corresponding coadjoint orbit.  The construction
used below implements this standard procedure for the local
$\mathfrak{su}(2|2)$ graded structure of the Hubbard model by replicating the
local Hilbert space $M$ times and projecting onto the completely
symmetric sector, thereby generating the sequence of
highest-weight representations required for the semiclassical
limit.

We introduce $M$ identical replicas of the physical fermion
operators
\begin{equation}
c^{(r)}_{i\sigma}, \qquad r=1,\dots,M ,
\end{equation}
which act on $M$ copies of the local Hilbert space.
The generalized Hubbard Hamiltonian is
\begin{align}
H^{(M)} =
- t \sum_{\langle ij\rangle\sigma}
\sum_{r=1}^{M}
\left(
c^{(r)\dagger}_{i\sigma} c^{(r)}_{j\sigma}
+ {\rm H.c.}
\right)
+
U
\sum_{i,r}
n^{(r)}_{i\uparrow} n^{(r)}_{i\downarrow},
\label{eq:M_replicated_ham}
\end{align}
where
\begin{equation}
n^{(r)}_{i\sigma} =
c^{(r)\dagger}_{i\sigma} c^{(r)}_{i\sigma}.
\end{equation}
To isolate a single irreducible representation from the tensor product of
$M$ copies of the local Hilbert space, we introduce the projector
$P_i^{(M)}$ onto the completely symmetric highest-weight irreducible
representation contained in
\begin{align}
    \mathcal H_i^{(M)}=\bigotimes_{r=1}^M \mathcal H_{i,(r)}.
\end{align}
The corresponding global projector is
\begin{align}
    P^{(M)}=\prod_i P_i^{(M)}
    . \label{eq:projection}
\end{align}
Here $P_i^{(M)}$ projects onto the subspace of $\mathcal H_i^{(M)}$
that is invariant under permutations of the replica index
$r=1,\dots,M$, i.e.,
\begin{align}
P_i^{(M)}=\frac{1}{M!}\sum_{\pi\in S_M} U_i(\pi),
\end{align}
with $U_i(\pi)$ implementing the permutation $\pi$ on the replicas at
site $i$, and $S_M$ being the symmetric group on $M$ elements. 
Because this action is local in $i$, the global projector
factorizes as in Eq.~\eqref{eq:projection}.
Rather than working on the full replicated Hilbert space, we define the
sequence of projected Hamiltonians
\begin{align}
\tilde{H}^{(M)} \equiv P^{(M)} H^{(M)} P^{(M)},    
\end{align}
where $H^{(M)}$ is the replicated Hamiltonian introduced in Eq.~\eqref{eq:M_replicated_ham}. For
$M=1$, this construction reduces to the original Hubbard model, while for
$M>1$ it defines an auxiliary sequence of Hamiltonians whose local degrees
of freedom transform in the completely symmetric representation of the
local $\mathfrak{su}(2|2)$ algebra.

This projected sequence provides the natural setting for the semiclassical
limit. 
Within this symmetric sector it is convenient to introduce
collective operators
\begin{equation}
S^{a}_i = \sum_{r=1}^{M} S^{a}_{i,(r)}, 
\qquad
\eta^{a}_i = \sum_{r=1}^{M} \eta^{a}_{i,(r)},
\end{equation}
where $a=x,y,z$, and
\begin{equation}
S^{a}_{i,(r)} =
\frac12
\sum_{\sg\sg'}
c^{(r)\dagger}_{i\sg}
\sigma^{a}_{\sg\sg'}
c^{(r)}_{i\sg'}
\end{equation}
are the spin operators of replica $r$, and the operators
$\eta^{a}_i$ generate the charge
$SU(2)_\eta$ symmetry.

In the symmetric highest-weight representation, the coherent-state
action is proportional to $M$, so quantum fluctuations are suppressed as
$M\to\infty$, and the saddle-point approximation becomes asymptotically
exact:
\begin{equation}
\langle S^{a}_i \rangle \sim M,
\qquad
\langle \eta^{a}_i \rangle \sim M ,
\end{equation}
with $a=x,y,z$.
This scaling has a natural geometric interpretation.
The coherent states introduced in Sec.~\ref{sec:path-int} parameterize a
super-coadjoint orbit of the graded algebra
$\mathfrak{su}(2|2)$ whose bosonic base manifold is
$S^2_{\rm spin}\times S^2_{\eta}$.
In the symmetric representation obtained by replicating the
local Hilbert space $M$ times, the symplectic form on this
orbit scales linearly with $M$.
Consequently the classical trajectory on this orbit becomes
exact in the limit $M\to\infty$.

The origin of this scaling can be seen directly by
evaluating the Hamiltonian in the symmetric coherent
state.
Because the Hamiltonian is a sum over replicas,
each replica contributes to the same expectation value.
We denote by $\tilde H_t^{(M)}$ and $\tilde H_U^{(M)}$ the projected
kinetic and interaction contributions, respectively.
For the hopping term we obtain
\begin{equation}
\langle \tilde{H}_t^{(M)} \rangle
=
- t
\sum_{\langle ij\rangle\sigma}
\sum_{r=1}^M
\langle
c^{(r)\dagger}_{i\sigma}c^{(r)}_{j\sigma}
+{\rm H.c.}
\rangle
=
M\,\langle H_t^{(1)} \rangle .
\end{equation}
The same additivity holds for the interaction term,
\begin{equation}
\langle \tilde{H}_U^{(M)} \rangle
=
U\sum_{i,r}
\langle
n^{(r)}_{i\uparrow}n^{(r)}_{i\downarrow}
\rangle
=
M\,\langle H_U^{(1)} \rangle .
\end{equation}
Thus the classical energy functional scales extensively,
\begin{equation}
E_M[\Omega,\psi]
=
M\,E_1[\Omega,\psi].
\end{equation}
The coherent-state path integral constructed in Sec.~\ref{sec:path-int}
can therefore be generalized straightforwardly to this
enlarged representation.
In the symmetric sector the coherent state is the
symmetric tensor product of $M$ fundamental coherent
states,
\begin{widetext}
\begin{equation}
|\Omega_c,\Omega_s;\psi\rangle_M
= 
\bigotimes_{r=1}^M
\left[
C_1(\Omega_c)\,|0\rangle_r
-\psi\Bigl(
C_\uparrow(\Omega_s)\,|\uparrow\rangle_r
+
C_\downarrow(\Omega_s)\,|\downarrow\rangle_r
\Bigr)
+
C_4(\Omega_c)\,|\uparrow\downarrow\rangle_r
\right],
\label{eq:coherent_M_tensor}
\end{equation}
\end{widetext}
Since all replica factors are identical, the tensor product state is manifestly invariant under any permutation of the replica indices $r=1,\dots,M$, and therefore already belongs to the completely symmetric sector of the replicated Hilbert space.
This construction makes explicit that all replicas share
the same set of coefficients $C_1,\ldots,C_4$, while the
tensor product is projected onto the completely symmetric
irreducible representation. The semiclassical limit is
then controlled by the representation size $M$.

The Berry phase and the full action
scale linearly with $M$,
\begin{equation}
S_{\rm eff}[\Omega,\psi] =
M\,S_0[\Omega,\psi].
\end{equation}
The partition function therefore takes the form
\begin{equation}
Z =
\int \mathcal D \Omega \mathcal D \psi\,
\exp\!\big( -M S_0[\Omega,\psi] \big).
\end{equation}
In the limit
$M\to\infty$,
the functional integral is dominated by the stationary
configuration of the action,
\begin{equation}
\frac{\delta S_0}{\delta \Omega}=0 ,
\end{equation}
which corresponds to 
the semiclassical equation of motion.
On the other hand, from the viewpoint of statistical mechanics, the Berry-phase term oscillates rapidly in the imaginary-time formalism in the large-$M$ limit.
As a result, only trajectories without imaginary-time dependence contribute to the partition function.
This corresponds to the semiclassical approximation employed
in Sec.~\ref{sec:semiclassical}.
Fluctuations around this saddle are suppressed by powers
of $1/M$, so that the semiclassical approximation becomes
exact in the large-$M$ limit.
The parameter $1/M$ therefore plays a role analogous to
$\hbar_{\rm eff}$ in ordinary semiclassical expansions,
\begin{equation}
\hbar_{\rm eff} \sim \frac{1}{M}.
\end{equation}
Note that in the half-filled large-$U$ limit, the projected Hamiltonian
$\tilde H^{(M)}$ reduces to an effective Heisenberg model
\begin{align}
    H_{\rm eff}^{(M)} = J \sum_{\langle ij\rangle}
\bm{S}_i \cdot \bm{S}_j
,
\end{align}
where $\bm{S}_i = \sum_{r=1}^M \bm{S}_{i,(r)}$ is the total spin
operator in the completely symmetric representation, corresponding to
spin $S=M/2$. 
Therefore, in this strong-coupling regime of the half-filled Hubbard model, the large-$M$ limit is equivalent
to the conventional large-$S$ limit.

Thus the present semiclassical formulation can be viewed
as the leading term of a systematic $1/M$ expansion of
the Hubbard model formulated in the completely symmetric
representation of the local $\mathfrak{su}(2|2)$ algebra.
This large-$M$ construction therefore provides a controlled
semiclassical limit of the Hubbard model that preserves the
underlying $\mathfrak{su}(2|2)$ graded algebraic structure.
In the following section we show that the same coherent-state
framework admits a natural generalization to multi-orbital
systems, where the graded-algebra structure becomes even more
transparent.

\section{Extension to multiorbital: Graded-algebra formulation}
\label{sec:multiorb}

The extension of the present construction to multi-orbital systems can
be formulated naturally using the $\mathbb{Z}_2$ grading associated with
fermion parity.
For clarity we focus on the local Hilbert space and suppress the site
index.

For $N$ spin–orbital flavors the local Hilbert space is the fermionic
Fock space
\begin{equation}
\mathcal H =
\mathrm{span}\left\{
(c_1^\dagger)^{n_1}\cdots (c_N^\dagger)^{n_N}|0\rangle
\right\},
\qquad
n_\gamma\in\{0,1\},
\end{equation}
whose dimension is $2^N$.
It carries a natural $\mathbb Z_2$ grading determined by the fermion
parity operator
\begin{equation}
\mathcal H
=
\mathcal H_{\bar 0}
\oplus
\mathcal H_{\bar 1},
\qquad
(-1)^{\hat n} |\varPsi\rangle
=
\pm |\varPsi\rangle,
\end{equation}
where $\hat n=\sum_\gamma c^\dagger_\gamma c_\gamma$.
The subspaces $\mathcal H_{\bar 0}$ and $\mathcal H_{\bar 1}$
contain states with even and odd fermion number,
respectively.
Both sectors have equal dimension
\begin{equation}
\dim \mathcal H_{\bar 0}
=
\dim \mathcal H_{\bar 1}
=
2^{N-1}.
\end{equation}
Operators acting on $\mathcal H$ inherit this grading.
Even operators preserve fermion parity, whereas odd operators change
it.
The full algebra of operators respecting the grading can therefore be
viewed as a Lie superalgebra acting on the graded vector space
$\mathcal H_{\bar 0}\oplus\mathcal H_{\bar 1}$.
In particular the fermionic creation and annihilation operators act as
odd generators that map
\begin{equation}
c^\dagger_\gamma :
\mathcal H_{\bar p} \rightarrow \mathcal H_{\overline{p+1}},
\qquad
c_\gamma :
\mathcal H_{\bar p} \rightarrow \mathcal H_{\overline{p+1}},
\qquad
p\in\{0,1\},
\end{equation}
and $\overline{p+1}\equiv p+1 \pmod 2$.

In the conventional fermionic path integral one introduces $N$
independent Grassmann variables associated with the $N$ fermionic
modes.
Here we instead exploit the parity grading and introduce a coherent
state with a single Grassmann coordinate.  Rather than assuming that
the bosonic parameters are shared by both parity sectors, we allow the
even and odd sectors to be parametrized independently.  Denoting the
corresponding bosonic coordinates by $\Omega_{\bar 0}$ and
$\Omega_{\bar 1}$, respectively, we write
\begin{equation}
|\Omega_{\bar 0},\Omega_{\bar 1};\psi\rangle
=
\sum_{\alpha=1}^{2^N}
C_\alpha(\Omega_{\bar 0},\Omega_{\bar 1})\,
\psi^{P_\alpha}
|\alpha\rangle ,
\end{equation}
where $|\alpha\rangle$ denotes the Fock basis,
$P_\alpha\in\{0,1\}$ is the fermion parity of the state,
and $\Omega_{\bar 0}$ and $\Omega_{\bar 1}$ parametrize rotations
within the even- and odd-parity sectors, respectively.
The Grassmann number is denoted by $\psi$.

Separating the two sectors explicitly gives
\begin{equation}
|\Omega_{\bar 0},\Omega_{\bar 1};\psi\rangle
=
|\alpha(\Omega_{\bar 0})\rangle
+
\psi\,|\beta(\Omega_{\bar 1})\rangle ,
\end{equation}
with
\begin{equation}
|\alpha(\Omega_{\bar 0})\rangle \in \mathcal H_{\bar 0},
\qquad
|\beta(\Omega_{\bar 1})\rangle \in \mathcal H_{\bar 1}.
\end{equation}
Thus the single Grassmann coordinate $\psi$ mediates transitions
between the even and odd sectors of the graded Hilbert space.
In this sense the construction represents a minimal Grassmann
parametrization of the fermionic Fock space consistent with fermion
parity.

For the single-orbital case ($N=2$) the local Hilbert space contains
four states
$
\{ |0\rangle, \;
|\uparrow\rangle, \;
|\downarrow\rangle, \;
|\uparrow\downarrow\rangle  \}
$.
These states naturally split into sectors of even and odd fermion
parity,
\begin{align}    
\mathcal H_{\bar 0}=\{|0\rangle,|\uparrow\downarrow\rangle\},
\qquad
\mathcal H_{\bar 1}=\{|\uparrow\rangle,|\downarrow\rangle\},
\end{align}
which are both two-dimensional.  In this case the coherent-state
construction reduces to the familiar $SU(2)\times SU(2)$ structure
discussed in the previous sections: one $SU(2)$ corresponds to spin rotations,
while the second corresponds to the $\eta$-pairing symmetry that mixes
empty and doubly occupied states.  The odd generators of the graded
algebra connect the two sectors and are represented by Grassmann
coordinates in the coherent-state path integral.

The graded structure of the local Hilbert space makes it natural to
consider operators built from fermionic bilinears.
A particularly important subset of these operators consists of the
quadratic fermion bilinears
\begin{align}
    Q_{\al\beta} =  c_\alpha^\dagger c_\beta, \ \ 
P^\dg_{\al\beta} = c_\alpha^\dagger c_\beta^\dagger,\ \ 
P_{\al\beta} = c_\beta c_\alpha ,
\end{align}
for $(\alpha,\beta=1,\dots,N)$
 These operators close under
commutation and generate the $SO(2N)$ Lie algebra
The number-conserving bilinears
$Q_{\alpha\beta}$
generate the algebra $U(N)=SU(N) \times U(1)$ corresponding to flavor
rotations and charge conservation.  The remaining generators $P_{\al\beta}^{(\dg)}$ are the
antisymmetric pair-creation and pair-annihilation operators
which change the fermion number by two and therefore connect
states within the same parity sector. 

For $N=2$ these operators reproduce the well-known decomposition
$SO(4)\simeq SU(2)_{\rm spin}\times SU(2)_{\eta}$,
where the first $SU(2)$ corresponds to spin rotations and the second
to the $\eta$-pairing symmetry that rotates between empty and doubly
occupied states.  For $N>2$ the algebra does not split into two
independent $SU(N)$ sectors; instead all generators combine into the
single algebra $SO(2N)$.  The familiar $SU(2)\times SU(2)$ structure
therefore appears as the special low-rank case $N=2$ of this more
general algebraic framework.

We denote by $L$ the number of Grassmann variables.
A natural explicit realization of the $L=1$ construction is obtained by
choosing the vacuum $|0\rangle$ as reference state and generating the
even sector with pair-creation operators.  Introducing an antisymmetric
complex matrix $\Omega_{\alpha\beta}=-\Omega_{\beta\alpha}$, we define
the even coherent state
\begin{equation}
|\Omega\rangle =
\exp\!\left(
\frac12\sum_{\alpha,\beta=1}^N
\Omega_{\alpha\beta}\,
c_\alpha^\dagger c_\beta^\dagger
\right)|0\rangle .
\end{equation}
The odd sector is then generated by acting with a single Grassmann
direction,
\begin{equation}
|\Omega,\psi\rangle
=
\left(
1
+
\psi
\sum_{\alpha=1}^{N}
\theta_{\alpha}\, c^{\dagger}_{\alpha}
\right)
\exp\!\left(
\frac{1}{2}
\sum_{\alpha,\beta=1}^{N}
\Omega_{\alpha\beta}\,
c^{\dagger}_{\alpha} c^{\dagger}_{\beta}
\right)
|0\rangle ,
\label{eq:super_coherent_state}
\end{equation}
where $\psi$ is the Grassmann variable and
$\theta=(\theta_1,\dots,\theta_N)$ is a complex $N$-component spinor.
This  generalization of Eq.~\eqref{eq:coherentnew} provides an explicit BCS-type representative of the minimal
$L=1$ super--coherent-state construction. In this explicit realization the odd state $|\beta\rangle$ depends on
both the pair amplitude $\Omega_{\alpha\beta}$ and the spinor
$\theta_\alpha$, so that the odd-sector coordinates may be viewed as
$\Omega_{\bar1}=(\Omega,\theta)$.

The explicit parametrization above also makes it possible to identify
the manifold over which these coherent states are defined, which is
important for constructing the resolution of the identity in the
path-integral formulation. The antisymmetric matrix
$\Omega_{\alpha\beta}$ contains $\binom{N}{2} = N(N-1)/2$ complex parameters and
provides a local coordinate chart for the orbit of the vacuum under
the pairing generators. In the fermionic Fock space the even and odd
parity sectors furnish the two chiral spinor representations of
$SO(2N)$, and the corresponding coherent states are obtained by acting
with the group generated by the quadratic bilinears
$c^\dagger_\alpha c_\beta$, $c^\dagger_\alpha c^\dagger_\beta$, and
$c_\beta c_\alpha$~\cite{RingSchuck1980,Perelomov1986}.

To identify the orbit, note that the number-conserving bilinears
$Q_{\alpha\beta}=c^\dagger_\alpha c_\beta$ generate the subalgebra
$U(N)\subset SO(2N)$, while the remaining generators are the pair
creation and pair annihilation operators. Since
$c_\alpha|0\rangle=0$, the vacuum is annihilated by all pair
annihilation operators, and it is also invariant under $U(N)$ because
$c^\dagger_\alpha c_\beta |0\rangle=0$. Therefore the isotropy
subgroup of the vacuum is $U(N)$, and the coherent-state manifold is
the coset
\begin{align}
SO(2N)/U(N),
\nonumber    
\end{align}
which has real dimension $N(N-1)$.

The spinor $\theta_\alpha$ appearing in the odd sector enters only
through the combination
$\psi \sum_\alpha \theta_\alpha c_\alpha^\dagger$.
Because the overall complex rescaling
$(\psi,\theta_\alpha)\rightarrow (\lambda^{-1}\psi,\lambda\theta_\alpha)$
leaves the state invariant, the coefficients $\theta_\alpha$ define a
point in the complex projective space $\mathbb{CP}^{N-1}$ (manifold of one-dimensional subspaces of the vector space $\mathbb{C}^{N}$).

Consequently, for this explicit $L=1$ realization, the bosonic part of
the super--coherent-state manifold is
\begin{align}
 \big(SO(2N)/U(N)\big)\times \mathbb{CP}^{N-1}.   
 \nonumber
\end{align}
For $N=2$ one has $SO(4)/U(2)\simeq S^2$ and
$\mathbb{CP}^1\simeq S^2$, so the construction reduces to the familiar
$S^2\times S^2$ manifold associated with the
$SU(2)_{\rm spin}\times SU(2)_{\eta}$ structure.  Integration over this manifold provides the resolution of the identity
associated with the corresponding coherent-state path integral.

The construction above corresponds to the minimal case $L=1$.
More generally one may partition the $N$ spin–orbital flavors into
blocks
\begin{equation}
N = K_1 + \cdots + K_L ,
\end{equation}
and associate one Grassmann coordinate with each block.
Each block then has a local Hilbert space of dimension $2^{K_a}$
with its own parity decomposition
\begin{equation}
\mathcal H^{(a)}
=
\mathcal H^{(a)}_{\bar 0}
\oplus
\mathcal H^{(a)}_{\bar 1},
\qquad
\dim \mathcal H^{(a)}_{\bar 0}
=
\dim \mathcal H^{(a)}_{\bar 1}
=
2^{K_a-1}.
\end{equation}
The local coherent state can then be written schematically as
\begin{equation}
|\Omega_{\bar 0},\{\Omega_{\bar 1}^{(a)}\};\{\psi_a\}\rangle
=
|\alpha(\Omega_{\bar 0})\rangle
+
\sum_{a=1}^{L}
\psi_a\, |\beta_a(\Omega_{\bar 1}^{(a)})\rangle
+\cdots ,
\end{equation}
where $|\alpha(\Omega_{\bar 0})\rangle \in \mathcal H_{\bar 0}$ denotes
the even-parity component, while
$|\beta_a(\Omega_{\bar 1}^{(a)})\rangle \in \mathcal H_{\bar 1}^{(a)}$
denotes the odd direction associated with block $a$.
The Grassmann variables $\psi_a$ therefore represent independent odd
directions associated with the different blocks, and the corresponding bosonic coordinates need not be shared between the
different odd directions.

The number $L$ of Grassmann coordinates therefore controls how many
fermionic degrees of freedom are treated explicitly.
In the extreme cases, $L = N$ corresponds to the standard fermionic
coherent-state path integral with $N$ independent Grassmann variables,
whereas $L = 1$ corresponds to the minimal construction used in this work,
in which the entire local Hilbert space is described by a single Grassmann
coordinate.

So far, we have considered generators composed only of $c_\alpha^\dagger$ and $c_\alpha^\dagger c_\beta^\dagger$ as shown in Eq.~\eqref{eq:super_coherent_state}.
More generally, one can also include products of three or more fermionic creation operators such as $c_\alpha^\dagger c_\beta^\dagger c_\gm^\dg$ as generators.
In the sector with $k$ fermionic operators, the number of generators is given by the binomial coefficient $\binom{N}{k}$.
When the set of generators is taken to be maximal, the total number of operators associated with each of the even and odd sectors is $2^{N-1}$.
Therefore, in the most general case, one can construct coherent states based on $SU(2^{N-1}) \times SU(2^{N-1})$.
Even in this case, it is sufficient to introduce only a single Grassmann variable ($L = 1$) in order to allow for the coexistence of the even and odd sectors.
Although, for multiorbital systems in real materials, the Hilbert space of dimension $2^{N}$ may be so large that a full exploration of the phase space becomes intractable, one can truncate the space depending on the specific details (e.g., the electron number), and the maximal-coherent-state approach can still be applied within this truncated Hilbert space.

\section{Mapping onto Majorana-Kondo lattice}
\label{sec:mapping}

\subsection{Hubbard to Majorana-Kondo}

In this section, we analyze the effective action without invoking  the semiclassical approximation.  
Within the path-integral representation, one may choose $C_1 = C_\uparrow$ and $C_\downarrow = C_4$ without loss of generality. 
This choice still provides a resolution of unity and leads to an alternative form of the effective Hamiltonian.  
Although this constraint on the $C$-coefficients may appear unnatural at first---since $C_{\uparrow,\downarrow}$ describe spin degrees of freedom, whereas $C_{1,4}$ correspond to doublon--holon degrees of freedom---the distinction between these degrees of freedom is in fact encoded in the Grassmann variables introduced to enforce the correct fermionic parity.

We begin with the action given by Eq.~\eqref{eq:Seff} with the constraints $C_1 = C_\uparrow$ and $C_\downarrow = C_4$ imposed:
\begin{align}
    &\mathscr S_{\rm eff} =  \sum_i \int_0^\beta \diff \tau \Big[ \la \al_i | \partial_\tau | \al_i \ra + \bar \psi_i \partial_\tau \psi_i
    \nonumber \\
    &\ \ 
    - \mu \bar \psi_i \psi_i 
    + (U-2\mu) (1-\bar\psi_i\psi_i) |C_\da(\Omega_i)|^2
    \Big] 
    \nonumber \\
    &\ 
    - t \int_0^\beta \diff \tau \sum_{\la ij \ra \sg}
    \Big[
    (Y_{i\sg}^*\psi_i + X_{i\sg}^* \bar \psi_i) (X_{j\sg}\psi_j + Y_{j\sg} \bar \psi_j)
    +{\rm conj.}
    \Big] ,
\end{align}
where $X$ and $Y$ are now written by $C_\sg$ as
\begin{equation}
\begin{aligned}
    &X_{i\ua} = |C_\ua(\Omega_i)|^2,
    \\
    &X_{i\da} = C_\ua^*(\Omega_i) C_\da(\Omega_i),
    \\
    &Y_{i\ua} = |C_\da(\Omega_i)|^2 = 1- X_{i\ua},
    \\
    &Y_{i\da} = - C_\ua^*(\Omega_i) C_\da(\Omega_i) = - X_{i\da}.
\end{aligned}
\end{equation}
It is notable that the above effective action can be reproduced by the following Hamiltonian
\begin{align}
&\mathscr H_{\rm eff} = -\mu \sum_i d_i^\dg d_i
+ (U-2\mu) \sum_i (1-d_i^\dg d_i) (\tfrac 1 2 - Q_i^z)
\nonumber \\
&\hspace{-2mm}
- t \sum_{\la ij \ra} 
\Big\{
\qty[ (\tfrac 1 2 \!-\! Q_i^z) d_i \!+\! (\tfrac 1 2 \!+\! Q_i^z) d_i^\dg ]
\qty[ (\tfrac 1 2 \!+\! Q_j^z) d_j \!+\! (\tfrac 1 2 \!-\! Q_j^z) d_j^\dg ]
\nonumber \\
&\ \ \ 
+ Q_i^- (-d_i + d_i^\dg) Q_j^+ (d_j - d_j^\dg)
\Big\}
+ {\rm H.c.} ,
\label{eq:dS_ham}
\end{align}
where $\bm{Q}_i$ is a localized pseudospin operator with spin $1/2$, distinct from the spin operators introduced previously.
If one approximates the quantum-mechanical operators by classical spins in the atomic limit ($t = 0$), the resulting partition function is equivalent to that in Eq.~\eqref{eq:atomic_z}.
Thus, our semiclassical scheme is associated with the Hubbard model after the nonlinear transformation in Eq.~\eqref{eq:dS_ham}, rather than the original one in Eq.~\eqref{eq:hubbard_original}.

Since the fermionic operators such as $d_i-d_i^\dg$ and $d_i+d_i^\dg$ are involved, the above form of the Hamiltonian suggests the introduction of Majorana fermions. 
More specifically, we define the Majorana fermion operators as
\begin{align}
\gm_{1i} &= d_i + d_i^\dg ,
\\
\gm_{2i} &= (d_i - d_i^\dg)/\imu  .
\end{align}
Using the relation
\begin{align}
    d_i^\dg d_i &= \frac 1 2 ( 1+ \imu \gm_{1i} \gm_{2i}) ,
\end{align}
we obtain 
\begin{align}
&\mathscr H_{\rm eff} 
= - \frac{U}{4} \sum_i \imu \gm_{1i} \gm_{2i}
- \frac{(U-2\mu)}{2} \sum_i (1 - \imu \gm_{1i} \gm_{2i} ) Q_i^z
\nonumber \\
&\ \ 
- t \sum_{\la ij \ra} 
\Big[
\imu \gm_{1i} \gm_{2j} Q_j^z - \imu \gm_{2i} \gm_{1j} Q_i^z
+ 2\imu \gm_{2i} \gm_{2j} (\bm Q_i\times \bm Q_j)^z
\Big] ,
\end{align}
where a constant term has been neglected.
We refer to this model as the Majorana–Kondo lattice, since the itinerant Majorana fermions $(\gamma_{1i}, \gamma_{2i})$ are coupled to quantum localized spin $1/2$ spins $\bm{Q}_i$. 
In this formulation, a three-body interaction emerges, involving two neighboring localized spins $(\bm{Q}_i, \bm{Q}_j)$ and the mobile Majorana fermions through the bilinear $\gamma_{2i} \gamma_{2j}$.

\subsection{Majorana-Kondo to Hubbard}

We can further rewrite the Hamiltonian for the Majorana-Kondo lattice in the previous subsection.
We introduce the Majorana representation of the quantum spins \cite{Tsvelik_book}:
\begin{align}
    Q_i^a &= - \frac{\imu}{4} \sum_{bc} \epsilon_{abc}  \gm_{b i} \gm_{c i} ,
\end{align}
where $a,b,c = x,y,z$.
The Hamiltonian is then given by
\begin{align}
\mathscr H_{\rm eff} 
&= - \frac{U}{4} \sum_i \imu \gm_{1i} \gm_{2i}
+ \frac{(U-2\mu)}{4} \sum_i (1 - \imu \gm_{1i} \gm_{2i} ) \imu \gm_{xi} \gm_{yi}
\nonumber \\
&\ \ \ 
- \frac{t}{2} \sum_{\la ij \ra} 
\Big[
\gm_{1i} \gm_{2j} \gm_{xj} \gm_{yj} - \gm_{2i} \gm_{1j} \gm_{xi} \gm_{yi}
\nonumber \\
&\hspace{14mm}
+ \imu \gm_{2i} \gm_{2j} \gm_{zi}\gm_{zj} (
\gm_{xi}\gm_{yj} 
- \gm_{yi}\gm_{xj} 
)
\Big] .
\end{align}
Now we define the composite fermion operator by
\begin{align}
    \Gamma_{a i} &= -2 \gm_{2i} Q_i^a
= \frac{\imu}{2} \sum_{bc} \epsilon_{abc} \gm_{2i} \gm_{b i} \gm_{c i} .
\label{eq:non-linear_Majorana}
\end{align}
There are the relations
\begin{equation}
\begin{aligned}
&\Gamma_{\mu i}^2  =  1 ,
\\
&\Gamma_{\mu i}^\dg  =  \Gamma_{\mu i} ,
\\
&\Gamma_{xi} \Gamma_{yi} = \gm_{xi} \gm_{yi} ,
\\
&\Gamma_{xi} \Gamma_{yi} \Gamma_{zi} = - \imu \gm_{2i} .
\end{aligned}
\end{equation}
Namely, $\gm_{2i}$ is represented by a product of $\Gamma_{a i}$.
The Hamiltonian then becomes
\begin{align}
&\hspace{-4mm}\mathscr H_{\rm eff} = \frac{U}{4} \sum_i \gm_{1i} \Gamma_{xi} \Gamma_{yi} \Gamma_{zi} + \frac{(U\!-\!2\mu)}{4} \sum_i (\imu \Gamma_{xi}\Gamma_{yi} - \imu \gm_{1i} \Gamma_{zi})
\nonumber \\
&\hspace{-2mm}  - \frac t 2 \sum_{\la ij \ra} \qty( -\imu \gm_{1i} \Gamma_{zj} + \imu \Gamma_{zi}\gm_{1j} + \imu \Gamma_{xi}\Gamma_{yj} - \imu \Gamma_{yi} \Gamma_{xj} )
.
\end{align}
Here, only the four kinds of Majorana operators $\gm_{1i}$ and $\Gamma_{a i}$ appear, and they satisfy the standard fermionic commutation relations, and the composite nature of $\Gamma_{ai}$ need not be taken into account apparently.
Let us now {\it define} the (complex) fermion operators by
\begin{equation}
\begin{aligned}
    c_{\ua i} &=( \Gamma_{zi} + \imu \gm_{1i} )/2 ,
    \\
    c_{\da i} &= ( \Gamma_{xi} + \imu \Gamma_{yi} )/2 .
\end{aligned}
\end{equation}
Then, we finally obtain
\begin{align}
    &\mathscr H_{\rm eff} = \frac U 4 \sum_{i} (2n_{\ua i}-1) (2n_{\da i}-1)
    \nonumber \\
    &
    + \frac{(U-2\mu)}{4} \sum_{i\sg} (2n_{\sg i} - 1)
    - t \sum_{\la ij \ra} (c_{\sg i}^\dg c_{\sg j} + c_{\sg j}^\dg c_{\sg i} )
    ,
\end{align}
which is shown to be equivalent to the original Hubbard model in Eq.~\eqref{eq:hubbard_original}.
This kind of Majorana representation of interactions has been considered in Refs.~\cite{Coleman95,Nilsson11,Nilsson14}.
A key step in deriving the Majorana--Kondo lattice Hamiltonian from the Hubbard model is the use of the nonlinear Majorana transformation defined in Eq.~\eqref{eq:non-linear_Majorana}.
We note, however, that even with this transformation, one cannot obtain the path-integral representation in Eq.~\eqref{eq:Seff}, which has no corresponding Hamiltonian formulation.

The explicit mapping from spinful fermions to a representation in terms of a localized pseudospin and a $d$ fermion is given by
\begin{equation}
\begin{aligned}
    Q_i^z &= \frac{1}{2} (1 - 2 n_{i\downarrow}), \\
    Q_i^x &= \frac{1}{2} \Bigl(
        c_{i\downarrow}^\dagger c_{i\uparrow} - c_{i\downarrow} c_\uparrow^\dagger
        + c_{i\downarrow}^\dagger c_{i\uparrow}^\dagger - c_{i\downarrow} c_{i\uparrow}
    \Bigr), \\
    Q_i^y &= \frac{-i}{2} \Bigl(
        c_{i\uparrow}^\dagger c_{i\downarrow} + c_{i\uparrow} c_{i\downarrow}^\dagger
        + c_{i\uparrow}^\dagger c_{i\downarrow}^\dagger + c_{i\uparrow} c_{i\downarrow}
    \Bigr), \\
    d_i &= i \Bigl[
        c_{i\uparrow}^\dagger n_{i\downarrow} + c_{i\uparrow} (n_{i\downarrow} - 1)
    \Bigr],
\end{aligned}
\end{equation}
which is identical to the transformation introduced in Ref.~\cite{Ostlund06}. 
Thus, we rederive the nonlinear fermionic transformation based on unconventional coherent states in the path-integral formalism.
An overview of the various representations of the Hubbard model discussed in this paper is summarized in Fig.~\ref{fig:overview}.

In principle, the representation of a model is immaterial when the problem can be solved exactly. In practice, however, exact solutions are rare, and one must rely on approximate schemes, such as semiclassical or mean-field approaches. In this context, the choice of representation can shape the range of physical phenomena that are accessible. It is therefore an open and interesting question to determine which aspects of solid-state systems can be captured when such approximations are formulated within the representation introduced here.

\section{Summary and Outlook}
\label{sec:summary}

By reconsidering the structure of fermionic coherent states, we have
formulated a semiclassical approximation scheme for the Hubbard model
based on a graded coherent-state representation of the local Hilbert
space.  The construction is guided by the principle of introducing only
a single Grassmann variable per site, which enforces the $\mathbb Z_2$
fermion-parity structure while encoding the remaining local degrees of
freedom in bosonic variables describing the spin and charge sectors.
The resulting coherent-state manifold is therefore
$S^2_{\rm spin}\times S^2_{\eta}$ rather than the full projective space
$\mathbb{CP}^3$ of the local Hilbert space.

To assess the validity of the proposed semiclassical approximation, we
have compared its predictions with exact solutions for the one-site and
two-site Hubbard models.  The semiclassical results reproduce the
qualitative features of the exact solutions, including their
temperature and filling dependences.  Just as classical spin models
provide useful approximations to quantum spin systems, we expect that
the present formulation can serve as a practical framework for studying
Hubbard-type models in regimes where exact analyses are intractable.

An important feature of the present formulation is that the classical
variables naturally encode both spin and charge degrees of freedom.
This makes the approach particularly well suited for semiclassical
descriptions of intertwined orders involving simultaneous spin and
charge symmetry breaking, such as the coexistence of magnetic order
with superconducting or charge-density-wave correlations.

The unconventional coherent states employed here also provide an
alternative representation of the Hubbard model.  By mapping the
constructed path-integral representation back to operator form, we
obtain transformations involving nonlinear fermionic operators that
naturally map the model onto a Kondo-lattice-like system composed of
itinerant fermions coupled to localized spins.

Several directions for future work are suggested by the present
framework.  A particularly interesting question is whether
$d$-wave superconductivity can be captured within this semiclassical
description.  Extending the approach to multi-site and two-dimensional
systems may involve Kosterlitz–Thouless–type physics associated with
vortex–antivortex excitations, and efficient numerical implementations
will likely be required to explore these regimes.

Another promising direction is the extension to nonequilibrium systems.
Semiclassical descriptions of out-of-equilibrium dynamics have already
proven successful in localized-electron systems
\cite{Dahlbom22,Pohle25}, and it is therefore natural to expect that the
present graded coherent-state formulation can be applied to study the
nonequilibrium dynamics of correlated fermionic systems.

Finally, the approach developed here can be extended to multi-orbital
systems and combined with first-principles electronic structure calculation
methods.  Although the resulting description is approximate, it offers
a computationally inexpensive framework capable of capturing the
essential qualitative physics of strongly correlated materials, thereby
providing guidance for more accurate theoretical and experimental
investigations.

\section*{Acknowledgement}

We would like to thank K. Hattori, R. Iwazaki, T. Miki, H. Shinaoka, and Y. Michishita
for useful discussions.
This work was supported by the
KAKENHI Grants No.~23K17668, No.~23K25827, No.~24K00578, No.~JP24K00583, and
by the Grant-in-Aid for Transformative Research Areas
(A) ``Correlation Design Science'' (KAKENHI Grant No.
25H01249) from JSPS of Japan. 
C.D.B. was supported by the US Department of Energy, Office of Science, Basic Energy Sciences, Materials Sciences and Engineering Division under Award No. DE-SC0018660.

\appendix

\section{Solution of a one-body problem}
\label{sec:solution_one_body}

\subsection{General formalism}

In evaluating partition function, we need to solve a generic one-body problem.
Below, we show the method of solution for the general one-body problem of both fermions and bosons in a unified manner.

In second quantization form, the creation and annihilation operators satisfy
\begin{align}
    &c_i c_j^\dagger - s c_j^\dagger c_i = \delta_{ij} ,
    \\
    &c_i c_j - s c_j c_i = 0 ,
\end{align}
with $i,j=1,\cdots, N$.
The sign $s=+$ is taken for bosons and $s=-$ for fermions. 
Let us consider the non-interacting Hamiltonian
\begin{align}
    H &= \sum_{i,j=1}^N \qty(
    E_{ij} c_i ^\dagger c_j + 
    \frac{1}{2} W_{ij} c_i^\dagger c_j^\dagger
    + \frac{1}{2} W_{ij}^* c_j c_i
    ) ,
\end{align}
where $E_{ij}^* = E_{ji} $ and $W_{ij} = sW_{ji}$ to ensure the Hermiticity of the Hamltonian $H^\dagger = H$.

We solve the problem by using a Green's function method.
First, we define the Heisenberg picture with imaginary time:
\begin{align}
    A(\tau) = e^{\tau H} A e^{-\tau H} ,
\end{align}
which satisfies
\begin{align}
    - \partial_\tau A(\tau) &=  [A,H](\tau) ,
    \\
    A(\tau)^\dagger &= A^\dagger (-\tau) .
\end{align}
The Heisenberg equations of motion for annihilation and creation operators are given by
\begin{align}
    -\partial_\tau c_i(\tau) &= \sum_j \qty[ E_{ij} c_j(\tau) + W_{ij} c_j^\dagger(\tau) ],
    \\
    -\partial_\tau c_i^\dagger(\tau) &= \sum_j \qty[ s E_{ji} c_j^\dagger(\tau) + W_{ji}^* c_j(\tau)  ]
    .
\end{align}
Now we define the $2N\times 2N$ Green's function matrix by
\begin{align}
    \check G(\tau) &= - \langle \mathcal T \Psi (\tau) \Psi^\dagger \rangle  ,
    \\
    \Psi &= \begin{pmatrix}
        c_1, \cdots, c_N, c_1^\dagger, \cdots , c_N^\dagger
    \end{pmatrix}^{\rm T} .
\end{align}
Here, the bracket denotes the grand-canonical ensemble average, and
$\beta = 1/T$ is the inverse temperature.
It follows that
$    \check{G}(\tau + \beta) = s\,\check{G}(\tau)$.
The Fourier transformation is defined by
\begin{align}
    \check G(i \omega_n) &= \int_0^\beta \diff  \tau \, \check G(\tau) \, e^{i \omega_n \tau},
    \\
    \check G(\tau) &= T \sum_n \check G(i\omega_n) \, e^{ - i \omega_n \tau} ,
\end{align}
where $\omega_n$ is the Matsubara frequency: $\omega_n = 2n \pi /\beta$ for boson and $\omega_n = (2n+1)\pi /\beta $ for fermion.
We thereby obtain
\begin{align}
    &\check G(i\omega_n)^{-1} = i \omega_n \check 1 - \check H ,
    \\
    &\check H = \begin{pmatrix}
        \hat E & \hat W \\
        \hat W^\dagger & s\hat E^{\rm T}
    \end{pmatrix}
    = \check V \check \Lambda \check V^{\dagger} ,
    \label{eq:diagonalization}
\end{align}
where we have defined the $2N\times 2N$ matrix $\check H$.
On the right-hand side of Eq.~\eqref{eq:diagonalization}, the Hamiltonian matrix is diagonalized to obtain the diagonal matrix $\check{\Lambda}$ and the matrix of eigenvectors $\check{V}$.
Now we use the formula
\begin{align}
    T\sum_n \frac{e^{i \omega_n 0^+}}{i\omega_n - x}
    = \frac{-s}{e^{\beta x} - s}
    ,
\end{align}
and then the equal time correlation is obtained as
\begin{align}
    \langle \Psi_k^\dagger \Psi_l  \rangle  &= - s G_{lk}(\tau=-0)
    = \sum_{\alpha=1}^{2N}V_{l\alpha} V^\dagger_{\alpha k}
    f(\Lambda_\alpha) ,
    \\
    f(x) &= \frac{1}{e^{\beta x} - s} ,
\end{align}
where $k,l=1,\cdots, 2N$ (including Nambu space).
Thus, diagonalizing the Hermitian matrix $\check{H}$ of size $2N \times 2N$ allows one to explicitly evaluate expectation values of physical quantities.
Note that originally the matrix representation of $H$ has the size of $2^N\times 2^N$.
Hence the matrix size is effectively much reduced, which is a characteristic of one-body problem with bilinear form of creation/annihilation operators.

Let us now derive the partition function $Z$ from the internal energy.
The Hamiltonian is rewritten as
\begin{align}
    H &= \frac 1 2 \Psi^\dagger \check H \Psi + \frac 1 2 {\rm Tr\,}\hat E .
\end{align}
The internal energy is given by
\begin{align}
    \langle H \rangle 
    &= \frac{1}{Z} \sum_x \langle x | H e^{-\beta H} |x\rangle
    = - \frac{\partial }{\partial \beta }\ln Z
    \\
    &= \frac 1 2 \sum_{\alpha=1}^{2N} \Lambda_\alpha f(\Lambda_\alpha)
     + \frac 1 2 {\rm Tr\,}\hat E ,
\end{align}
where $\sum_x 1 = 2^N$.
The partition function is then obtained by integrating the internal energy as
\begin{align}
    Z &= 
    \exp \qty( - \frac \beta 2 {\rm Tr\,}\hat E ) \prod_{\alpha=1}^{2N} \qty(1 - s e^{-\beta \Lambda_\alpha})^{-s/2} .
\end{align}
In the above formalism, the Hilbert space, i.e., the Fock-space representation of the basis states, is not explicitly involved.

\subsection{
Exact spectrum of the fermionic two-site quadratic Hamiltonian}
\label{sec:solution_one_body_2}

While in the above we derive a generic problem with $N$ degrees of feedom, the $N=2$ case for fermions can be more intuitively analyzed.
Specifically,
we derive Eq.~\eqref{eq:Z_two_site_HF} by computing the exact
two-site partition function of the quadratic $d$-fermion problem and expanding
$\ln Z$ to second order in the hopping $T$ and pairing $\Delta$ amplitudes.

We consider the Hamiltonian
\begin{equation}
H_{\rm eff}
=
-\varepsilon\,(n_1+n_2)
+
\Big(
T\, d_1^\dagger d_2 + T^\ast d_2^\dagger d_1
+\Delta\, d_1^\dagger d_2^\dagger + \Delta^\ast d_2 d_1
\Big),
\label{eq:Heff_app}
\end{equation}
where $n_i=d_i^\dagger d_i$.
The Hilbert space has four basis states
$
|0\rangle,\;
|1\rangle=d_1^\dagger|0\rangle,\;
|2\rangle=d_2^\dagger|0\rangle,\;
|12\rangle=d_1^\dagger d_2^\dagger|0\rangle.
$
Since $H_{\rm eff}$ preserves fermion parity, it splits into two $2\times2$ blocks:

\paragraph{Odd-parity block \texorpdfstring{$\{|1\rangle,|2\rangle\}$}{\{|1>,|2>\}}.}
Pairing terms vanish and one finds
\begin{equation}
H_{\rm odd}
=
\begin{pmatrix}
-\varepsilon & T \\
T^\ast & -\varepsilon
\end{pmatrix},
\ \ \ 
E_{\rm odd}^{\pm}=-\varepsilon\pm |T|.
\label{eq:Hodd_app}
\end{equation}

\paragraph{Even-parity block \texorpdfstring{$\{|0\rangle,|12\rangle\}$}{\{|0>,|12>\}}.}
Hopping terms vanish and one finds
\begin{equation}
H_{\rm even}
=
\begin{pmatrix}
0 & \Delta \\
\Delta^\ast & -2\varepsilon
\end{pmatrix},
\ \ 
E_{\rm even}^{\pm}
=
-\varepsilon\pm \sqrt{\varepsilon^2+|\Delta|^2}.
\label{eq:Heven_app}
\end{equation}
Therefore the exact partition function is
\begin{align}
&Z(T,\Delta)
\equiv
\Tr\,e^{-\beta H_{\rm eff}}
=
\sum_{s=\pm} e^{-\beta E_{\rm odd}^{s}}
+
\sum_{s=\pm} e^{-\beta E_{\rm even}^{s}}
\nonumber\\
&=
2e^{\beta\varepsilon}\cosh(\beta|T|)
+
2e^{\beta\varepsilon}\cosh\!\Big(\beta\sqrt{\varepsilon^2+|\Delta|^2}\Big).
\label{eq:Zexact_app}
\end{align}
Specifically, the unperturbed result at $T=\Delta=0$ is
\begin{equation}
Z(0,0)
=
\big(1+e^{\beta\varepsilon}\big)^2,
\label{eq:Z0_app}
\end{equation}
which reproduces the partition function of the two isolated fermion systems.

\bibliography{hubbard}

\end{document}